\documentclass[a4paper,12pt]{article}
\pdfoutput=1
\usepackage{amssymb}
\usepackage{amsmath}
\usepackage{epsfig}
\usepackage{slashed}
\usepackage{lineno}
\usepackage{setspace}
\newcommand{\dbl}{}

\usepackage[top=15mm,bottom=20mm,left=15mm,right=15mm]{geometry}
\usepackage[backend=bibtex]{biblatex}
\AtEveryBibitem{\clearfield{doi}}
\AtEveryBibitem{\clearfield{url}}

\newcommand{\be}{\begin{equation}}
\newcommand{\ee}{\end{equation}}
\newcommand{\bea}{\begin{eqnarray}}
\newcommand{\eea}{\end{eqnarray}}

\newcommand{\authors}{
\noindent
Sz.\ Borsanyi$^{1}$,
Z.\ Fodor$^{1,2,3}$,
K.-H.\ Kampert$^{1}$,
S.\ D.\ Katz$^{3,4}$, 
T. Kawanai$^{2}$,
T.\ G.\ Kovacs$^{5}$,
S.\ W.\ Mages$^{2}$,
A.\ Pasztor$^{1}$,
F.\ Pittler$^{3,4}$,
J.\ Redondo$^{6,7}$,
A.\ Ringwald$^{8}$,
K.\ K.\ Szabo$^{1,2}$
\vspace*{2cm}

\noindent
$^{1}$\ Department of Physics, University of Wuppertal, D-42119 Wuppertal, Germany\\
$^{2}$\ J\"ulich Supercomputing Centre, Forschungszentrum J\"ulich, D-52428 J\"ulich, Germany\\
$^{3}$\ Institute for Theoretical Physics, E\"otv\"os University, H-1117 Budapest, Hungary\\
$^{4}$\ MTA-ELTE Lend\"ulet Lattice Gauge Theory Research Group, H-1117 Budapest, Hungary\\
$^{5}$\ Institute for Nuclear Research of the Hungarian Academy of Sciences, H-4026 Debrecen, Hungary\\
$^{6}$\ University of Zaragoza, E-50009 Zaragoza, Spain\\
$^{7}$\ Max Planck Institut f\"ur Physik, D-80803, Germany\\
$^{8}$\ Deutsches Elektronen-Synchrotron DESY, D-22607 Hamburg, Germany
\vspace*{3cm}
}

\newcommand{\arxabstract}{
\noindent We present a full result for the equation of state (EoS) in 2+1+1 
(up/down, strange and charm quarks are present) flavour lattice QCD. We 
extend this analysis and give the equation of state in 2+1+1+1 flavour 
QCD. In order to describe the evolution of the universe from 
temperatures several hundreds of GeV to the MeV scale we also 
include the known effects of the electroweak theory and give the 
effective degree of freedoms. As another application of lattice QCD we 
calculate the topological susceptibility ($\chi$) up to the few GeV 
temperature region. These two results, EoS and $\chi$, can be used to 
predict the dark matter axion's mass in the post-inflation scenario 
and/or give the relationship between the axion's mass and the universal 
axionic angle, which acts as a initial condition of our universe.

}

\begin{document}

\dbl 

\thispagestyle{empty}
\hspace*{14cm}DESY 16-105
\vspace*{1cm}
\begin{center}
{\huge\bf Lattice QCD for Cosmology}
\end{center}
\vspace*{2cm}

\authors

\arxabstract

\clearpage
\begin{refsection}


{\bf The well-established theories of the strong interaction (QCD) and 
the electroweak theory determine the evolution of the early universe. 
The Hubble rate and the relationship between the age of the 
universe and its temperature ($T$) are determined by the equation of 
state (EoS). Since QCD is highly non-perturbative, the calculation of the 
EoS is a particularly difficult task. The only systematic way to carry 
out this calculation is based on lattice QCD. Here we present complete 
results of such a calculation. QCD, unlike the rest of the Standard 
Model, is surprisingly symmetric under time reversal, leading to a 
serious fine tuning problem. The most attractive solution for this is a 
new particle, the axion --a promising dark matter candidate. 
Assuming that axions are the dominant component of dark matter we 
determine the axion mass. The key quantities of the calculation 
are the previously mentioned EoS and the temperature dependence of the 
topological susceptibility ($\chi(T)$) of QCD, a quantity 
notoriously difficult to calculate. Determining $\chi(T)$ 
was believed to be impossible in the most 
relevant high temperature region, however an understanding of the deeper 
structure of the vacuum by splitting it into different sectors and 
re-defining the fermionic determinants has led to its fully controlled 
calculation. Thus, our two-fold prediction helps most 
cosmological calculations to describe the evolution of the early 
universe by using the fully controlled EoS and may be decisive for 
guiding experiments looking for dark matter axions. In the next couple 
of years, it should be possible to confirm or rule out
post-inflation axions experimentally~\cite{SI} if the axion's mass is or is not found to be 
as predicted here. Alternatively, in a pre-inflation scenario our calculation 
determines the universal axionic angle that corresponds to 
the initial condition of our universe. }

The Euclidean Lagrangian for the theory of the strong interaction is 
${\cal L}_{\rm QCD} = 1/(2g^2)\rm{Tr}F_{\mu\nu}F_{\mu\nu} + \sum_f{\bar 
\psi_f}[\gamma_\mu (\partial_\mu + i A_\mu) + m_f ] \psi_f$, where $g$ 
is the QCD coupling constant, $m_f$ represent the quark masses and $f$ 
runs over the four quark flavors.
In QCD $F_{\mu\nu}=\partial_\mu A_\nu - 
\partial_\nu A_\mu+[A_\mu,A_\nu]$, where $A_\mu$ is a Hermitian 
$3{\times}3$ matrix field denoting the gauge fields. The $\psi_f$ are 
Dirac-spinor fields representing the quarks and carry a ``colour'' 
index, which runs from 1 to 3. The form of the action is (almost) 
unambiguously defined by the condition of gauge invariance and 
renormalizability. For many quantities, the only systematic way to solve 
this theory is to discretize it on a Euclidean space-time lattice, 
determine the partition function stochastically and using smaller and 
smaller lattice spacings to extrapolate to the continuum limit (the limit 
with vanishing lattice spacing). In this paper, we use this lattice 
formulation to determine the EoS of QCD and the topological 
susceptibility for low temperatures up to very high temperatures. 
Finally, as an application, we combine them to present our results on 
the dark matter axion mass.

Our most important qualitative knowledge about the QCD transition is 
that it is an analytic crossover~\cite{Aoki:2006we}, thus no 
cosmological relics are expected. Outside the narrow temperature range 
of the transition we know that the Hubble rate and the relationship 
between temperature and the age of the early universe can be described by 
a radiation-dominated EoS. The calculation of the EoS 
is a challenging task~\cite{SI}, the determination of the 
continuum limit at large temperatures is particularly difficult.

In our lattice QCD setup we used 2+1 or 2+1+1 flavours of staggered 
fermions with four steps of stout-smearing. For the gluonic sector 
tree-level improved gauge fields were used. The quark masses are set to 
their physical values, however we use degenerate up and down quark 
masses and the small effect of isospin breaking is included analytically. 
The continuum limit is taken using three, 
four or five lattice spacings with temporal lattice extensions of 
$N_T$=6, 8, 10, 12 and 16. In addition to dynamical staggered 
simulations we also used dynamical overlap simulations with 2+1 flavours 
down to physical masses. The inclusion of an odd number of flavours was 
a non-trivial task, however this setup was required for the 
determination of $\chi(T)$ at large temperatures in the several GeV 
region~\cite{SI}.

Charm quarks start to contribute to the equation of state above 300~MeV. 
Therefore up to 250~MeV we used 2+1 flavours of dynamical quarks. 
Connecting the 2+1 and the 2+1+1 flavour results at 250~MeV can be done 
smoothly. For large temperatures the step-scaling method for the 
equation of state of Ref.~\cite{Borsanyi:2012ve} was applied. We 
determined the EoS with complete control over all sources of systematics 
all the way to the GeV scale.

Two different methods were used to set the overall scale in order to 
determine the equation of state. One of them took the pion decay 
constant the other applied the $w_0$ scale~\cite{Borsanyi:2012zs}. 32
different analyses (e.g. the two different scale setting 
procedures, different 
interpolations, keeping or omitting the coarsest lattice)  
entered our histogram 
method~\cite{Durr:2008zz,Borsanyi:2014jba} to estimate systematic 
errors. We also calculated the goodness of the fit Q and 
weights based on the Akaike information criterion 
AICc~\cite{Borsanyi:2014jba} and we looked at the unweighted or weighted 
results. This provided the systematic errors on our findings. In the low 
temperature region we compared our results with the prediction of 
the Hadron 
Resonance Gas (HRG) approximation and found perfect agreement. This HRG 
approach is used to parametrize the equation of state for small 
temperatures. In addition, we used the hard thermal loop 
approach~\cite{Andersen:2010wu} to extend the EoS to high
temperatures.

In order to have a complete description of the thermal evolution of the early 
universe we supplement our QCD calculation for the EoS by 
including the rest of the Standard Model particles 
(leptons, bottom and top quarks, $W$, $Z$, Higgs bosons) 
and results on the electroweak transition~\cite{SI}. As a consequence, the final result on the EoS covers 
four orders of magnitude in temperature from MeV to several 
hundred GeV.

\begin{figure}
    \centering
    \includegraphics*[height=10cm,angle=-90]{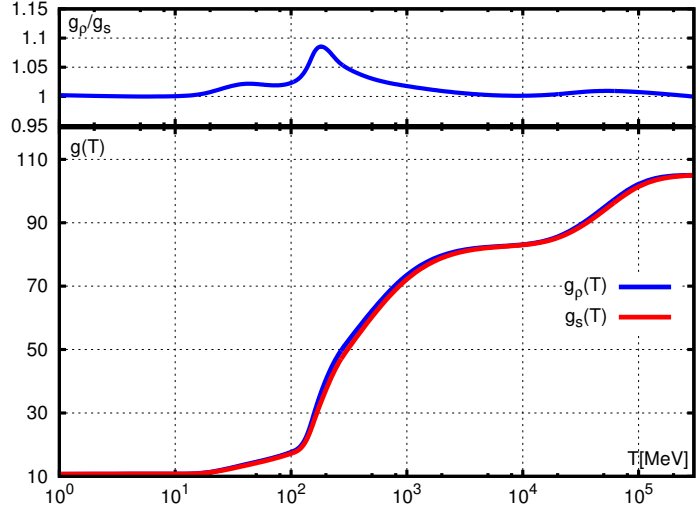}\hspace*{.1cm}
    \caption
    {\dbl
	\label{fi:eos} The effective degrees of freedom for the energy density
	($g_\rho$) and for the entropy density ($g_s$).  The line width is
	chosen to be the same as our error bars at the vicinity of the QCD
	transition where we have the largest uncertainties. At temperatures
	$T<1$~MeV the equilibrium equation of state becomes irrelevant for
	cosmology, because of neutrino decoupling. The EoS comes from our
	calculation up to $T=100$~GeV. At higher temperatures the electroweak
	transition becomes relevant and we use the results of
	Ref.~\cite{Laine:2015kra}.  Note that for temperatures around the QCD
	scale non-perturbative QCD effects reduce $g_\rho$ and $g_s$ by 10-15\%
	compared to the ideal gas limit, an approximation which is often used
	in cosmology. For useful parametrizations 
        for the QCD regime or for the whole temperature range see~\cite{SI}.
    }
\end{figure}

Figure~\ref{fi:eos} shows the result for the effective numbers of degrees 
of freedom as a function of temperature. The widths of the lines 
represent the uncertainties. The tabulated data are also presented 
in~\cite{SI}.
Both the figure and the 
data can be used (similarly to Figure~22.3 of 
Ref.~\cite{Agashe:2014kda}) to describe the Hubble rate and the 
relationship between temperature and the age of the universe in a very 
broad temperature range.

We now turn to the determination of another 
cosmologically important quantity, $\chi(T)$. 
In general the Lagrangian of QCD should have a term
proportional to ${\cal L}_{\rm 
Q}=1/(32\pi^2) \epsilon_{\mu\nu\rho\sigma}F_{\mu\nu}F_{\rho\sigma}$, the 
four-dimensional integral of which is called the topological charge. This 
term violates the combined charge-conjugation and parity symmetry (CP). 
The surprising experimental observation is that the proportionality 
factor of this term $\theta$ is unnaturally small. 
This is known as the strong CP problem.
A particularly 
attractive solution to this fundamental problem is the so-called 
Peccei-Quinn mechanism~\cite{Peccei:1977hh}. One introduces an additional (pseudo-)scalar 
U(1) symmetric field. The underlying Peccei-Quinn U(1) symmetry is 
spontaneously broken --which can happen pre-inflation or 
post-inflation-- and an axion field $A$ acts as a (pseudo-)Goldstone 
boson of the broken symmetry \cite{Weinberg:1977ma,Wilczek:1977pj}. Due to the 
chiral anomaly the axion also couples to ${\cal L}_{\rm Q}$. As a 
consequence, the original potential of the axion field with its U(1) 
symmetry breaking gets tilted and has its minimum where 
$(\theta+A/f_A)=0$. This sets the coefficient in front of ${\cal L}_{\rm 
Q}$ to zero and solves the strong CP problem. Furthermore, the free 
parameter $f_A$ determines the mass of the axion at vanishing or 
non-vanishing temperatures by $m_A^2=\chi/f_A^2$. Here $\chi$ is the 
susceptibility of the topological charge. We determined its value at 
$T=0$, which~\cite{SI} turned out to be 
$\chi(T=0)=0.0245(24)(12)/{\rm fm}^4$ in the isospin symmetric case,
where the first error is statistical, the second is systematic. For the 
error budget see~\cite{SI}. Isospin breaking results in
a small, 12\% correction~\cite{SI}, thus the physical value is
$\chi(T=0)=0.0216(21)(11)/{\rm fm}^4=[75.6(1.8)(0.9) {\rm MeV}]^4$.

In an earlier study of ours \cite{Borsanyi:2015cka} we looked at 
$\chi(T)$ in the quenched approximation. We provided a result within 
the quenched framework and reached a temperature about half to one third 
of the necessary temperatures for axion cosmology (a similar study with 
somewhat less control over the systematics is \cite{Berkowitz:2015aua}). 
To obtain a complete result one should use dynamical quarks with 
physical masses.  Dynamical configuration production is, however, about 
three orders of magnitude more expensive and the $\chi(T)$ values are 
several orders of magnitude smaller than in the quenched case. Due to 
cutoff effects~\cite{SI} the continuum limit is far more difficult to 
carry out in dynamical QCD than in the pure gauge theory 
\cite{Borsanyi:2015cka}. All in all we estimate that the brute-force 
approach to provide a complete result on $\chi(T)$ in the relevant 
temperature region would be at least ten orders of magnitude more 
expensive than the result of \cite{Borsanyi:2015cka}. 
Ref.~\cite{Bonati:2015vqz} used this approach, however it turned out to 
be quite difficult. As we will show our result for $\chi(T)$ is many 
orders of magnitude smaller than that of Ref.~\cite{Bonati:2015vqz} in 
the cosmologically relevant  temperature region.

Whilst writing up our results, a paper~\cite{Petreczky:2016vrs} appeared 
with findings similar to ours, for which two remarks are in order.
A common set of criteria for assessing
the reliability of lattice results
was set by FLAG and published in Ref.~\cite{Aoki:2013ldr}.
They introduced for some quantities a measure indicating how far the continuum extrapolated result is from the 
lattice data. 
Firstly, in order to control the continuum extrapolation 
they demand to have a maximum difference of a few percent between the 
finest lattice data and the continuum extrapolated result. (In other 
words, no extrapolation with large cutoff effects are allowed.) In 
Ref.~\cite{Petreczky:2016vrs} the cutoff effects on the direct 
topological susceptibility $\chi(T)$ measurements are orders of
magnitude larger than the FLAG requirement (even for 
secondary quantities such as $\chi(T)^{1/4}$ or for the exponent they 
are large), whereas here we show how to keep these cutoff effects on 
$\chi(T)$ to the ${\cal O}$(10\%) level. Secondly, FLAG requires three 
or more lattice spacings for controlled continuum extrapolation results. 
When attempting to determine $\chi(T)$ for approximately T=3~GeV 
Ref.~\cite{Petreczky:2016vrs} fulfills this condition for the continuum 
extrapolation in a narrow temperature region between 250 to 330~MeV. 
However, no direct measurements are carried out at large temperatures, 
only an extrapolation using the behaviour in this narrow range, whereas 
we show how to determine $\chi(T)$ all the way into the physically 
relevant 3~GeV region.

Since cutoff effects are huge and the CPU demand is tremendous one faces 
significant physics challenges and needs dramatic algorithmic 
developments in a large-scale simulation setup in order to provide a 
complete answer for $\chi(T)$, thus resolve the tension between e.g. 
the two References \cite{Bonati:2015vqz} and \cite{Petreczky:2016vrs} 
(or other analytic techniques).

The huge computational demand and the physics issue behind the 
determination of $\chi(T)$ has two main sources. a.) The tiny 
topological susceptibility needs extremely long simulation threads to 
observe enough changes of the topological sectors and b.) In high 
temperature lattice QCD the most widely used actions are based on 
staggered quarks. When dealing with topological observables staggered 
quarks have very large cutoff effects.

We solve both problems and determine the continuum result for 
$\chi(T)$ for the entire temperature range of interest.

{\it ad a.} We summarize our solution to problem a, which we call 
``fixed $Q$ integration'' (for a detailed discussion see 
\cite{SI}). As a first step one takes a starting-point in the quark 
mass-temperature space, at which it is possible to determine $\chi$ 
using conventional methods. One such point could be e.g. the quenched 
theory at some low temperature (below the phase transition), or 
alternatively relatively large quark masses at not too high temperatures could also be 
used in practice.  Next we determine, at this point, the weights of the 
various topological sectors in a given volume. This is done by taking 
derivatives with respect to the gauge coupling and $m_f$ and then 
by integrating these parameters one can calculate the change in the weights 
of the various topological sectors all the way to the new point. This 
provides the weights of the topological sectors and therefore $\chi$ 
in a new point. The technique is similar to the so-called integral 
method used to determine the equation of state~\cite{SI}.

The CPU costs of the conventional technique scale as $T^8$, whereas the 
new ``fixed $Q$ integration'' method scales as $T^0$. The gain in 
CPU time is tremendous. This efficient technique is used to obtain the 
final result for $\chi(T)$. Since we work with continuum extrapolated 
quantities both for the ratios in the starting-point as well as for 
their changes, one can in principle use any action in the procedure, 
we will use here overlap~\cite{SI} and/or staggered actions.

{\it ad b.} We summarize our solution to problem b, which we call 
``eigenvalue reweighting''. According to the index theorem the continuum 
Dirac operator has exact zero modes with definite chirality on Q$\neq$0 
gauge configurations. Staggered fermions lack these exact zero modes, 
which introduces large to very large lattice artefacts when determining 
$\chi(T)$. We developed a reweighting technique, which resolves this 
problem. The method is based on the determination of the lowest 
eigenvalues of the Dirac operator and using the ratios of the 
eigenvalues and the mass in a reweighting procedure. The basic idea is 
to substitute the smallest eigenvalue (the would be zero modes) for a 
given configuration of a given topological charge with its continuum 
value and reweight according to it. Note that if we have a zero mode, 
the smallest eigenvalue should be $m_f$ in the continuum with massive 
fermions, whereas for staggered quarks it can be significantly higher. 
We correct for this by reweighting the configurations with the ratio of 
the lowest continuum and the corresponding staggered eigenvalue. The 
topological charge is measured by the Wilson-flow method.
The details and a discussion on the validity of the 
technique is presented in ~\cite{SI}. Overlap fermions do not
suffer from this problem and have exact zero modes. Even though
they are computationally too expensive to be used for the entire project
(e.g. the integration in the gauge coupling requires tens of millions of 
trajectories~\cite{SI}) the mass integration down to the physical
point has been done with overlap quarks. 

\begin{figure}[h]
    \centering
    \includegraphics*{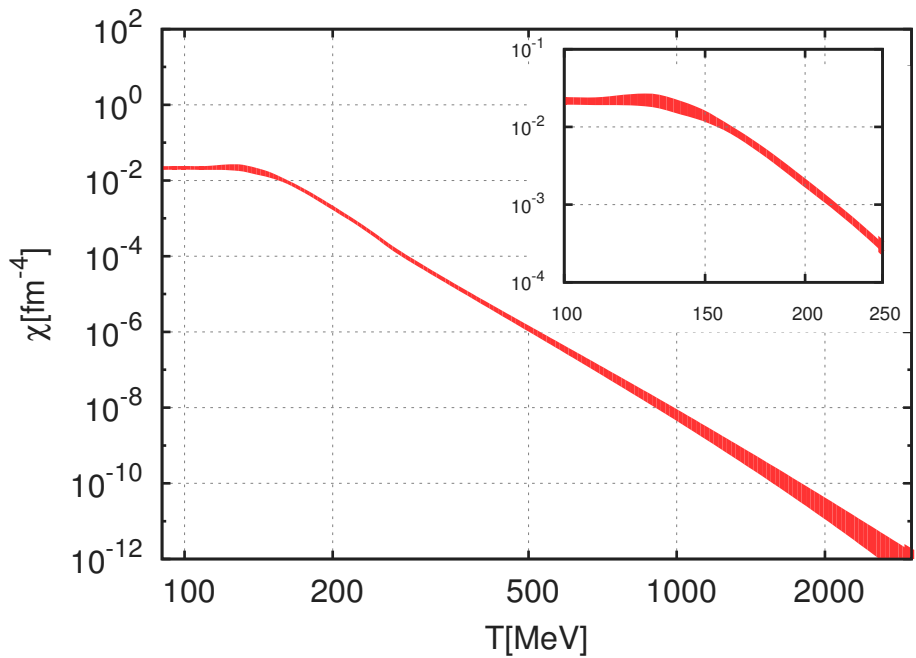}
    \caption
    {\dbl
	\label{fi:chi}
	Continuum limit of 
	$\chi(T)$. The insert shows the behaviour around the transition 
	temperature. The width of the line represents the 
        combined statistical and systematic errors. 
        The dilute instanton gas approximation (DIGA) predicts a 
	power behaviour of $T^{-b}$ with $b$=8.16, which is
	confirmed by the lattice result for temperatures above $\sim 1$ GeV.
    }
\end{figure}

\begin{figure}[h]
    \centering
    \includegraphics*{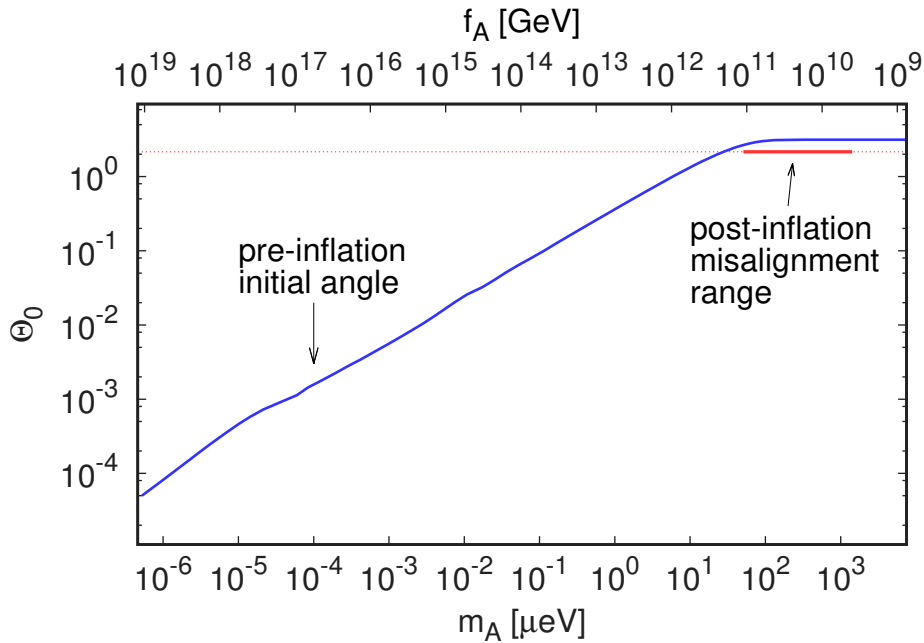}
    \caption
    {\dbl
	\label{fi:axion} Relation between the axion's mass and the initial
	angle $\theta_0$ in the pre-inflation scenario. The post-inflation
	scenario corresponds to $\theta_0=2.155$ with a strict lower bound on
	the axion's mass of $m_A$=28(2)$\mu eV$. The thick red line shows our
	result on the axion's mass for the post-inflation case. E.g.
	$m_A$=50(4)$\mu eV$ if one assumes that axions from the misalignment
	mechanism contributes 50\% to dark matter. Our final estimate is
	$m_A$=50-1500$\mu eV$ (the upper bound assumes that only 1\% is the
	contribution of the misalignment mechanism the rest comes from other
	sources e.g. topological defects). For an experimental setup
        to detect post-inflationary axions see~\cite{SI}. The slight bend around $m_A\sim
	10^{-5}$ $\mu$eV corresponds to an oscillation temperature at the QCD
	transition \cite{Aoki:2009sc,Borsanyi:2010bp}.
    }
\end{figure}

Through combining these methods one can determine $\chi(T)$ by 
controlling all the systematic uncertainties (see
Figure~\ref{fi:chi}). Therefore, removing several thousand percent 
cutoff effects of previous approaches leaving a very mild ${\cal 
O}(10\%)$ continuum extrapolation to be performed. In addition, the 
direct determination of $\chi(T)$ all the way up to 3~GeV means that 
one does not have to rely on the dilute instanton gas approximation 
(DIGA). Note that {\it a posteriori} the exponent predicted by
DIGA turned out to be compatible with our finding but its prefactor is off by an
order of magnitude, similar to the quenched case.

As a possible application for these two cosmologically relevant lattice 
QCD results, we show how to calculate the amount of axionic dark matter 
and how it can be used to determine the mass of the axion. As we have 
seen, $\chi(T)$ is a rapidly decreasing function of the temperature.
Thus, at high temperature $m_A$ (which is proportional to 
$\chi(T)^{1/2}$) is small. In fact, much smaller than the Hubble 
expansion rate of the universe at that time or temperature ($H(T)$). The 
axion does not feel the tilt in the Peccei-Quinn Mexican hat type 
potential yet and it is effectively massless and frozen by the Hubble 
friction. As the Universe expands the temperature decreases, $\chi(T)$ 
increases and the axion mass also increases. In the meantime, the Hubble 
expansion rate --given by our equation of state-- decreases. As the 
temperature decreases to $T_{\rm osc}$ the axion mass is of the same 
order as the Hubble constant ($T_{\rm osc}$ is defined by $3H(T_{\rm 
osc})=m_A(T_{\rm osc})$). Around this time the axion field rolls down 
the potential, starts to oscillate around the tilted minimum and the 
axion number density increases to a nonzero value, thus axions as dark 
matter are produced. The details of this production mechanism, 
usually called misalignment, are quite well known (see e.g. 
\cite{Wantz:2009it,SI}).

In a post-inflationary scenario the initial value of the angle $\theta$ 
takes all values between -$\pi$ and $\pi$, whereas in the pre-inflationary 
scenario only one $\theta_0$ angle contributes (all other values are 
inflated away). One should also mention that during the U(1) symmetry 
breaking topological strings appear which decay and also produce dark 
matter axions. In the pre-inflationary scenario they are inflated away. 
However, in the post-inflationary framework their role is more 
important. This sort of axion production mechanism is less well-understood
and in our final results it is necessary to make some assumptions.

The possible consequences of our results on the predictions of the 
amount of axion dark matter can be seen in
Figure~\ref{fi:axion}. Here we also study cases, for which the dark matter 
axions are produced from the decay of unstable axionic strings (see the 
discussion in the figure's caption). For the pre-inflationary 
Peccei-Quinn symmetry breaking scenario the axion mass determines the 
initial condition $\theta_0$ of our universe.
\vspace*{0.5cm}

{\bf Acknowledgments.} We thank M. Dierigl, M. Giordano, S. Krieg, D. Nogradi and B. Toth for
useful discussions. This project was funded by the DFG grant SFB/TR55, and 
by OTKA under grant OTKA-K-113034. 
The work of J.R. is supported by the Ramon y Cajal Fellowship 2012-10597
and FPA2015-65745-P (MINECO/FEDER).
The computations were performed on JUQUEEN at Forschungszentrum
J\"ulich (FZJ), on SuperMUC at Leibniz Supercomputing Centre in
M\"unchen, on Hazel Hen at the High Performance Computing Center in Stuttgart, on QPACE 
in Wuppertal and on GPU clusters in Wuppertal and Budapest.
\vspace*{0.5cm}

{\bf Author contributions.} SB and SM developed the fixed sector integral,
TGK and KS the eigenvalue reweighting, FP the odd flavour overlap techniques, 
respectively. SB, SK, TGK, TK, SM, AP, FP and KS wrote the
necessary codes, carried out the runs and determined the EoS and $\chi(T)$. 
JR, AP and AR calculated the DIGA prediction. K-HK and JR and AR worked
out the experimental setup. ZF wrote the main paper and coordinated the project.
\vspace*{0.5cm}

\printbibliography

\end{refsection}

\nolinenumbers

\singlespacing
\begin{refsection}

\clearpage

\setcounter{table}{0}
\setcounter{figure}{0}
\setcounter{page}{1}

\renewcommand\thefigure{S\arabic{figure}}
\renewcommand\thetable{S\arabic{table}}
\renewcommand{\theequation}{S\arabic{equation}}
\renewcommand\thesection{S\arabic{section}}

\thispagestyle{empty}
\vspace*{1cm}
\begin{center}
{\bf \huge Supplementary Information}\\
{\large\bf Lattice QCD for Cosmology}
\end{center}
\vspace*{1.5cm}

\authors

In the following sections we provide details of the work presented in the main
paper. In Section \ref{se:som_zero} we summarize the simulation setup for our
staggered lattice QCD calculations. This is the basis for the determination of
the equation of state and one of the two key elements of the topological
susceptibility calculations. Section \ref{sec:latticeeos} contains the
technical details for the EoS focusing on the lattice QCD part, Section
\ref{sec:perteos} presents the perturbative methods to determine the QCD
equation of state.  Section \ref{sec:eosresult} lists the complete results for
the equation of state both in the $n_f=2+1+1$ and in the $n_f=2+1+1+1$
frameworks. For cosmological applications we give the effective degrees of
freedom for a wide temperature range, starting from 1~MeV all the way up to
$500$~GeV. 

In Section \ref{se:som_ovalgo} we discuss our simulations with overlap fermions using
even and odd number of flavors. The line of constant physics is determined. In
order to reduce cutoff effects for the topological susceptibility we introduced
a eigenvalue reweighting method, which is presented in Section \ref{se:som_rw}. This
method significantly extends the quark-mass vs. temperature parameter space,
which can be reached by lattice QCD. To reach even higher temperatures a second
method was invented: the fixed sector integral method.
The method is illustrated first in the quenched theory Section \ref{se:som_ym}.
In the next two Sections \ref{se:som_st}, \ref{se:som_ov} we apply the method for the real physical case
using staggered and overlap fermions. It is interesting to mention that for the determination
of the topological susceptibility overlap fermions turned out to be the less
CPU-demanding fermion formulation. The non-perturbative lattice findings are
compared with those of the dilute instanton gas approximation.

In Section \ref{se:som_axion} we use the equation of state and the topological
susceptibility results to make predictions for axion cosmology. Both the
pre-inflation and the post-inflation Peccei-Quinn symmetry breaking scenarios
are discussed. In Section \ref{se:som_exp} 
we present experimental setups, which could explore axions in the predicted post-inflation mass range.

\newpage

\section{Staggered simulations}
\label{se:som_zero}

For the majority of the results in this paper we use a four flavor staggered
action with 4 levels of stout smearing. The action parameters are given in
\cite{Bellwied:2015lba}.  The quark masses and the lattice spacing are
functions of the gauge coupling:
\begin{align}
    \label{eq:oldlcp}
    m_s= m_{s}^{st}(\beta),\quad m_{ud}= R\cdot m_{s}^{st}(\beta), \quad m_c=C\cdot m_s^{st}(\beta), \quad a= a^{st}(\beta),
\end{align}
these sets of functions are called the Lines of Constant Physics (LCP), and
are also given in \cite{Bellwied:2015lba}. For the quark mass ratios we use
$1/R=27.63$ and $C=11.85$, which are in good agreement with recent large scale
lattice QCD simulations \cite{Durr:2010vn,Durr:2010aw}. 

In addition to the algorithms mentioned in \cite{Bellwied:2015lba} we now used
a force gradient time integrator \cite{Clark:2011ir,Yin:2011np} to generate
gauge configurations. 

Throughout this paper the index $f$ labels the quark flavors,
$f=\{ud,s,c\}$, and $N_t$ and $N_s$ are the number of lattice points in the
temporal and spatial directions. The temperature $T$ is introduced in the
fixed-$N_t$ approach of thermodynamics, ie. $T=(aN_t)^{-1}$, which can be
changed  by the parameter $\beta$ while $N_t$ and $N_s$ are fixed. The quark masses
$m_f$ are given in lattice units, if not indicated otherwise.

Two different sets of staggered ensembles are used in this paper: a physical
$n_f=2+1+1$ flavor simulation set and a three flavor symmetric $n_f=3+1$
simulation set. In the following we describe them in more detail.

\subsection{Physical point $n_f=2+1+1$ simulations}

The lattice geometries and statistics of the $n_f=2+1+1$ simulations at zero and
finite temperature, are described in \cite{Bellwied:2015lba}. The quark masses
are set to their physical values. In a few cases we increased the statistics of
the existing ensembles. 

On these ensembles we evaluated the clover definition of the topological charge
$Q$ after applying a Wilson-flow \cite{Luscher:2010iy}. We used an adaptive
step size integration scheme to reduce the computational time. For the finite
temperature ensembles we chose a flow time of $(8T^2)^{-1}$, whereas at zero
temperature $t=w_0^2$, where the $w_0$ scale is defined in
\cite{Borsanyi:2012zs}. In the systematic error analyses we usually allow for a
variation in the flow time. As was shown in our quenched study at finite
temperature \cite{Borsanyi:2015cka} the susceptibility reaches a plateau for
large flow times. The above choices are nicely in this plateau region even on
coarse lattices. The so obtained charge is not necessarily an integer number.
To evaluate the topological susceptibility we usually considered both
definitions: with and without rounding the topological charge, the difference
between the two is used to estimate systematic errors.  When the $\langle Q^2
\rangle$ was large, or close to the continuum limit the rounding did not change
the results significantly.

For some of the finite temperature ensembles we also calculated the eigenvalues
and eigenvectors of the staggered Dirac-operator, $D_{st}$. For this purpose we
used a variant of the symmetric Krylov-Schur algorithm described in
\cite{Hernandez:2005:SSF}.  The chirality of these eigenmodes was determined
using the staggered taste-singlet $\gamma_5$-matrix, which is described in
\cite{Durr:2013gp}.

\subsection{Topological susceptibility in the vacuum}

\begin{figure}[h]
    \centering
    \includegraphics*{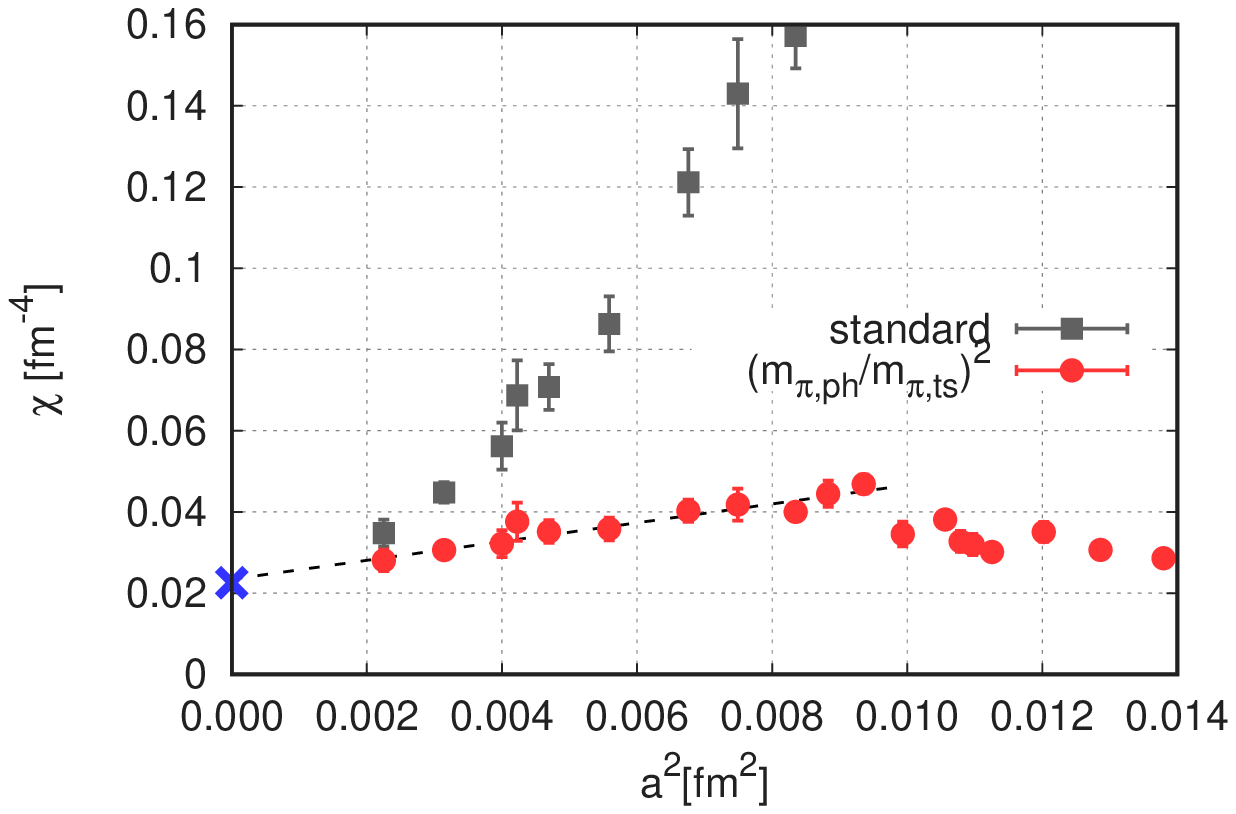}
    \caption
    {
	\label{fi:zeroT}
	Lattice spacing dependence of the zero temperature topological
	susceptibility. The grey squares are obtained with the standard
	approach, the red circles after dividing by the taste singlet
	pion mass squared.  The line is a linear fit. The blue cross
	corresponds to leading order chiral perturbation theory. The plot
	shows $n_f=2+1+1$ flavor staggered simulations at zero temperature.
    }
\end{figure}

Due to the index theorem, in the background of a gauge field with non-zero
topological charge $Q$, the Dirac operator has at least $|Q|$ exact zero
eigenvalues \cite{Atiyah:1968mp}. These zero-modes have chirality $+1$ or $-1$.
This is, however, true only in the continuum theory. On the lattice, due to
cut-off effects, a non-chiral Dirac operator, like the staggered operator, does
not have exact zero eigenvalues, only close to zero small eigenvalues. Also the
magnitude of the chirality of these would-be zero modes is smaller than unity.
In the continuum limit of the lattice model the would-be zero modes become zero
modes with chirality of unit magnitude. However, at any nonzero lattice spacing
the lack of exact zero modes results in cut-off effects that can be
unexpectedly large for some observables. This is shown for the topological
susceptibility in Figure \ref{fi:zeroT}, where $\chi$ is plotted as function of
the lattice spacing squared. The result changes almost an order of magnitude
by moving from the coarsest to the finest lattice spacing on the plot.

At zero temperature $\chi$ is proportional to the pion mass squared in the
continuum. On the lattice with staggered fermions it is expected, that $\chi$
will be proportional to the mass squared of the taste singlet pion. For the
staggered chiral perturbation theory analysis of $\chi$, see
\cite{Bazavov:2010xr}.  Thus it is natural to rescale $\chi$ with the
pseudo-Goldstone mass squared over the taste-singlet pion mass squared.  Since
in our $n_f=2+1+1$ simulations physical pion pseudo-Goldstone masses are used,
in Figure \ref{fi:zeroT} we plot $\chi$ multiplied by
$(m_{\pi,phys}/m_{\pi,ts})^2$, where $m_{\pi,ts}$ is the taste-single pion mass
\cite{Golterman:1985dz,Ishizuka:1993mt}.  The data shows much smaller cut-off
effects, than without multiplication and a nice $a^2$-scaling sets in starting
from a lattice spacing of about $0.1$~fm. The continuum extrapolated value is
\begin{align}
    \label{eq:chi0}
    \chi(T=0)= 0.0245(24)_{\mathrm{stat}}(03)_{\mathrm{flow}}(12)_{\mathrm{cont}}/\mathrm{fm}^4,
\end{align}
the first error is statistical. The second, systematic error comes from varying
the definition of the charge, i.e. the flow time at which the charge is
measured. The third error comes
from changing the upper limit of the lattice spacing range in the fit. According
to leading order chiral perturbation theory
\begin{align}
    \Sigma/\chi= 2m_{ud}^{-1} + m_s^{-1} + \dots
\end{align}
where $\Sigma$ is the condensate in the chiral limit and the ellipses stand for higher order terms. Using the values for
quark masses and the condensate from \cite{Durr:2010vn,Durr:2013goa} we obtain
$\chi_{\mathrm{LO}}=0.0224(12)/\mathrm{fm}^4$ in the isospin symmetric limit, which
is in good agreement with Equation \eqref{eq:chi0}. For isospin corrections
see Section \ref{se:som_ana}.

For the high temperature region the cut-off effects are not supposed to be
described by chiral perturbation theory.  Other techniques are required to get
the large cut-off effects under control, such a technique is presented in
Section \ref{se:som_rw}.

\subsection{Three flavor symmetric $n_f=3+1$ simulations}
\label{se:nf3}

As it will be described in later Sections as an intermediate step we used
results from simulations at the three flavor symmetric point, i.e. where all the
light-quark masses are set to the physical strange quark mass
$m_{s}^{st}(\beta)$ and the ratio of the charm and the 3 degenerate flavour
masses is $C=11.85$.

\begin{figure}[h]
    \centering
    \includegraphics*{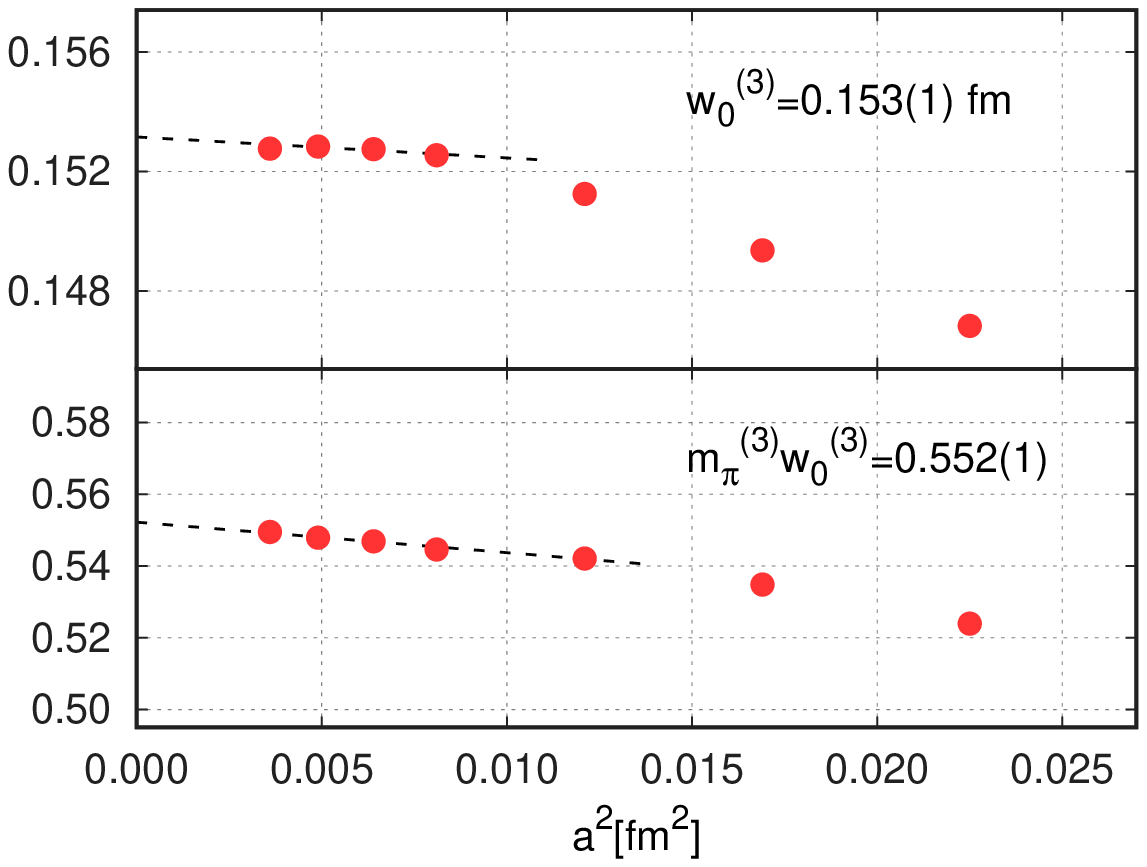}
    \caption{
	\label{fi:nf3st}
	Continuum extrapolations of the $w_0$-scale in lattice units (up) and
	$m_\pi w_0$ (down). The plots show $n_f=3+1$ flavor staggered
	simulations at zero temperature.
    }
\end{figure}

In this theory for each gauge coupling $\beta$ one can measure the pseudoscalar
mass $m_\pi$ and the Wilson-flow based $w_0$-scale.  The dimensionless product
$m_\pi w_0$ as well as the $w_0$ in physical units, i.e. $w_0 a^{st}(\beta)$,
have well defined continuum limits, since the $n_f=3+1$ and $n_f=2+1+1$
theories differ only in the masses of quarks, that do not play a role in the
ultraviolet behaviour of those theories.  So we end up with the continuum value
of the three flavor pseudoscalar mass $m_{\pi}^{(3)}$, and that of the
$w_0$-scale $w_{0}^{(3)}$.  We performed $n_f=3+1$ flavor zero temperature
simulations in $64\times32^3$ volumes for seven lattice spacings ranging
between $a=0.15$ and $0.06$ fm. We measured $w_0$ and $m_\pi w_0$ and performed
a continuum extrapolation. This is shown in Figure \ref{fi:nf3st}. We obtain
the continuum values
\begin{align}
    \label{eq:nf3st}
    w_0^{(3)}=0.153(1)\text{ fm}\quad\text{and}\quad m_\pi^{(3)}=712(5) \text{MeV}.
\end{align}

In the $n_f=3+1$ theory we performed finite temperature simulations and
measured the same observables as in the $n_f=2+1+1$ case. Additionally, for the
new integral method described in Section \ref{se:som_st}, we generated
configurations at fixed topology. This was achieved by measuring the
topological charge after each Hybrid Monte-Carlo trajectory and rejecting the
configuration in case of topology change. The typical acceptance rates were
$40$\% on the coarsest lattices and higher on the finer ones. The finite
temperature ensembles, unconstrained and fixed topology, are listed in the
analysis section, Section \ref{se:som_ana}.

\section{\label{sec:latticeeos}Lattice methods for the equation of state} 

For 2+1 dynamical flavors with physical quark masses we calculated the
equation of state in Refs.~\cite{Borsanyi:2010cj,Borsanyi:2013bia}.  This
result was later confirmed in Ref.~\cite{Bazavov:2014pvz}.  The additional
contribution of the charm quark has since been estimated by several authors
using perturbation theory \cite{Laine:2006cp}, partially quenched lattice
simulations \cite{Cheng:2007wu,Levkova:2009gq} and from simulations with four
non-degenerate quarks, but unphysical masses
\cite{Burger:2013hia,Bazavov:2013pra}.  However, the final results must come
from a dynamical simulation where all quark masses assume their physical
values \cite{Borsanyi:2012vn}.

We meet this challenge by using the 2+1+1 flavor staggered action (with 4 levels
of stout smearing) that we introduced in Section \ref{se:som_zero}. 
The action parameters, the bare quark masses and the tuning procedure, as well
as the lattice geometries and  statistics have been given in detail in
Ref.~\cite{Bellwied:2015lba}.

We calculate the EoS, ie. the temperature dependence of the pressure $p$,
energy density $\rho$ and entropy density $s$, from the trace anomaly $I(T)$.
This latter is defined as $I=\rho-3p$, and on the lattice it is given by the
following formula:
\begin{equation}
\frac{I(T)}{T^4}=\frac{\rho-3p}{T^4}= N_t^4
R_\beta
\left[
\frac{\partial}{\partial \beta}
+\sum_f R_f m_f \frac{\partial}{\partial m_f}
\right]\frac{\log Z[m_u,m_d,m_s,...]}{N_t N_s^3}
\label{eq:traceaexpression}
\end{equation}
with
\begin{equation}
R_\beta = - \frac{d \beta}{d\log a}\qquad\mathrm{and}\qquad
R_f= \frac{d\log m_f}{d\beta}, \quad f=u,d,s,\dots\,.
\end{equation}
  
The derivatives of $\log Z$ with respect to the gauge coupling $\beta$ and the
quark masses $m_f$ are easily accessible observables on the lattice: they are
the gauge action $S_g$ and the chiral condensate, respectively.

The $R_\beta$ and $R_f$ functions describe the running of the coupling and the
mass. They can be obtained by differentiating the LCP in Eq.~\eqref{eq:oldlcp}.
Since $R_\beta$ appears as a factor in front of the final result, the
systematics of the determination of the LCP directly distorts the trace
anomaly. To estimate this uncertainty we use two different LCP's, one based on
the $w_0$-scale and another other on the pion decay constant $f_\pi$, which are
supposed to give the same continuum limit, but can differ by lattice artefacts.
We calculate $R_\beta$ both from the $w_0$ and the $f_\pi$-based LCP and
include the difference in the systematic error estimate.  Let us mention, that
the $R_\beta$ and $R_f$ functions are universal at low orders of perturbation
theory: e.g. for QCD with $n_f$ flavors we have
$R_\beta=12\beta_0+72\beta_1/\beta+\mathcal{O}(\beta^{-2})$ with
$\beta_0=(33-2n_f)/48\pi^2$ and $\beta_1=(306-38n_f)/768\pi^2$. 

\begin{figure}[ht]
\centerline{\includegraphics[width=3.4in]{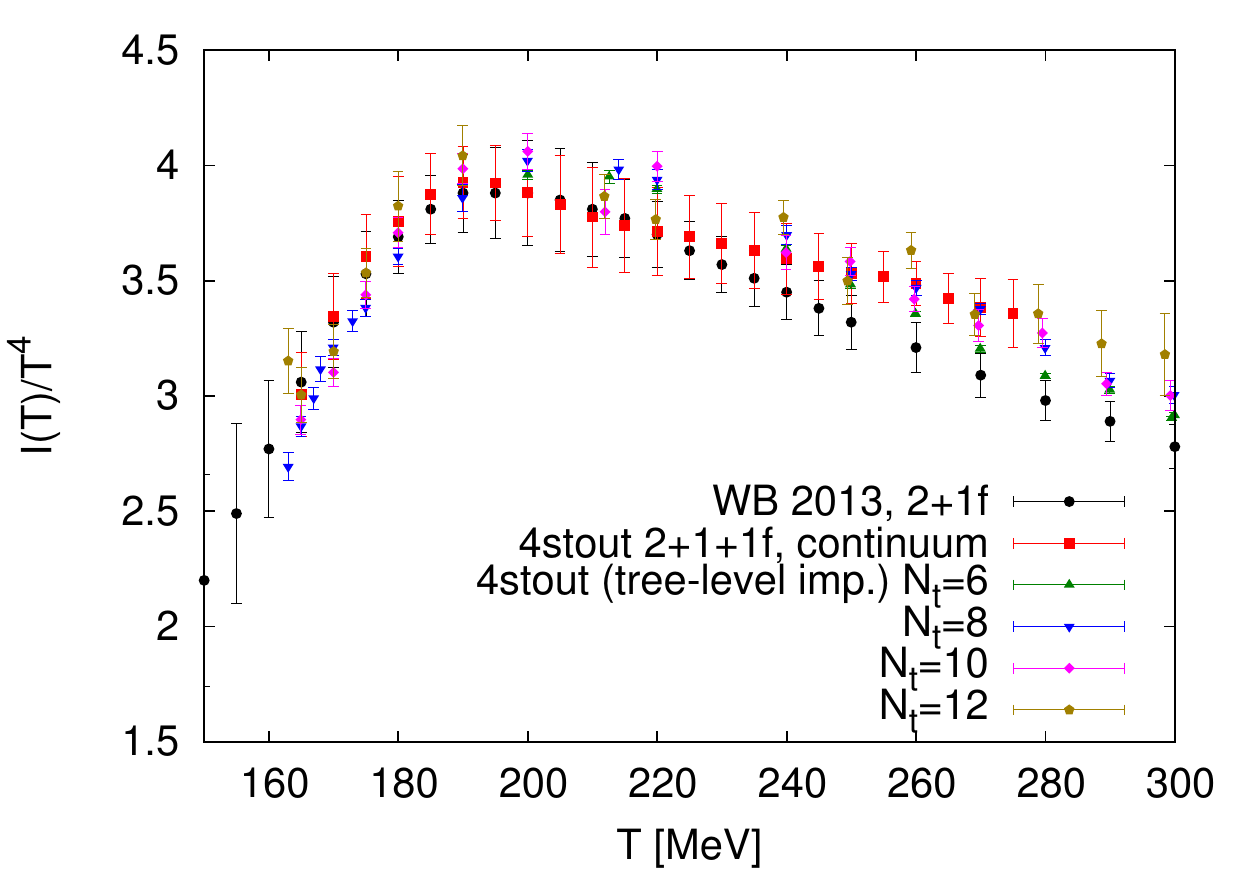}\hspace{1cm}
\includegraphics[width=3.4in]{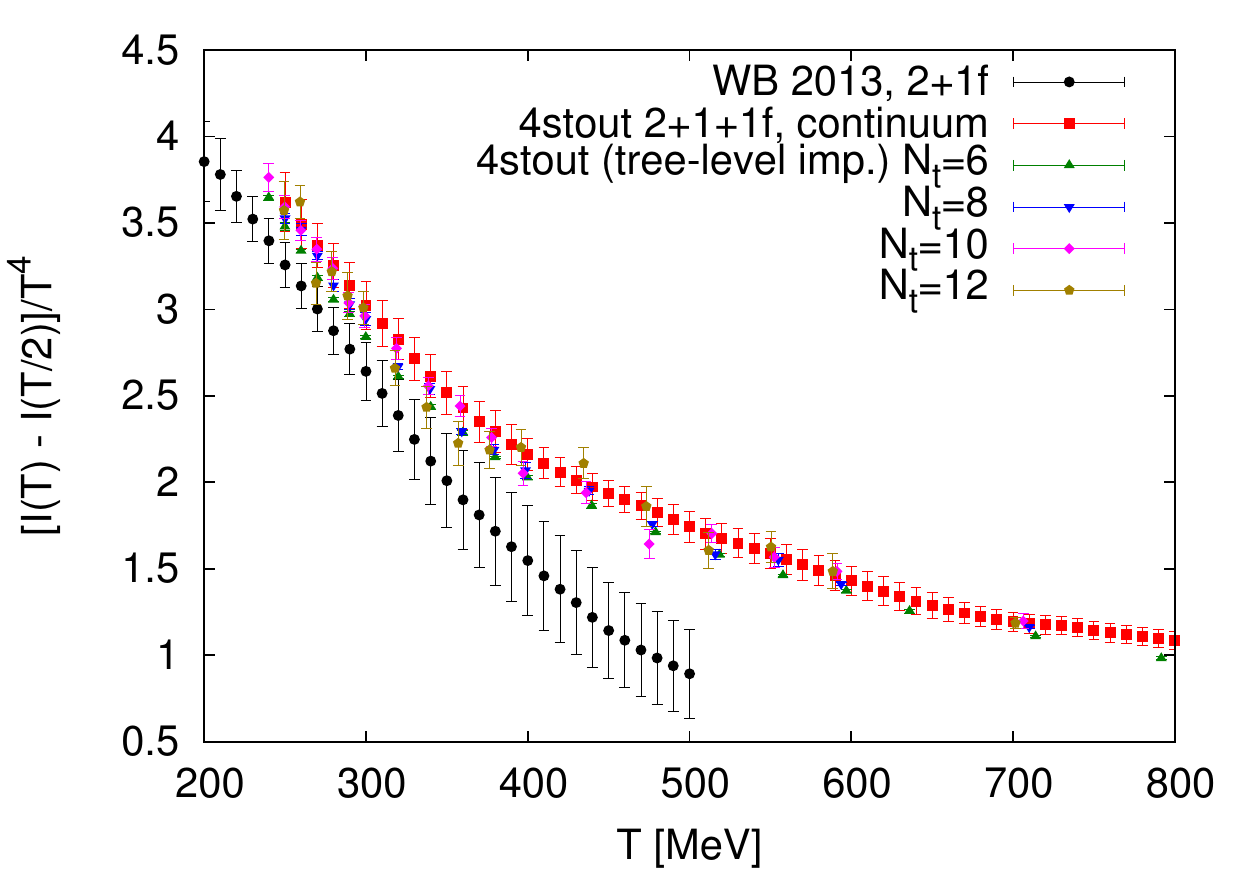}}
\caption{
The trace anomaly renormalized with zero temperature simulations (left panel)
and the subtracted trace anomaly (right panel) in the 2+1+1 and 2+1 flavor theories. 
For $T<300$~MeV the two results agree within our uncertainty.
We also show the individual lattice resolutions $(N_t=6\dots12)$ that
contribute to the continuum limit.
\label{fig:trah}
}
\end{figure}

There is an additive, temperature independent divergence in the trace
anomaly. In the standard procedure, that we also followed in
Ref.~\cite{Borsanyi:2010cj}, each finite temperature simulation is
accompanied by a zero temperature ensemble. The trace anomaly is then 
calculated on both sets of configurations, their difference is the physical
result. This defines a renormalization scheme where the zero temperature
pressure and energy density both vanish. Using the configurations already
introduced in Ref.~\cite{Bellwied:2015lba} and applying the standard
method that we also described in Ref.~\cite{Borsanyi:2013bia} we calculated
the trace anomaly, as shown in the left panel of Fig.~\ref{fig:trah}.

This strategy is bound to fail at high temperatures. Since the temperature on
the lattice is given by $T=(aN_t)^{-1}$, high temperatures can only be reached
with very fine lattices. As the lattice spacing is reduced, the autocorrelation
times for zero temperature simulations rise, and the costs of these simulations
explode beyond feasibility. Notice, however, that the renormalization constant
is accessible not only from zero temperature simulations, but from any finite
temperature data point that was taken using the same gauge coupling and quark
masses.

In our approach we generate a renormalization ensemble for each finite
temperature ensemble at exactly half of its temperature with the same physical
volume and matching bare parameters. The trace anomaly difference is then
divergence-free.  We have tested this idea in the quark-less theory in
Refs.~\cite{Endrodi:2007tq,Borsanyi:2012ve}.

Thus we determine the quantity $[I(T)-I(T/2)]/T^4$ in the temperature
range 250\dots 1000~MeV for four resolutions $N_t=6,8,10$~and~12. The volumes
are selected such that $LT_c\gtrsim2$. In Fig.~\ref{fig:trah} we show this
subtracted trace anomaly and its continuum limit. 

We now turn to our systematic uncertainties. We use the histogram method
to estimate the systematic errors, this means that we analyze our data
in various plausible ways and form a histogram of the results. The median
gives a mean, the central 68\% defines the systematic error \cite{Durr:2008zz}.
Here we use uniform weights dropping the continuum results where the fit
quality of the continuum limit was below 0.1. For the details of the analysis
we largely follow our earlier work in Ref.~\cite{Bellwied:2015lba}: we
interpolated the lattice data in temperature and then we performed a continuum
extrapolation in $1/N_t^2$ temperature by temperature.  The error bars in
Fig.~\ref{fig:trah} combine the statistical and systematic errors, the latter
we estimate by varying the scale setting prescription ($w_0$-based or
$f_\pi$-based scale setting), the observable (subtracted trace anomaly or its
reciprocal), the interpolation, and whether or not we use tree-level
improvement prior to the continuum extrapolation \cite{Borsanyi:2010cj}.

Then we use the trace anomaly data with zero temperature renormalization from the left panel of
Fig.~\ref{fig:trah} and extend it towards lower temperatures from our 
already established 2+1 flavor equation of state result.  Then we can calculate
$I(T)/T^4$ from the continuum limit of $[I(T)-I(T/2)]/T^4$ using the formula:
\begin{equation}
\frac{I(T)}{T^4}=
\sum_{k=0}^{n-1}2^{-4k}\left.\frac{I\left(T/2^k\right)-I\left(T/2^{k+1}\right)}{\left(T/2^k\right)^4}\right.
+
2^{-4n}\left.\frac{I\left(T/2^n\right)}{\left(T/2^n\right)^4}\right.\,,
\end{equation}
where $n$ is the smallest non-negative integer with $T/2^n <250~\mathrm{MeV}$.

The temperature integral of $I(T)/T^5$ gives the normalized pressure
$p(T)/T^4$.  The energy density and entropy density are then
obtained from the thermodynamic relations: $\rho=I+3p$ and $sT = \rho+p$,
respectively.


\section{\label{sec:perteos}The QCD equation of state in the perturbative regime}

\subsection{Massless perturbation theory} 

\begin{figure}[ht]
\centerline{\includegraphics[width=3.4in]{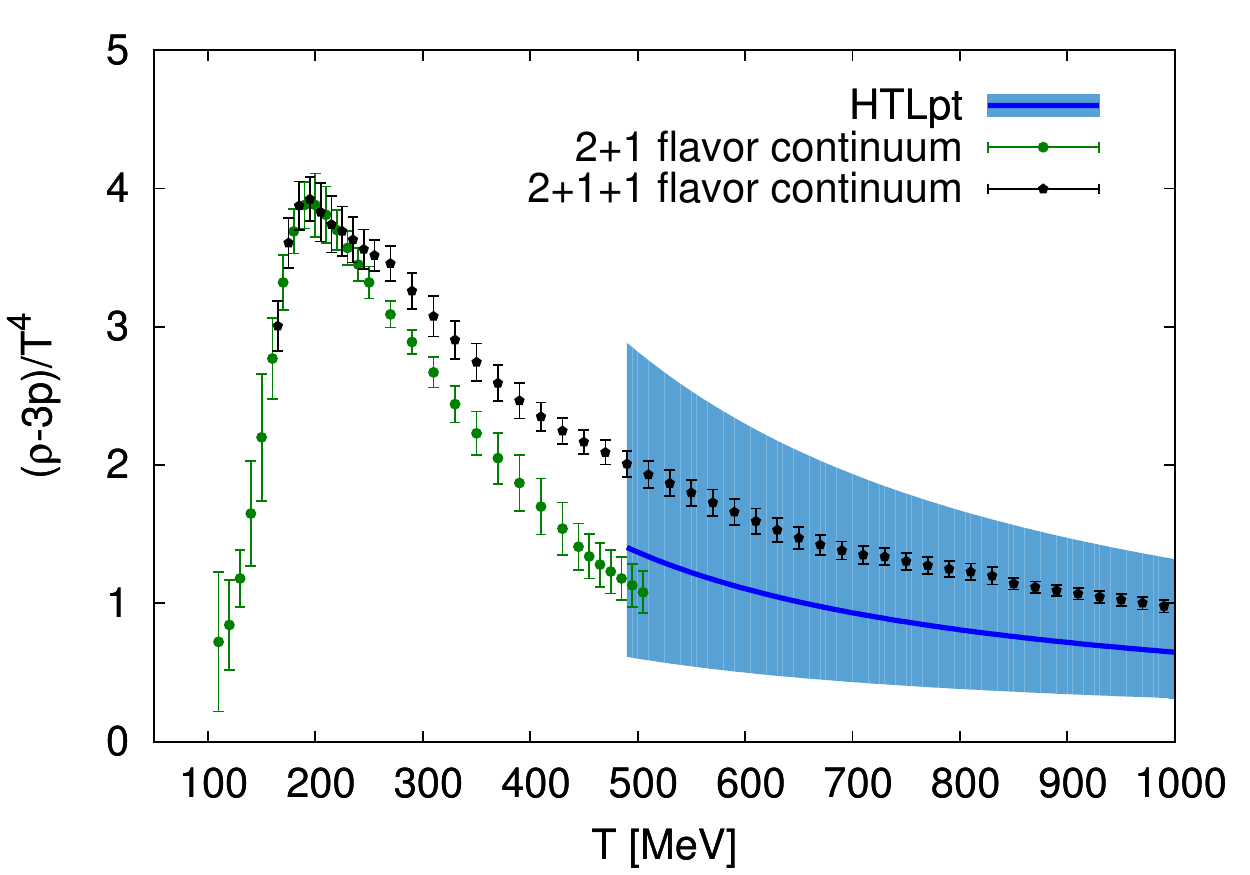}\hspace{1cm}
\includegraphics[width=3.4in]{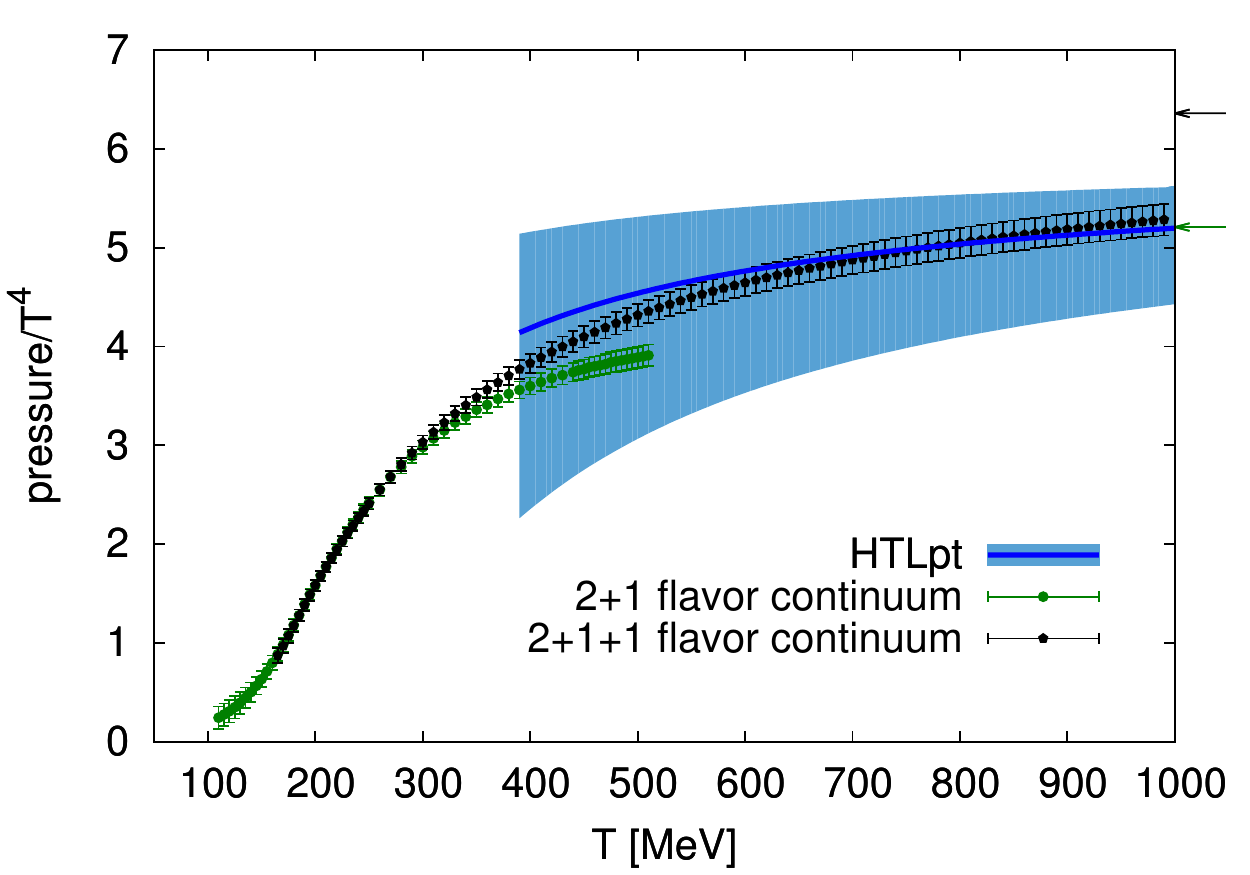}}
\caption{
The QCD trace anomaly and pressure in the 2+1+1 and 2+1 flavor theories. We also give 
the four flavor NNLO HTL result at high temperatures
\cite{Andersen:2010wu}. 
\label{fig:pressure}
}
\end{figure}

Recent progress in Hard Thermal Loop (HTL) perturbation theory has provided for
a next-to-next-to-leading-order (NNLO) result for the free energy, which is in
fair agreement with our data both for the 2+1 theory \cite{Andersen:2010wu} and
also for the 2+1+1 flavor theory for high enough temperatures.  The trace
anomaly and the pressure are shown in Fig.~\ref{fig:pressure} for both cases.

\begin{figure}[ht]
\begin{center}
\includegraphics[width=5in]{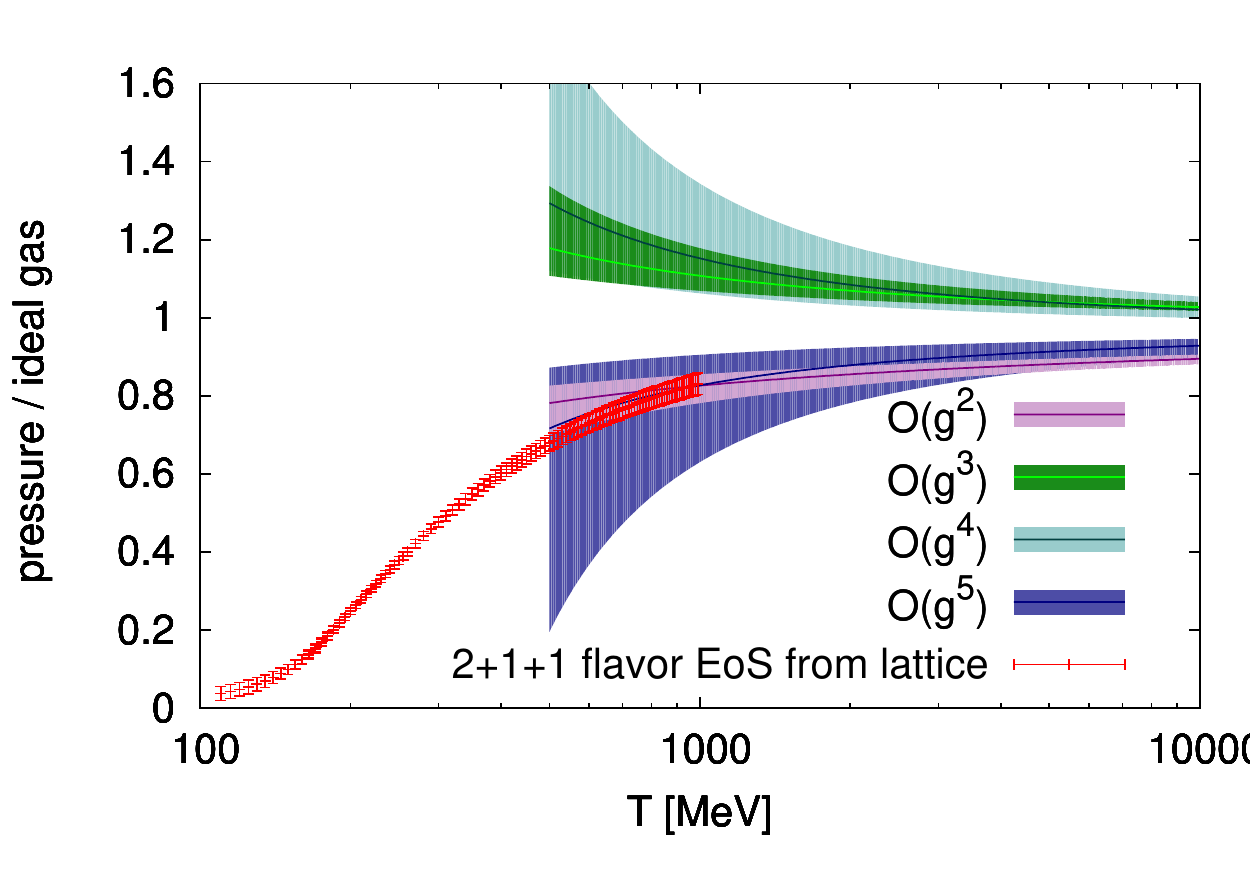}
\end{center}
\caption{\label{fig:pcmp}
The QCD pressure for 2+1+1 flavors together with 
various orders of conventional perturbation theory. The renormalization
scale is varied in the range $\mu=(1\dots  4)\pi T $. The middle lines
correspond to $\mu=2\pi T$.
}
\end{figure}

We also show a comparison to the results of conventional perturbation theory
with four massless quarks in Fig.~\ref{fig:pcmp}. Here $g=\sqrt{4\pi\alpha_s}$
is the QCD coupling constant. The completely known fifth
order \cite{Zhai:1995ac} result is in good agreement with the lattice data. Note, that
whereas the perturbative result treats even the charm quark massless, the lattice EoS
includes the mass effects correctly.

\subsection{Charm mass threshold in the QCD equation of state} 

Thanks to the lattice data that we have generated, we can present
non-perturbative results for the charm quark contribution. It is instructive to
study the inclusion of the charm quark in detail. This way we can design an
analytical technique for the inclusion of the bottom quark, for which the
standard formulation of lattice QCD is computationally not feasible.

The quark mass threshold for the charm quark entering the EoS has already been
estimated in Ref.~\cite{Laine:2006cp}.  There, the
effect of a heavy quark was calculated to a low order of perturbation theory.
This effect was expressed as a pressure ratio between QCD with three light and
one heavy flavor and QCD with only three light flavors.  When that paper was
completed the lattice result for the QCD equation of state was not yet
available, but the perturbative methods were already in an advanced state.

\begin{figure}[ht]
\begin{center}
\includegraphics[width=3in,angle=270]{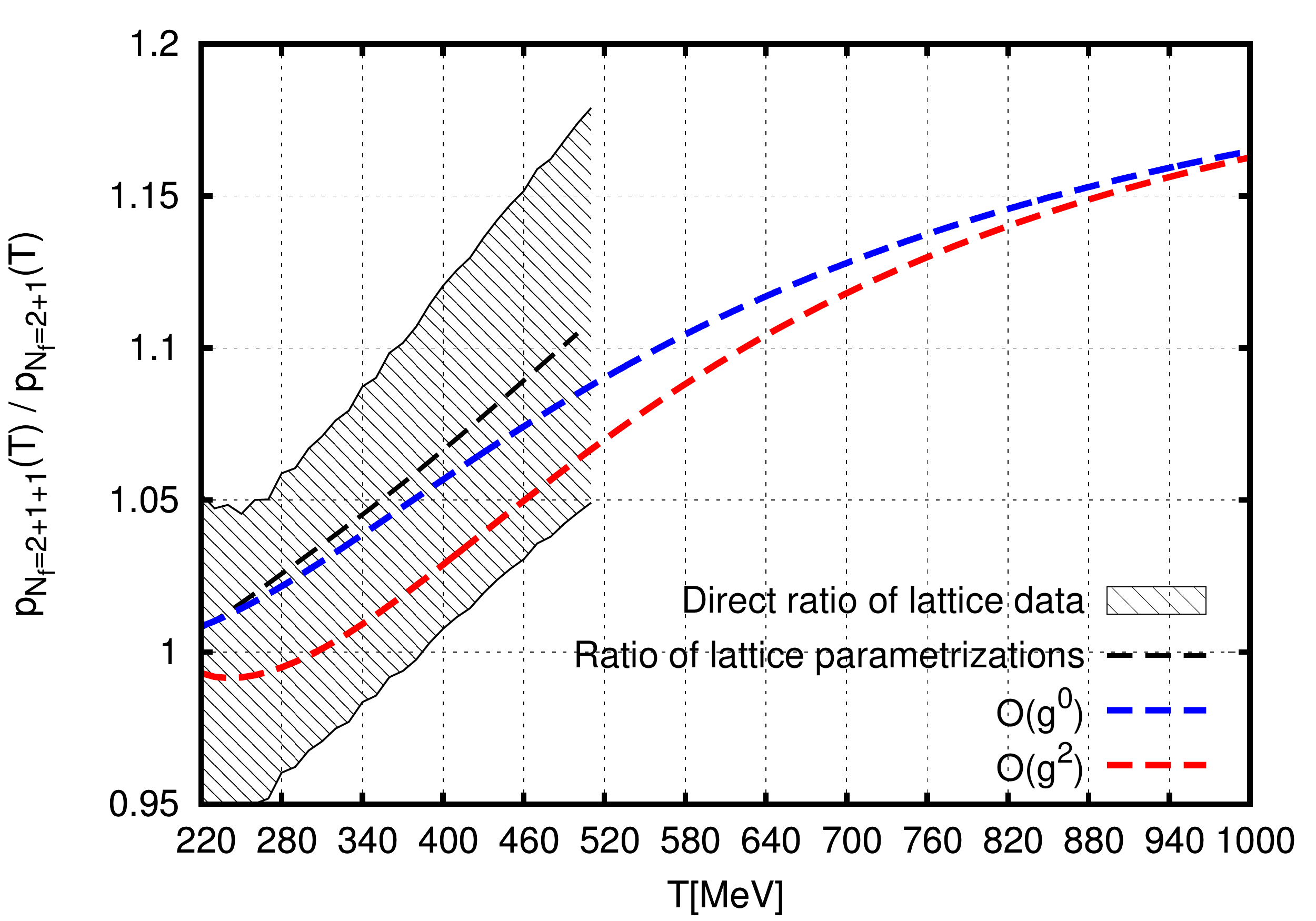}
\end{center}
\caption{\label{fig:charm_effect}
The ratio of the pressure between 2+1+1 flavor and 2+1 flavor theories from our 
lattice simulations. Note that the parametrization smoothly approaches 1 as we
further decrease the temperature (see Eq.~(\ref{eq:param211})).
We also show two perturbative estimates for the ratio of the pressure
functions. The tree-level ($\mathcal{O}(g^0)$) estimate (see text),
and the $\mathcal{O}(g^2)$ estimate of Ref.~\cite{Laine:2006cp}. Both of these estimates show agreement with the lattice data within our accuracy.
}
\end{figure}

Despite the known difficulties of perturbation theory the estimate of
Ref.~\cite{Laine:2006cp} is very close to our lattice result if we plot
the ratio of the pressure with and without the charm quark included. 
We show our lattice data together with the perturbative
estimate in Fig.~\ref{fig:charm_effect}.

Though the individual values for the 2+1+1 and 2+1 flavor pressures of
\cite{Laine:2006cp} are not very accurate, their ratio describes well 
the lattice result.  This is true both for the leading and for the
next-to-leading order results (See Fig.~\ref{fig:charm_effect}).

The tree-level charm correction is given by
\begin{equation}
\frac{p^{(2+1+1)}(T)}{p^{(2+1)}(T)}=\frac{SB(3) + F_Q(m_c/T)}{SB(3)}\,
\label{eq:charmcorr3}
\end{equation}
where $SB(n_f)$ is the Stefan Boltzmann limit of the $n_f$ flavor theory,
and $F_Q(m/T) T^4$ is the free energy density
of a free quark field with mass $m$. In this paper we used the
$\overline{MS}$ mass $m_c(m_c)=1.29$~GeV \cite{Agashe:2014kda}.

Order $g^2$ in the ratio of Fig.~\ref{fig:charm_effect} starts to be important
correction below a temperature of about $2 - 3 T_c^{QCD}$ temperature. Near
$2T_c$ the difference between the two approximations is $3\%$. The difference
reduces to 0.2\% at 1~GeV up to which point we have lattice data.


\subsection{\label{sec:bottom}Bottom mass threshold in the QCD equation of state} 

In the previous discussion we saw that even the tree-level quark mass
threshold gives a correct estimate for the equation of state. This
allows us to introduce the bottom threshold along the same lines.

First, we remark that one can write the charm threshold relative to
the $2+1+1$ flavor theory:
\begin{equation}
\frac{p^{(2+1+1)}(T)}{\left.p^{(2+1+1)}(T)\right|_{m_c=0}} = \frac{SB(3) + F_Q(m_c/T)}{SB(4)}\,.
\label{eq:charmcorr}
\end{equation}
The error of not using the $g^2$ order
is about 0.2\% of the total QCD contribution at 1 GeV. 

From the lattice data we have $p^{(2+1+1)}(T)$. Using Eq.~(\ref{eq:charmcorr})
we can calculate the QCD pressure for the theory with four light quarks.
Perturbation theory can provide just that, at least, if the temperature
is high enough. In the following we demonstrate that it is possible
to use a perturbative formula that matches our lattice-based 
$\left.p^{(2+1+1)}(T)\right|_{m_c=0}$ already from approx. 500 MeV, i.e.
below the bottom threshold.

The perturbative results have a mild dependence on the choice of the $\Lambda$
parameter, here we use the standard $\Lambda_{\overline{MS}}=290$~MeV value for
$n_f=4$ \cite{Agashe:2014kda}.

\begin{figure}[ht]
\centerline{
\includegraphics[width=3.4in]{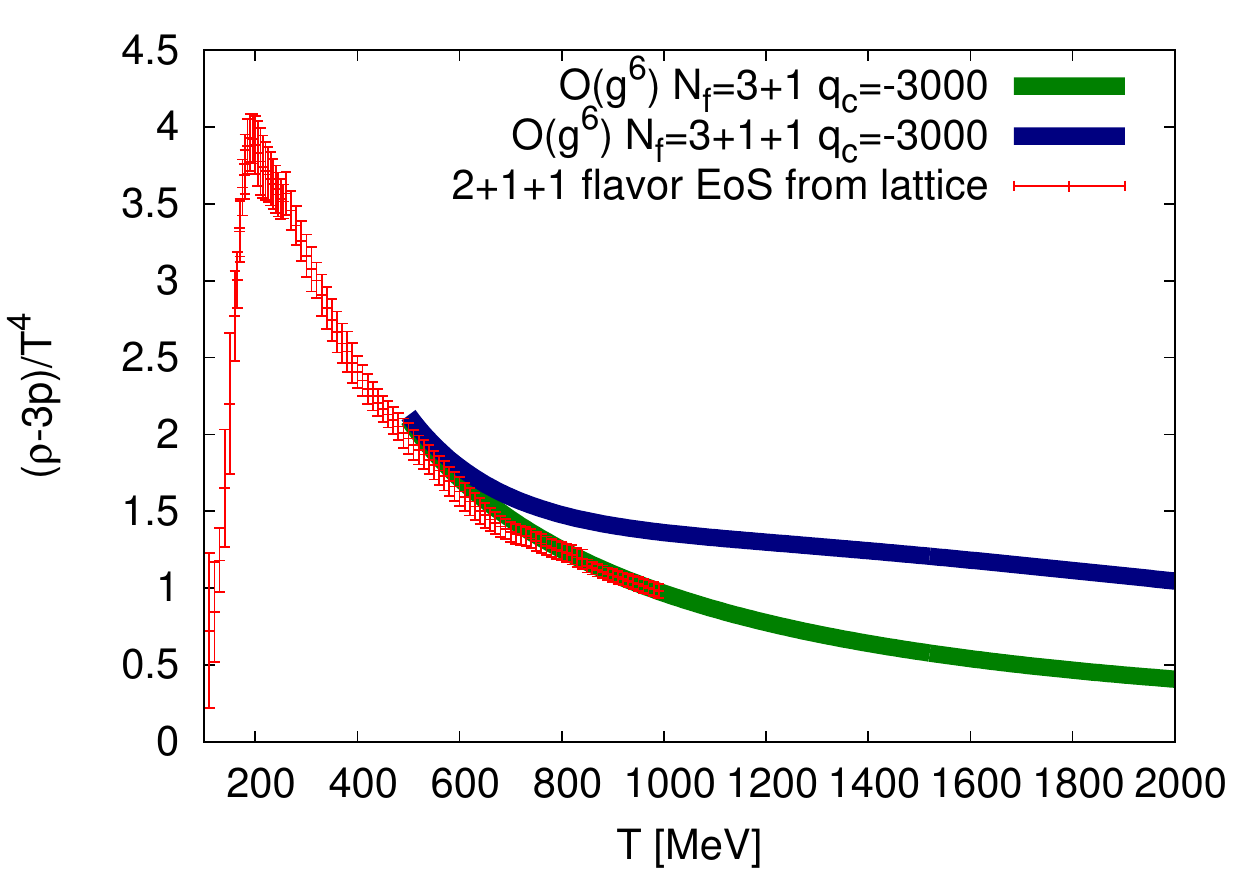}
\includegraphics[width=3.4in]{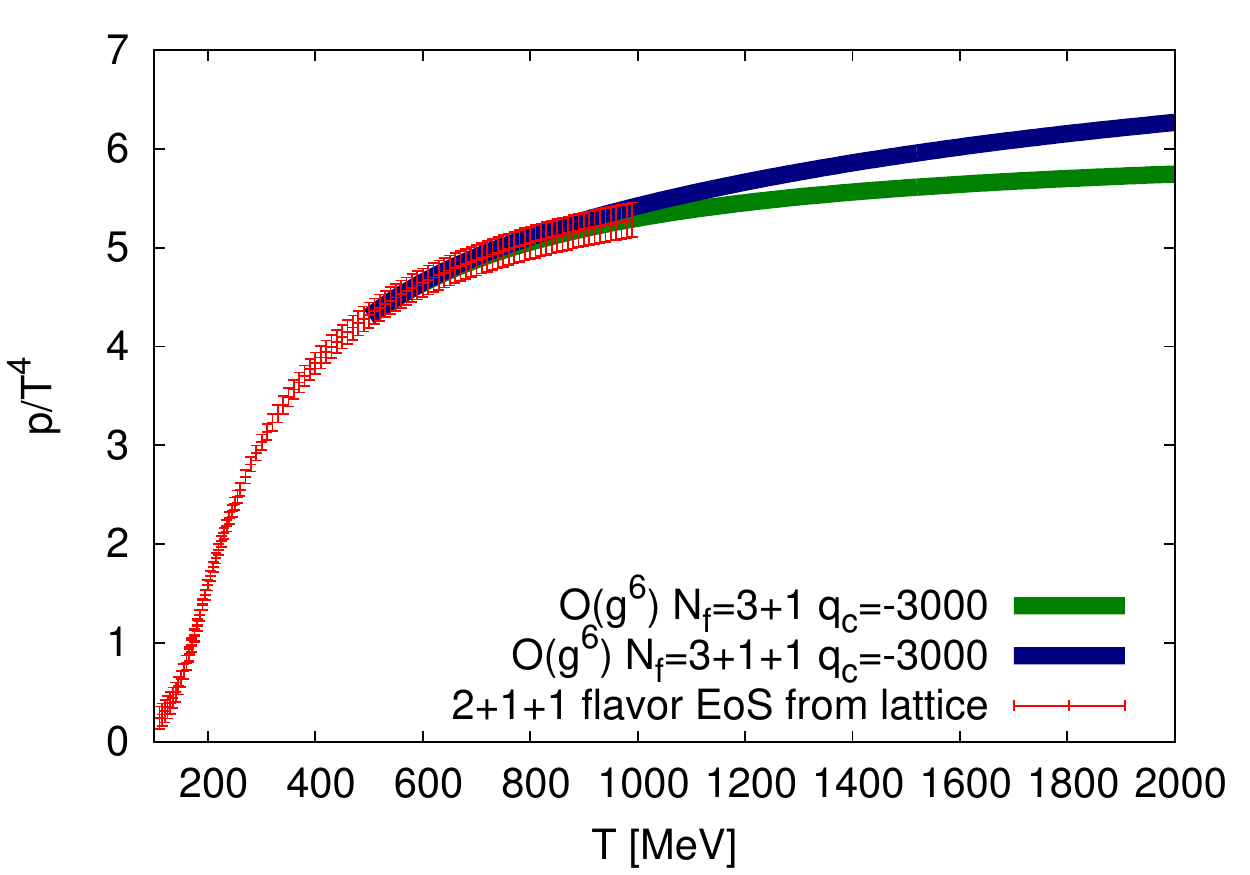}
}
\caption{\label{fig:withmass}
The lattice result for the 2+1+1 flavor QCD pressure together with the fitted
value of the $g^6$ order. We included the charm mass at tree-level. 
The perturbative result agrees with the data from about 500~MeV temperature.
Using the same fitted coefficient we also calculated the effect of the bottom
quark with the same method.  The blue curve shows the EoS including
the bottom contribution.
}

\end{figure}

The highest fully known order for the QCD pressure is $g^6\log g$
\cite{Kajantie:2002wa}.  The coefficient of the $g^6$ order is not known
analytically, but the missing term, $q_c$ (following the notation of
Ref.~\cite{Kajantie:2002wa}) can be fitted against lattice data. We fix
the remormalization scale to $\mu=2\pi T$.
This fitting method has already been applied for the Yang-Mills theory
\cite{Kajantie:2002wa,Borsanyi:2012ve}.
The order $g^6$ result describes our pressure data at 1 GeV within errors if
$-3400 < q_c < -2600$, for the trace anomaly we have $-3200 < q_c < -2800$. We
propagate this uncertainty into the perturbative result, keeping the range
$2700 < -q_c < 3200$.  We show the fitted curves for the central choice,
$q_c=-3000$, in Fig.~\ref{fig:withmass}. The result in the plot has already
been converted to the case of a massive charm. Notice, that both for
the trace anomaly and for the pressure the $\mathcal{O}(g^6)$
perturbative result follows the lattice data already from 500~MeV.  

Having a pressure and trace anomaly function that is valid for the 2+1+1
flavor theory, that agrees with the lattice data below 1 GeV and provide
a perturbatively correct continuation towards high temperatures we can
proceed to include the effect of the bottom quark.
The tree-level correction for the bottom quark reads
\begin{equation}
p^{(2+1+1+1)}(T) = p^{(2+1+1)}(T) \frac{SB(4) + F_Q(m_b/T)}{SB(4)}\,
\label{eq:bottomcorr}
\end{equation}
where $m_b(m_b)=4.18$~GeV is the bottom mass \cite{Agashe:2014kda}.
Eq.~(\ref{eq:bottomcorr}) works beyond the bottom threshold, too, since the
ratio of the perturbative massless 4 and 5 flavor pressure is, to a very good
approximation, equal to the ratio of the Stefan-Boltzmann limits. Comparing the
free energy up to order $g^5$ this statement holds to 0.3\% accuracy in the
entire relevant temperature range. (For earlier formulations of this idea see
Refs.~\cite{Karsch:2000ps,Hindmarsh:2005ix}.)

Since the massless four flavor perturbative result is used as a starting point
for both heavy quarks we have a fair approximation between 500~MeV and 10 GeV
with one analytical formula. 



\section{\label{sec:eosresult}Equation of state for 2+1+1 flavor and 2+1+1+1 flavor QCD and for the whole Standard Model}  

It has been a longstanding challenge to determine the pressure,
energy density, and the number of effective degrees of freedom as a function of
the temperature from first principles.
This is the equation of state of the universe. Cosmology
requires this information over a temperature range of many orders of magnitude,
ranging from beyond the electroweak scale down to the 
MeV scale \cite{Srednicki:1988ce,Hindmarsh:2005ix}.

As the Universe covers this broad temperature range it passes though several
epochs, each with a different dominant interaction. We restrict our study here
to the Standard Model of particle physics.
At the high end of the temperature range of interest there is the
electroweak phase transition at a temperature of about 160~GeV
\cite{DOnofrio:2015mpa}.  The equation of state of
the electroweak transition has been worked out in
Refs.~\cite{Laine:2006cp,Laine:2013raa,Laine:2015kra} with the intent to
provide a description for the entire Standard Model. For this reason the
contribution from all other degrees of freedom (i.e. up to the bottom quark)
had to be estimated. While the photon, neutrinos and leptons can easily be
described as practically free particles, the QCD part requires a
non-perturbative approach. This was not available when
Refs.~\cite{Laine:2006cp,Laine:2013raa} were published. 

In this paper we add the last missing piece to the cosmological equation of
state: the QCD contribution. In this section we give the results of our
efforts for the 2+1+1 and 2+1+1+1 flavor theories separately.
Finally, we combine all the elements of the Standard Model and present the
number of effective degrees of freedom from the energy density and entropy
($g_{\rho}(T)$ and $g_{s}(T)$) in the full temperature range.


\subsection{The 2+1+1 flavor QCD equation of state} 

\begin{figure}[ht]
\centerline{\includegraphics[width=3.4in]{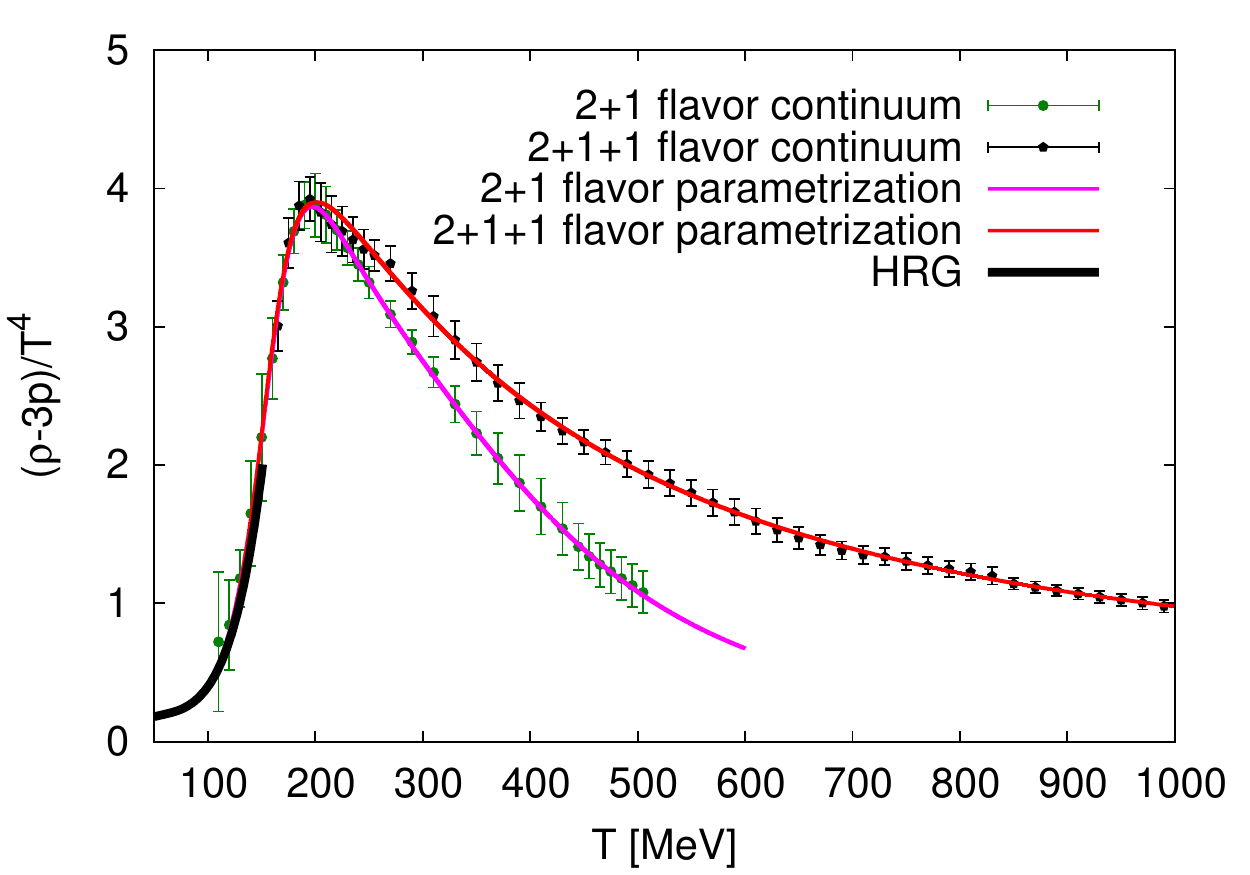}\hspace{3mm}
\includegraphics[width=3.4in]{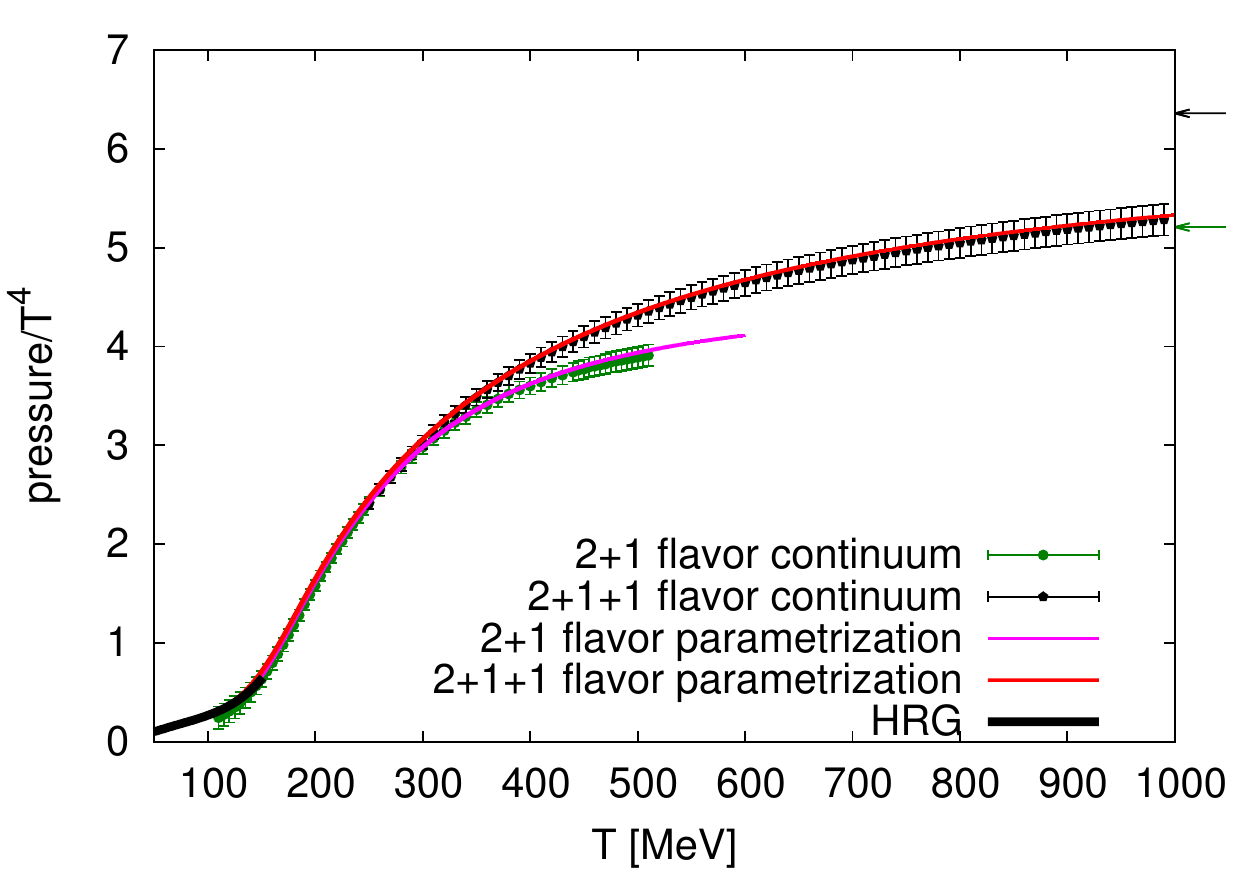}}
\caption{
The QCD trace anomaly and pressure in the 2+1+1 and 2+1 flavor theories in
our parametrization Eq.~(\ref{eq:param211}). We also show the Hadron Resonance Gas model's prediction for comparison.
\label{fig:param}
}
\end{figure}

Now we show the complete result obtained from $n_f=2+1+1$ lattice QCD.
Figure~\ref{fig:param} depicts the trace anomaly (left panel) and pressure (right panel). For comparison the 2+1 flavor results are also shown.

Plotting $p/T^4$ (which is the normalized free energy density),
we can compare our result to other approaches. At low temperatures
the Hadron Resonance Gas model (using the 2014 PDG spectrum) gives a good
description of the lattice data. This was already observed in
Ref.~\cite{Borsanyi:2013bia}.

In Ref.~\cite{Borsanyi:2013bia} we gave a simple parametrization for the
2+1 flavor equation of state. Here we update the 2+1 flavor parameters and
provide a parametrization that covers the
100-1000~MeV temperature range and describes the 2+1+1 lattice data,
i.e. including the effect of the charm quark. As before, the parametrizing
formula reads
\begin{equation}
\frac{I(T)}{T^4} = \exp(-h_1/t-h_2/t^2)\cdot\left( h_0 + f_0 \frac{ \tanh(f_1\cdot t+f_2)+1}{1+g_1\cdot t + g_2 \cdot
t^2} \right),
\label{eq:param211}
\end{equation}
with $t=T/200~\mathrm{MeV}$. The parameters are given in
Table.~\ref{tab:coeffs}, the resulting curves are shown in
Fig.~\ref{fig:param}.
For completeness the $n_f=2+1$ parametrization is also shown.

\begin{table}
\begin{center}
\begin{tabular}{|c||c|c|c|c|c|c|c|c|}
\hline
&$h_0$&$h_1$&$h_2$&$f_0$&$f_1$&$f_2$&$g_1$&$g_2$\\
\hline
\textbf{2+1+1 flavors} & 0.353 & -1.04 & 0.534 & 1.75 & 6.80 & -5.18 & 0.525 & 0.160\\
\textbf{2+1 flavors}   & -0.00433 & -1.00 & -0.288 & 0.293 & 6.10  & -4.90 & -0.787 & 0.289\\
\hline
\end{tabular}
\end{center}
\caption{\label{tab:coeffs}
Constants for our parametrization of the trace anomaly in Eq.~(\ref{eq:param211}).
}
\end{table}


\subsection{The 2+1+1+1 flavor QCD equation of state}

Here we present our final result on 2+1+1+1 flavor QCD. The bottom threshold
has been added as described in Sec.~\ref{sec:bottom}.  We use the 2+1 flavor
lattice results up to 250 MeV, 2+1+1 flavor data up to 500 MeV. In the range
500\dots1000~MeV we observed that our $\mathcal{O}(\alpha_s^3)$ order
perturbative result agrees very well with the 2+1+1 flavor lattice data.  This
justifies the use of the $\mathcal{O}(\alpha_s^3)$ formula to include the
effect of the bottom quark as described in Sec.~\ref{sec:bottom}.  The effect
of the bottom quark starts at a temperature of about 600 MeV. See
Fig.~\ref{fig:withmass}.

\begin{figure}[ht]
\begin{center}
\includegraphics[width=3.4in]{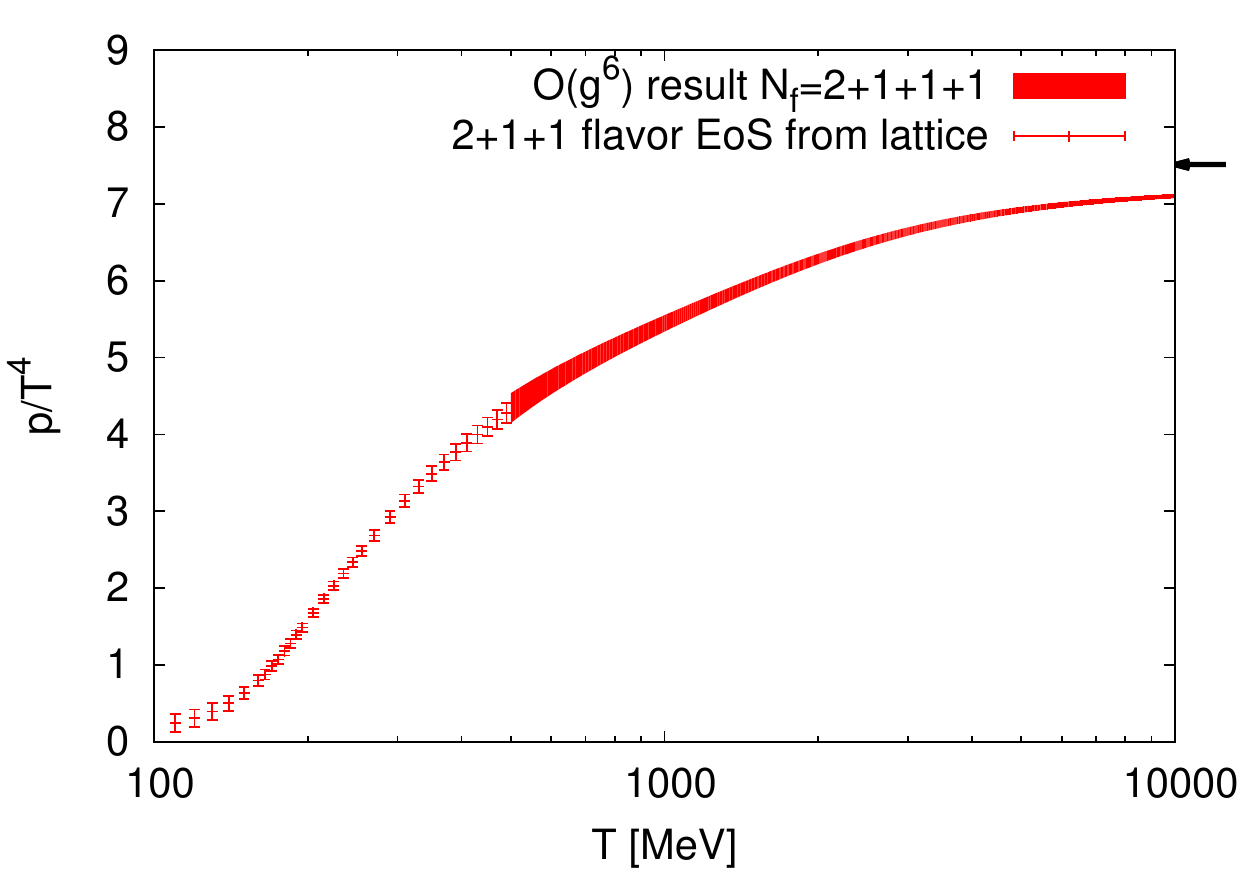}\hspace{3mm}
\includegraphics[width=3.4in]{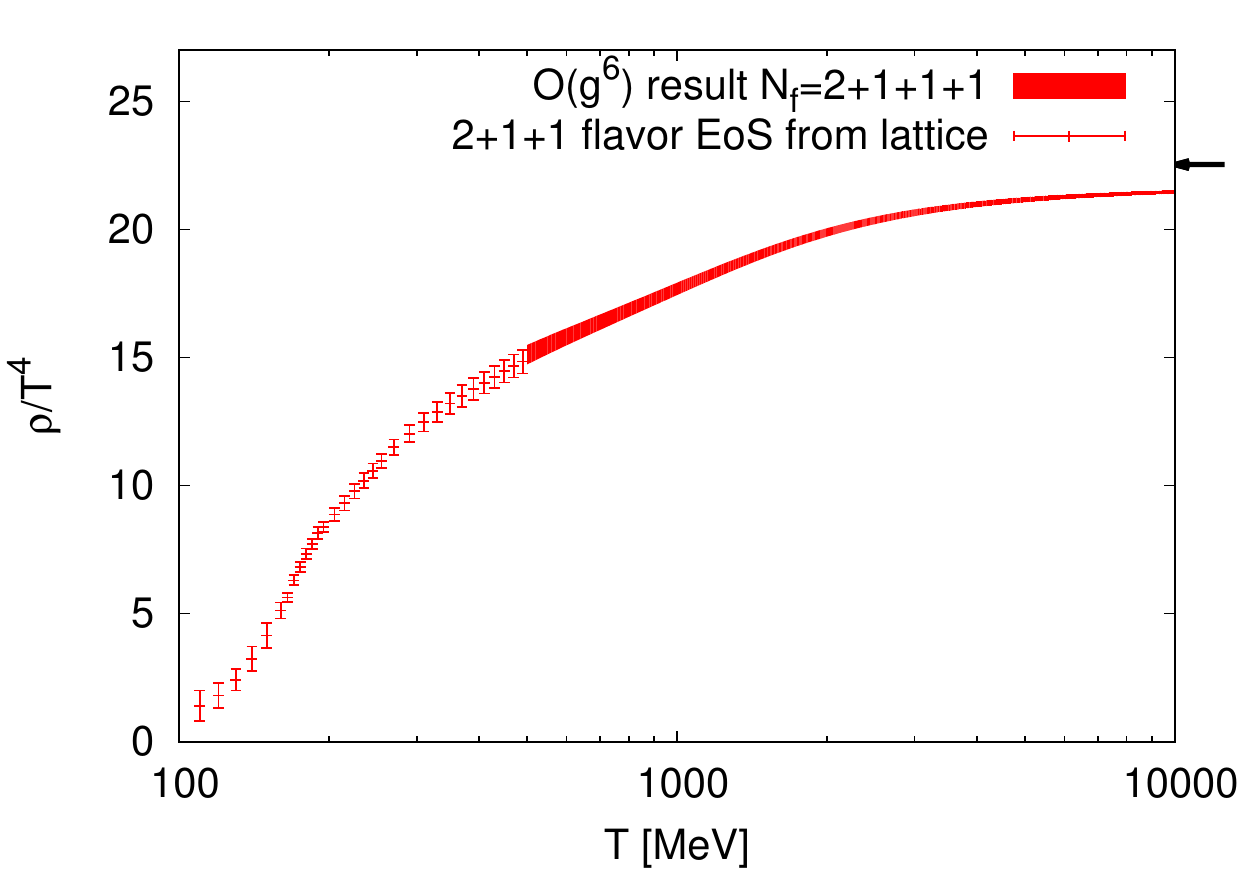}\\
\includegraphics[width=3.4in]{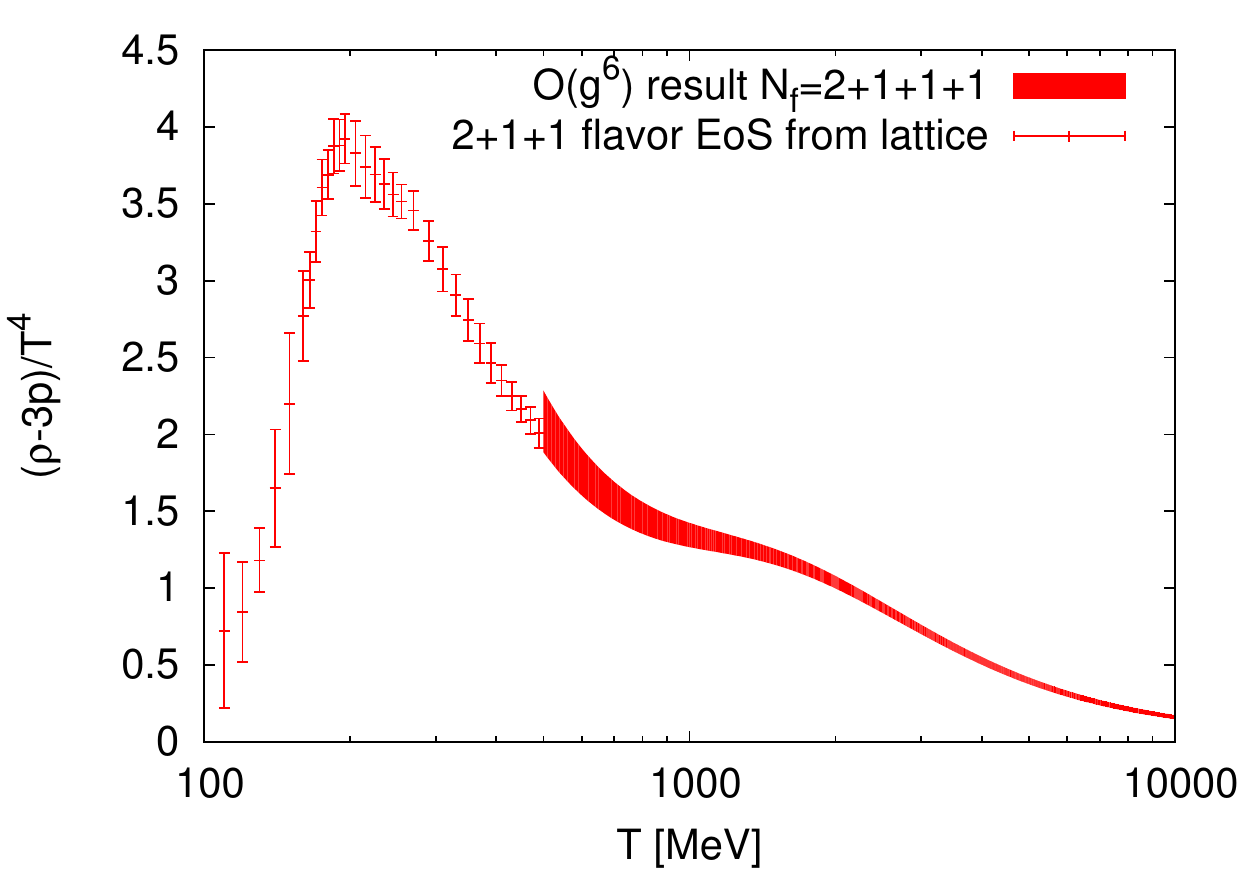}
\end{center}
\caption{
The QCD pressure, energy density and the trace anomaly in the 2+1+1+1 theory.
\label{fig:bottomeos}
}
\end{figure}

Because of its large mass the top quark can only be included in the framework
of the electroweak theory. Thus the calculation of the bottom quark's effect
completes the discussion of the QCD contribution to the equation of state. The
resulting thermodynamic functions are shown in Fig.~\ref{fig:bottomeos}.

\subsection{The Standard Model}

Now we are in the position to construct an equation of state that gives
a good description over the entire temperature range of the Standard Model.

In particular, we compute the effective number of degrees of freedom. This is
defined by the energy density or the entropy normalized by the
Stefan-Boltzmann limit of a single scalar field:
\begin{equation}
g_{\rho}(T)=\rho(T)\frac{30}{\pi^2T^4}\,,\qquad
g_{s}(T)=s(T) \frac{45}{2\pi^2T^3}\,.
\label{eq:gdef}
\end{equation}
(These quantities are not to be confused with the strong coupling constant of the previous section.)

The final result for the pressure is the combination of the following contributions,
the energy density and the entropy density can be obtained from the pressure
using the thermodynamical identities
\begin{equation}
\rho(T) = 3p(T)  +T^5 \frac{d p(T)/T^4}{dT}\,,\qquad
s(T) T = \rho(T)+ p(T)\,.
\end{equation}

\textit{a) Photons and neutrinos}\\
We treat these light particles in their Stefan-Boltzmann limit, assuming three
generations of left-handed neutrinos. In this paper we work out the equilibrium
equation of state, the neutrinos give a trivial
contribution of $p_\nu/T^4 = \frac{7}{8} \cdot 2 \cdot 3 \cdot \pi^2/90$, for the photons
we have $p_\gamma/T^4 = \pi^2/45$.

\textit{b) Charged leptons}\\
We sum the free energy of the non-interacting leptons with the formula
\begin{equation}
p/T^4 = \frac{1}{2\pi^2} \sum_i g_i \left(\frac{m_i}{T}\right)^2 \sum_{k=1}^{\infty} \frac{(-1)^k}{k^2} K_2\left(\frac{m_ik}{T}\right)\,,
\label{eq:fermionsum}
\end{equation}
where $K_2$ is a modified Bessel function of the second kind, and $g_i$ is the 
spin degeneracy factor, $g_i=4$ for leptons, and $m_i$ is its mass. The right hand side of Eq.~(\ref{eq:fermionsum}) with $g_i=12$ gives the free quark contribution $F_Q(m_i/T)$ already introduced
in Eq.~(\ref{eq:charmcorr3}).

\textit{c) Light hadrons}\\
In Refs.~\cite{Borsanyi:2013bia,Bellwied:2015lba} we have tested the Hadron Gas
Model's prediction in detail. Here we use this model's prediction up to a
temperature of 120~MeV. From that point we switch over to the lattice result.

\textit{d) QCD}\\
We have continuum extrapolated lattice results for the equation of state up to
a temperature of 1~GeV. We have found these as a combination of 2+1 and 2+1+1 flavor
simulations.  The lattice data can be smoothly continued with $\alpha_s^3$
order perturbative result, where one analytically unknown parameter was fitted
to our data.
In Fig.~\ref{fig:withmass} we showed that the sixth order 
(highest order with one fitted coefficient) perturbative result gives a good
description of both the pressure and the trace anomaly.
This pressure function is the basis of our result at high temperatures.
We included the bottom threshold using the method described in
Sec.~\ref{sec:bottom}. The full 2+1+1+1 flavor QCD contribution we show
in Fig.~\ref{fig:bottomeos}.

\textit{e) $W^\pm, Z^0$ and the Higgs boson}
The bosonic version of Eq.~(\ref{eq:fermionsum}) can be used as a first estimate:
\begin{equation}
p/T^4 = \frac{1}{2\pi^2} \sum_i g_i \left(\frac{m_i}{T}\right)^2 \sum_{k=1}^{\infty} \frac{1}{k^2} K_2\left(\frac{mk}{T}\right)\,.
\end{equation}
Ref.~\cite{Laine:2006cp} goes beyond this and adds the one-loop electroweak corrections.
The one loop corrections become noticeable at the
temperature of approximately $T\gtrsim 90~\mathrm{GeV}$. For this correction
we use the data of Ref.~\cite{Laine:2006cp}.

\textit{f) The electroweak transition}
For the electroweak epoch  we quote the results of Ref.~\cite{Laine:2015kra}.
They use perturbation theory, dimensional reduction \cite{Kajantie:1995dw} and
the results of 3D simulations to estimate the equation of state of the Standard
Model near the electroweak transition.  This is a continuation of the earlier
work \cite{Laine:2006cp}.  Although at the time continuum extrapolated
electroweak lattice input was not yet available \cite{DOnofrio:2014kta}, the
final continuum extrapolation shows a very mild lattice spacing dependence
\cite{DOnofrio:2015mpa}. 

Adding all components from a) to f) we arrive at our final result for 
$g_\rho(T)$ and $g_s(T)$ that we plot in the main text. Here we give a cubic spline
parametrization for $g_\rho(T)$ and the ratio of $g_\rho(T)$ and $g_s(T)$, see Table~\ref{tab:spline}.

\begin{table}
\begin{center}
\begin{tabular}{|c|c|c|}
\hline
$\log_{10} T/\mathrm{MeV}$&$g_\rho(T)$ & $g_\rho(T)/g_s(T)$\\
\hline
0.00 & 10.71 & 1.00228\\
0.50 & 10.74 & 1.00029\\
1.00 & 10.76 & 1.00048\\
1.25 & 11.09 & 1.00505\\
1.60 & 13.68 & 1.02159\\
2.00 & 17.61 & 1.02324\\
2.15 & 24.07 & 1.05423\\
2.20 & 29.84 & 1.07578\\
2.40 & 47.83 & 1.06118\\
2.50 & 53.04 & 1.04690\\
3.00 & 73.48 & 1.01778\\
4.00 & 83.10 & 1.00123\\
4.30 & 85.56 & 1.00389\\
4.60 & 91.97 & 1.00887\\
5.00 & 102.17 & 1.00750\\
5.45 & 104.98 & 1.00023\\
\hline
\end{tabular}
\end{center}
\caption{\label{tab:spline}
Data set on the logarithmic scale that can be used with simple cubic spline interpolation
to find a parametrization for the entire Standard Model. 
The spline's typical deviation from $g_\rho$ is about 1\%, and 0.3\% for the ratio.
}
\end{table}

\begin{figure}[ht]
\centerline{\includegraphics[height=3.4in,angle=270]{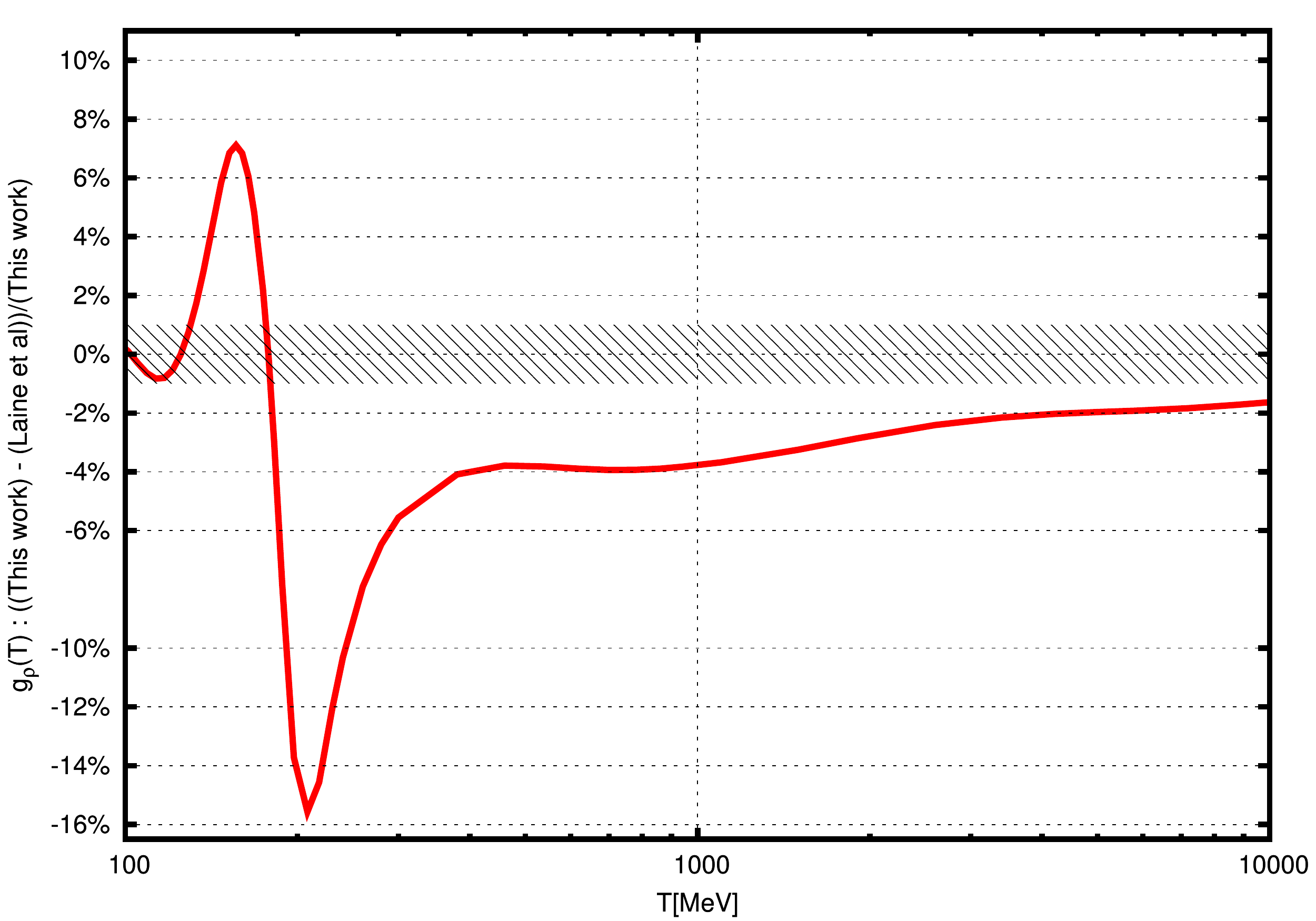}}
\caption{
\label{fig:vs_laine}
Our results and a previous estimate \cite{Laine:2015kra}.
We computed the relative difference of the two results.
Before and after the QCD epoch there is agreement. The discrepancy in the range
between 0.1 and 10~GeV is explained by the lack of lattice QCD input in
\cite{Laine:2015kra}. The dashed region around $\pm 1$\% indicates the
systematics of our parametrization.
}
\end{figure}

Ref.~\cite{Laine:2015kra} has constructed an equation of state of the universe.
In their work only the electroweak theory was based on lattice simulations.
Here we replace the earlier perturbative deliberations on the QCD epoch by
fully controlled lattice QCD result, which we conveniently parametrize. In
Fig.~\ref{fig:vs_laine} we show this non-perturbative effect by comparing our
result to the published data set in Ref.~\cite{Laine:2015kra}. 

Our final result for the quantities in Eq.~(\ref{eq:gdef}), and their ratios
is shown in Fig.~\ref{fi:eos} of the main paper.

\section{Overlap simulations}
\label{se:som_ovalgo}

One of the results of this paper, namely the mass dependence of the topological
susceptibility, is obtained using overlap fermions.  In this Section we give a
short summary of the numerical simulations with overlap quarks used in this
work. We also describe our method to determine the Lines of Constant Physics
with overlap fermions.

Our setup is based on the one used in
Refs.~\cite{Borsanyi:2012xf,Borsanyi:2015zva}.  For completeness we give a
brief summary here:
\begin{itemize}

    \item tree level Symanzik improved gauge action with gauge coupling
	parameter $\beta$.

    \item three flavors of overlap quarks.  The sign function in the Dirac
	operator $D_{ov}$ is computed using the Zolotarev approximation. The Dirac operator is
	constructed from a Wilson operator $D_{W}$ with mass $-1.3$.  The quark
	fields are coupled to two step HEX smeared \cite{Durr:2007cy} gauge
	fields.

    \item two flavors of Wilson fermions using the above $D_{W}$ operator.

    \item two scalar fields with mass $0.54$.

\end{itemize}
The extra Wilson-fermion fields are required to fix the topological charge
\cite{Fukaya:2006vs} and to avoid difficulties when topology changing is
required \cite{Fodor:2003bh,Egri:2005cx}.  These fields are irrelevant in the
continuum limit.  Note, that their action does not constrain the topology but
suppresses the probability of the low lying modes of $D_W$. In a continuous
update algorithm, no eigenmode of $D_W$ can cross zero, which is equivalent to
having a fixed topology.  The role of the boson fields is to cancel the
ultraviolet modes of the extra Wilson fermions. These boson fields are also irrelevant in
the continuum limit.

\subsection{Odd flavor algorithm}

The main difference to the works in \cite{Borsanyi:2012xf,Borsanyi:2015zva} is, 
that here we use $n_f=2+1$ flavors instead of $n_f=2$. The simulations are done
using the standard Hybrid Monte Carlo (HMC) algorithm.  For the strange quark
we use the chiral decomposition suggested first in Ref.~\cite{Bode:1999dd}
and later in \cite{DeGrand:2006ws}. The square of the Hermitian Dirac operator
$H_{ov}^2=(\gamma_5 D_{ov})^2$ can be decomposed as $H^2_{\pm}$=$P_\pm H_{ov}^2
P_\pm$ with $P_\pm=(1 \pm \gamma_5)/2$. It is trivial to show that the
determinant of the one flavor Dirac operator is $\det D_{ov} \sim \det
H^2_{\pm}$ where the proportionality constants depend only on the topological
charge.  Since we do our runs at fixed topology, these constants can be factored
out from the partition function and become irrelevant.  Therefore a
straightforward HMC with either of the $H^2_{\pm}$ operators corresponds to
simulating a single flavor at fixed topology. The actual implementation is very
simple: the pseudo-fermion generated at the beginning of each trajectory of the HMC
has to be projected to one of the chirality sectors.

\begin{figure}[h]
    \centering
    \includegraphics*{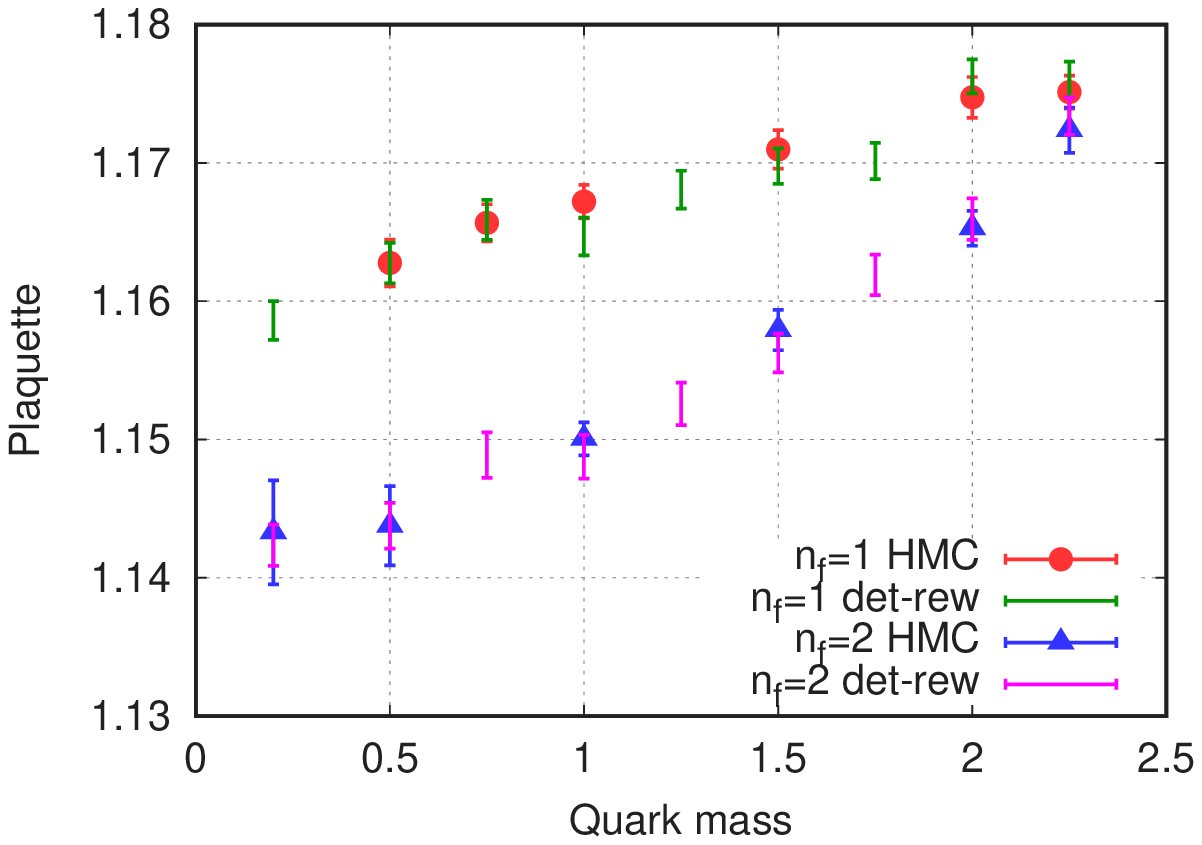}
    \caption{
	\label{fi:1p1test}
	Comparison of the plaquette obtained by two different updating
	algorithms: the HMC algorithm and determinant reweighting. Shown are
	the average plaquette in the simulation as a function of the quark mass
	with both updating algorithms. The plot shows $n_f=1$ and $n_f=2$ flavor
	overlap simulations on a test lattice of size $4^4$.
    }
\end{figure}

We tested the simulation code by comparing its results to a brute force update,
in which the configurations are picked from a pure gauge heatbath and
subsequently reweighted by the exact determinant. Since the determinant
calculation is expensive, we ran the test on a small lattice, $4^4$. We found a
perfect agreement between the two updating algorithms, as shown in Figure
\ref{fi:1p1test}.  We did tests by running the code with two copies of
$n_f=1$ fields and comparing the results obtained with our previous code for
$n_f=2$ and agreement was found in this case, too.
	
\subsection{Determination of the LCP}

We now show how to determine the LCP for physical quark masses in the overlap
formalism.  It usually requires zero temperature simulations at the physical point,
which is for overlap quarks prohibitively expensive with todays computer resources.
Fortunately, these are not needed. Here we present an alternative, cost efficient
strategy to determine the physical LCP with overlap quarks, based on the physical LCP,
that is already known in a different, less expensive fermion formulation. In our
case this will be an LCP with staggered fermions, which we have described in
the previous Section, see Equation \eqref{eq:oldlcp}.

The idea is to use the inexpensive three-flavor symmetric $n_f=3$ theory as a
bridge between the staggered and the overlap LCP's. In Subsection \ref{se:nf3}
we already determined the pion mass and the $w_0$ scale in the $n_f=3$ theory in the
continuum limit using staggered quark simulations\footnote
{
    Note, that the staggered theory also contained the charm quark, whereas the overlap
    simulations not. The continuum values of $m_{\pi}^{(3)}$ and $w_0^{(3)}$ are expected
    to be insensitive to the presence of the charm.
    
}. This can be used to construct an $n_f=3$ overlap LCP, by tuning the quark
mass for each gauge coupling so that $m_{\pi} w_0\equiv
m_{\pi}^{(3)}w_{0}^{(3)}$, and in this way we get the quark mass function
$m_{s}^{ov}(\beta)$. This function is of course not the same as
$m_s^{st}(\beta)$, as they are obtained with different fermion
discretizations.  The important point is, that both define the same physics,
e.g. they both give the same $m_{\pi} w_0$ in the continuum limit.  To close the
definition of this three-flavor overlap LCP one has to measure $w_0^{ov}(\beta)$,
the $w_0$-scale
as a function of the coupling.
From this non-physical LCP one can get a physical $n_f=2+1$ LCP with overlap
fermions as follows
\begin{align}
    \label{eq:newlcp}
    m_s= m_{s}^{ov}(\beta), \quad m_{ud}= R\cdot m_{s}^{ov}(\beta), \quad a= w_{0}^{(3)}/w_{0}^{ov}(\beta).
\end{align}
Here the value of the lattice spacing at the physical point was obtained by
dividing the three-flavor continuum value of $w_0$ in physical units by the
dimensionless three-flavor $w_0^{ov}$-scale measured in the overlap
simulations.

\begin{figure}[h]
    \centering
    \includegraphics*{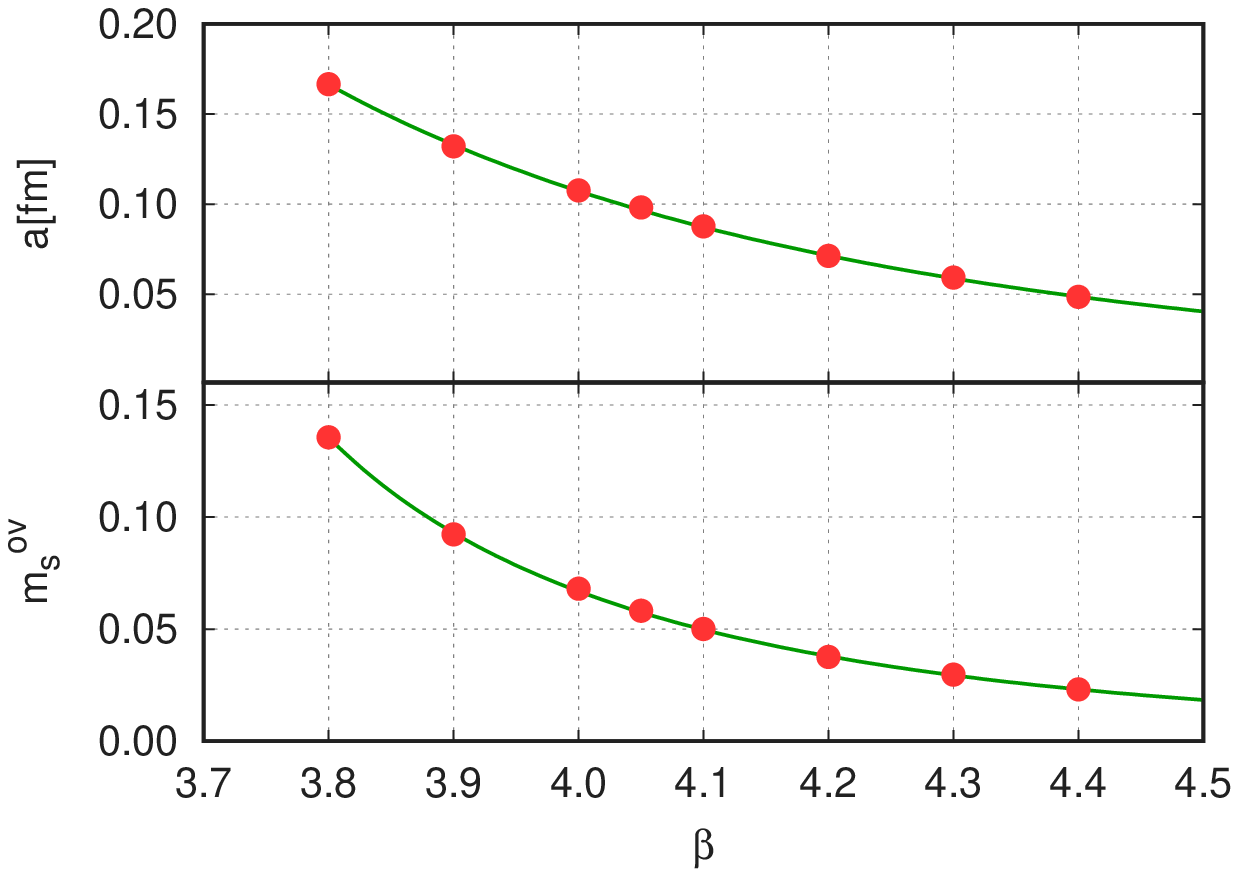}
    \caption
    {
	\label{fi:nf3ov}Lines of constant physics in the physical $n_f=2+1$
	flavor theory with overlap fermions. The upper plot shows the lattice
	spacing, the lower plot the strange mass parameter, both as the
	function of $\beta$. The light mass can be obtained as $m_{ud}=R\cdot m_s^{ov}$. The errors
	are smaller than the symbol size, the lines are smooth interpolations between the points.
    }
\end{figure}

For the $n_f=3$ flavor overlap LCP we performed simulations at parameters
listed in Table \ref{ta:ovalg}. From those we determined the quark mass by
interpolating to the point, where $m_\pi w_0=0.552$. Then we used the formulas
of Equation \eqref{eq:newlcp} to obtain the lattice spacing and quark mass
parameters for each $\beta$. The results are shown in Figure \ref{fi:nf3ov}.
Finally we fitted a three-parameter curve to these points to interpolate to
$\beta$ values, where no simulations were performed. These interpolations are
also shown in Figure \ref{fi:nf3ov}.

\begin{table}[h]
    \centering
    \begin{tabular}{|c|c|c|c|}
	\hline
	$\beta$ & $m$ & $N_t\times N_s$ & ntraj \\
	\hline
	3.80 & 0.150,0.130 & $16\times32$ & 1000 \\
	3.90 & 0.120,0.100 & $16\times32$ & 1500 \\
	4.00 & 0.090,0.070 & $16\times32$ & 2000 \\
	4.05 & 0.070,0.055 & $16\times32$ & 1200 \\
	4.10 & 0.042,0.032 & $24\times48$ & 2200 \\
	4.20 & 0.042,0.032 & $24\times48$ & 2000 \\
	4.30 & 0.030,0.025 & $32\times64$ & 1600 \\
	4.40 & 0.020,0.030 & $32\times64$ & 1400 \\
	\hline
    \end{tabular}
    \caption{
	\label{ta:ovalg}Gauge coupling parameter, quark mass, lattice size and number of trajectories
	for the three flavor overlap simulations at zero temperature, that were used to determine
	the LCP.
    }
\end{table}

\section{Eigenvalue reweighting technique}
\label{se:som_rw}

In the present Section we demonstrate how cut-off effects arise in the
topological susceptibility with staggered quarks and present a method to
efficiently suppress them. 

The cut-off effects are strongly related to the zero-modes. To understand their
importance, we first note that in the quark determinant for every zero-mode
each dynamical flavor
contributes a factor $m_f$, the corresponding quark mass.  In
this way gauge configurations with zero modes are strongly suppressed in the
path integral, especially if the quark masses are small. Due to the index
theorem, this also implies that light dynamical quarks strongly suppress higher
topological sectors and thus the topological susceptibility.

On the lattice, however, there can be strong cut-off effects in this
suppression. This is because the suppression factor is not $2m_f$ but
$2m_f+i\lambda_0$, where $\lambda_0$ is the given would-be zero mode of the
staggered Dirac operator, $D_{st}$\footnote
{
    Note, that our normalization of $D_{st}$ is
    such, that in the free field continuum limit it approaches
    $2\slashed{\partial}$.
}.The lack of exact zero modes can thus introduce
strong cut-off effects and slow convergence to the continuum limit.  Indeed, as
long as the typical would-be zero eigenvalues are comparable to or larger than
the lattice bare quark mass $m_f$, higher topological sectors are much less
suppressed on the lattice than in the continuum.

To improve the situation, even at finite lattice spacing we can identify the
would-be zero modes and restore their continuum weight in the path integral.
In case of rooted staggered quarks this amounts to a reweighting of each
configuration with a weight factor 
\begin{align}
    w[U]= \prod_{f} \prod_{n=1}^{2|Q[U]|} \prod_{\sigma=\pm}
    \left( \frac{2m_f}{\sigma i\lambda_n[U] + 2m_f} \right)^{n_f/4}
    \label{eq:reweight}
\end{align}
where the second product runs over the would-be zero eigenvalues of the
staggered Dirac operator with positive imaginary part. The third product takes
into account the $i\lambda\to -i\lambda$ symmetry of the eigenvalue spectrum.
The $n_f/4$ factor takes rooting into account, the factor $2$ next to $|Q|$
together with the $\pm$ symmetry make up for the fact that in the continuum
limit the staggered zero modes become four-fold degenerate \cite{Durr:2004as}.

Let us now turn to the most important part of the reweighting: the definition
of the would-be zero modes.  Since we are interested in the topological
susceptibility, we identify the number of these modes with the magnitude of the
topological charge $2|Q|$ as obtained from the gauge field after using the
Wilson flow, see Section \ref{se:som_zero}.  We investigated two specific
choices for the would-be zero modes. In the first approach we took the $2|Q|$
eigenmodes that have the largest magnitude of chirality among the eigenmodes
with the appropriate sign of chirality, positive if $Q<0$ and negative if
$Q>0$.  In the second approach we took the $2|Q|$ eigenmodes with smallest
magnitude. These two approaches are equivalent in the continuum limit, where
zero-modes are exactly at zero and their chirality is unity. In practical
simulations they give very similar results, we use the second approach in our
analysis.

Since in the continuum limit the would-be zero eigenvalues get closer to zero,
the reweighting factors tend to unity and in the continuum limit we recover the
original Dirac operator. In this way, even at finite lattice spacings the
proper suppression of higher topological charge sectors is restored and cut-off
effects are strongly reduced resulting in much faster convergence in the
continuum limit. For completeness let us note, that the above modification
corresponds to a non-local modification of the path integral\footnote
{
    In this respect it stands on a footing similar to another method, which
    also modifies the quark determinant and which we also use in our staggered
    simulations: determinant rooting. As of today there is ample theoretical and
    numerical evidence for the correctness of the staggered rooting. See \cite{Durr:2005ax} and
    its follow ups.
}. In the following we provide several pieces of numerical evidence for the correctness
of the approach.

\begin{figure}[h]
    \centering
    \includegraphics{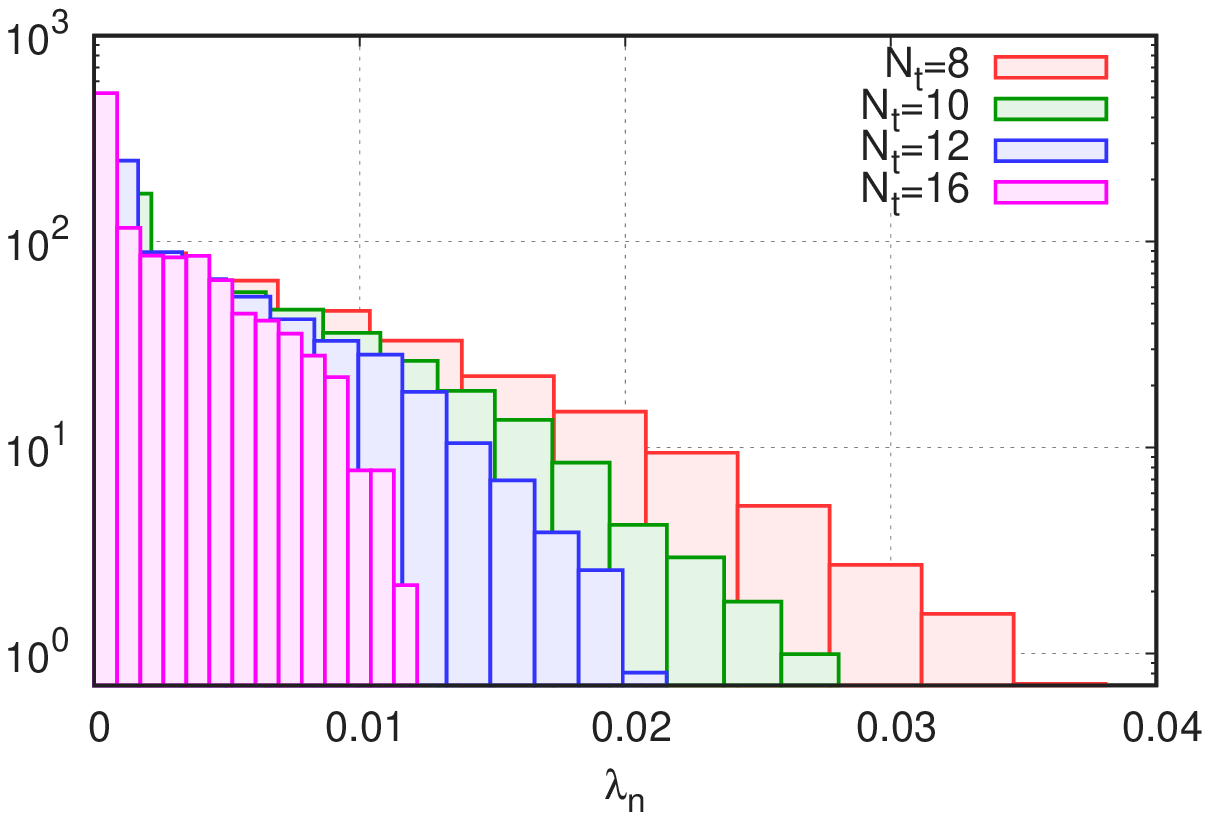}
    \caption
    {
	\label{fi:zm_dist} The probability distribution of the eigenvalues
	corresponding to the would-be zero modes obtained using the chirality
	method described in the text. The different colors refer to different
	lattice spacings. The plot shows $n_f=2+1+1$ flavor staggered
	simulations at a temperature of $T=240$ MeV.
    }
\end{figure}

\begin{figure}[p]
    \centering
    \includegraphics{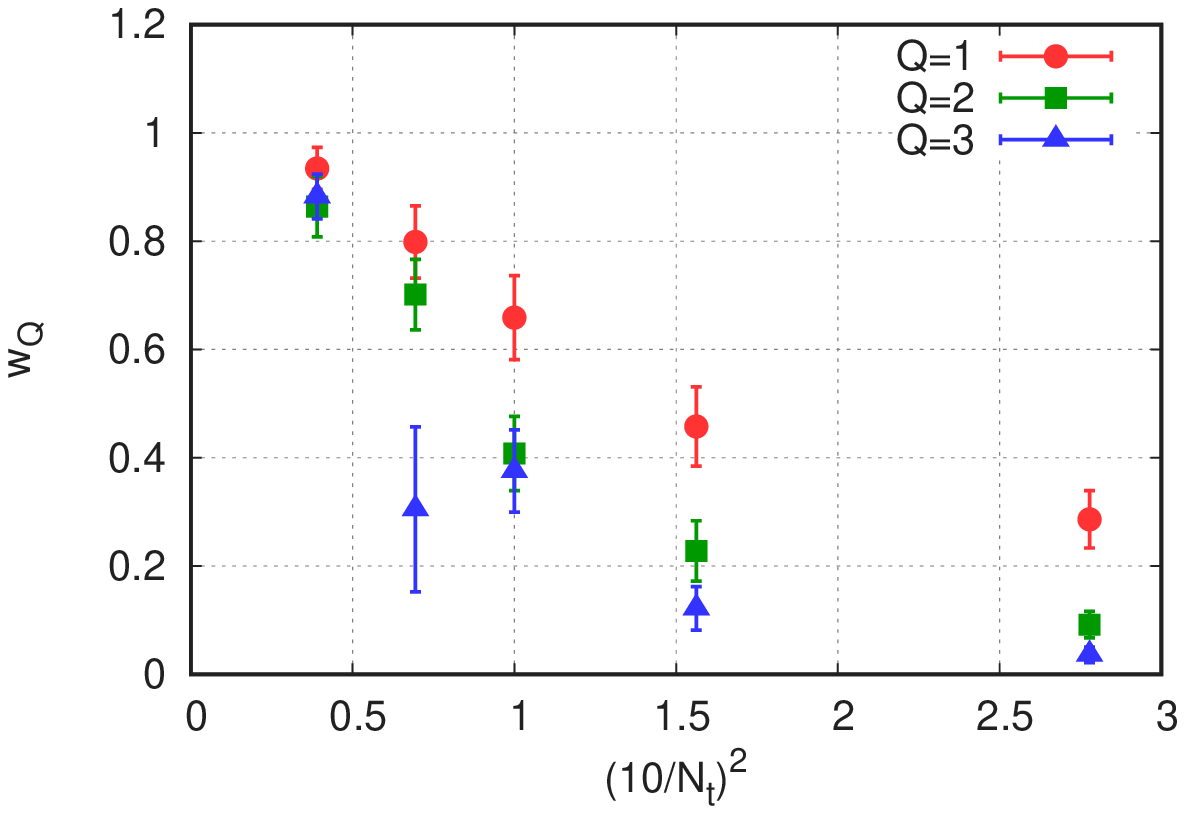}
    \caption
    {
	\label{fi:wq1} Expectation value of the weight factors in different
	topological sectors, $\langle w \rangle_Q$, as the function of the
	lattice spacing squared. The plot shows $n_f=3+1$ flavor staggered
	simulations at a temperature of $T=300$ MeV.
    }
\end{figure}

In Figure \ref{fi:zm_dist} we plot the distribution of the eigenvalues
corresponding to the would-be zero modes at a temperature of $T=240$~MeV for
different lattice spacings. The distributions get narrower and their center
moves towards zero as the lattice spacing is decreased.  In Figure \ref{fi:wq1}
we show the expectation value of the reweighting factors in the first few
topological sectors. In the continuum limit $\langle w \rangle_Q=1$ should be
fulfilled in each sector. The results nicely converge to 1. 

\begin{figure}[p]
    \centering
    \includegraphics{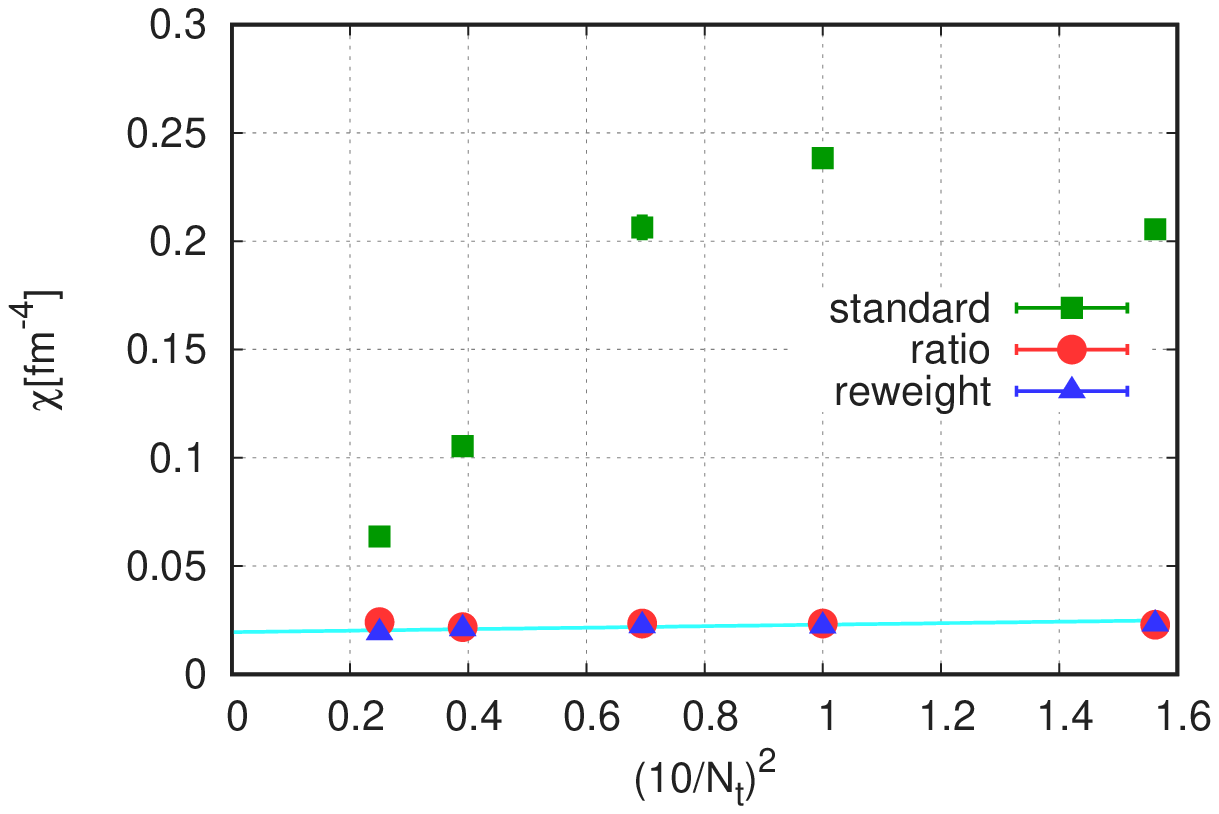}
    \caption
    {
	\label{fi:T150} Lattice spacing dependence of the topological
	susceptibility obtained from three different methods described in the
	text: standard, ratio and reweighting. For the last method a continuum
	extrapolation is also shown.  At this relatively small temperature the
	standard (``brute force'') method still cannot provide three lattice
	spacings, which extrapolate to the proper continuum limit, though they
	correspond to very fine lattices with $N_t=12,16$ and $20$.  The plot
	shows $n_f=2+1+1$ flavor staggered simulations at a temperature of $T=150$ MeV.
    }
\end{figure}

\begin{figure}[p]
    \centering
    \includegraphics{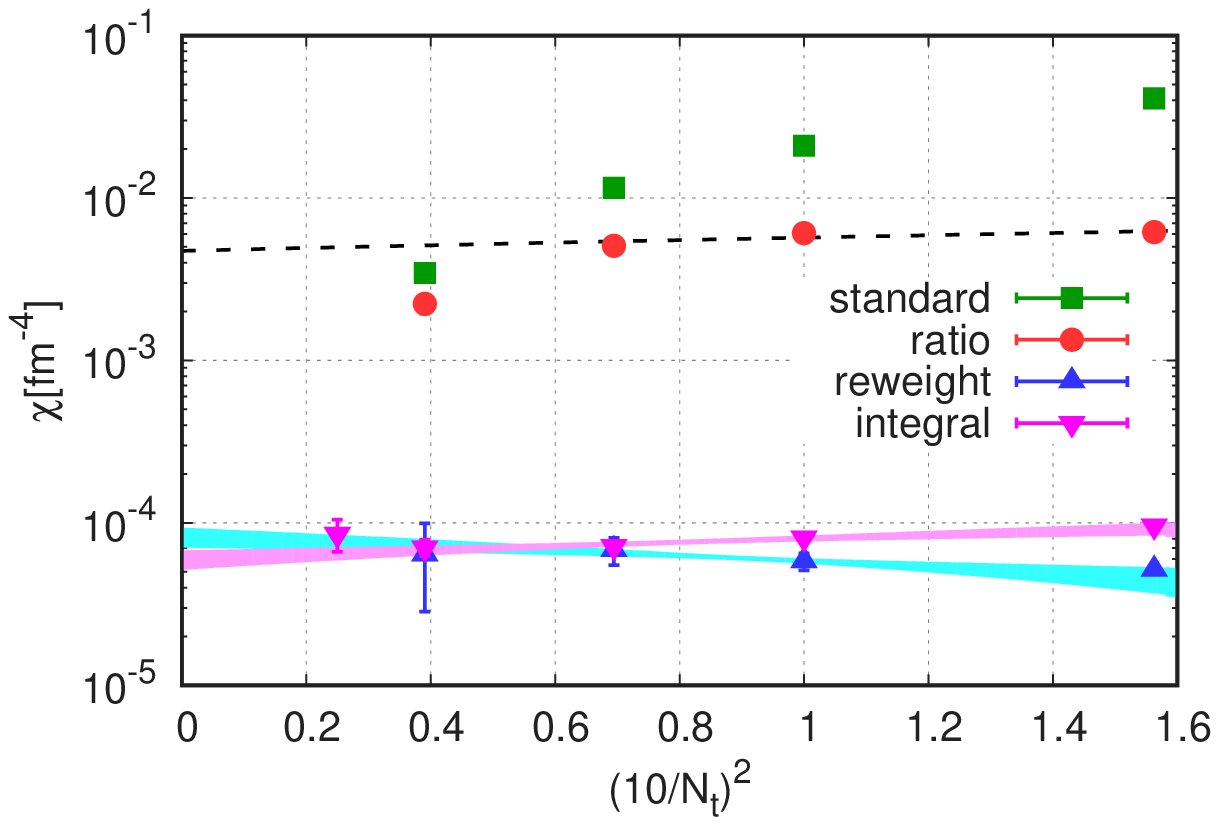}
    \caption
    {
	\label{fi:T300} Lattice spacing dependence of the topological
	susceptibility obtained from four different methods described in the
	text: standard, ratio, reweighting and integral. For the ratio method a
	misleading continuum extrapolation using $N_t=8,10$ and $12$ is shown
	with dashed line. For the reweighting and integral methods continuum
	extrapolations are shown with bands.  The plot shows $n_f=2+1+1$ flavor staggered
	simulations at a temperature of $T=300$ MeV.
    }
\end{figure}

In most of our runs, especially at large temperatures and small quark masses,
the weights were much smaller than $1$. As a result there are orders of
magnitude differences between the topological susceptibility with and without
reweighting. It is therefore important to illustrate how the brute force
approach breaks down if the lattice spacing is large and how the correct result
is recovered for very small lattice spacings.  In the following show two
examples, Figures \ref{fi:T150} and \ref{fi:T300}, where the standard method
produces cut-off effects so large, that a reliable continuum extrapolation is
not possible.  In contrast the lattice spacing dependence of the reweighted
results is much milder. To make sure that the reweighted results are in the
$a^2$-scaling regime, for both cases we present a non-standard approach to
determine the topological susceptibility and compare their results to those of
the reweighting approach.

In the first case, Figure \ref{fi:T150} the temperature is just at the
transition point, $T=150$ MeV, where we expect to get a value close to the zero
temperature susceptibility.  This suggests that in this case the cut-off
effects of the standard method can be largely eliminated by performing the
continuum limit of the ratio $\chi(T,a)/\chi(T=0,a)$, where the finite
temperature result is divided by the zero temperature one at the same lattice
spacing. We call this approach ``ratio method'', see e.g. \cite{Bonati:2015vqz}.
As it can be seen in the Figure, this is indeed the case. The so obtained
continuum extrapolation is nicely consistent with the value from the
reweighting approach.

In the second case, Figure \ref{fi:T300}, we have a temperature well above the
transition, $T=300$ MeV. We see again, that the standard method produces
results with large cut-off effects.  The ratio method seems to perform better,
however the apparent scaling is misleading. Although a nice continuum
extrapolation can be done from lattice spacings $N_t=8,10$ and $12$, the
$N_t=16$ result is much below the extrapolation curve. The reweighting produces
a result that is an order of magnitude smaller. In Sections \ref{se:som_ym},
\ref{se:som_st} and \ref{se:som_ov} we introduce a new method, called
``integral method'', which is tailored for large temperatures. The so obtained
result, where no reweighting is applied, agrees nicely with the reweighted one
in the continuum limit.

These results provide numerical evidence for our expectations: the reweighting
does not only produce a correct continuum limit, it also eliminates the large
cut-off effects of staggered fermions.

\section{Fixed sector integral in the pure gauge theory}
\label{se:som_ym}

To measure the topological susceptibility in the conventional way one needs
configurations in sectors with non-zero topology. This gets increasingly
difficult for two reasons: first with increasing the temperature the weight of
the non-zero topology sectors decreases rapidly with the temperature. The
necessary computer time increases with the inverse of the susceptibility.
Secondly, as the lattice spacing is decreased, the change between topological
sectors gets very improbable, see e.g. \cite{Schaefer:2010hu}, which increases
the computer time demand even more.

In this section we present a novel way to measure the topological
susceptibility, which is especially useful for large temperatures.  To
illustrate the new approach in this section we will work in the pure gauge
theory. The gauge configurations are generated with a probability proportional
to $\exp(-\beta S_g)$, where $\beta$ is the gauge coupling parameter and $S_g$
is the gauge action. In the next two sections we apply the method for staggered
and overlap setups.

\subsection{A novel approach}

There are many proposals to increase the tunneling between the topological
sectors, see e.g. \cite{Luscher:2011kk,Laio:2015era,Mages:2015scv}, however, we do not
pursue these here.  Instead we forbid topological sector changes and
determine the relative weight of the topological sectors by measuring the $Q$
dependence of certain observables.\footnote{A few hours after the submission of the present paper to the arXiv a paper appeared by J.~Frison et al.
discussing essentially the same method~\cite{Frison:2016vuc}, though only in the quenched approximation
using coarse lattices.} More precisely we consider the following
differentials:
\begin{align}
    \label{eq:bQ}
    b_Q\equiv \frac{d\log Z_Q/Z_0}{d\log T}=
    \frac{d\beta}{d\log a} \langle S_g \rangle_{Q-0},
\end{align}
where $Z_Q$ is the partition function of the system restricted to topological
sector $Q$. In the continuum limit the sectors are unambiguously defined,
however, on the lattice several different definitions are possible, for our
particular choice on $Z_Q$ see the next subsection. In Equation \eqref{eq:bQ} we
also introduced the notation $\langle O \rangle_{Q-0}= \langle O \rangle_{Q} -
\langle O \rangle_{0}$ for the difference of the expectation of an observable between the
sectors $Q$ and $0$.  Equation \eqref{eq:bQ} gives a renormalized quantity, the
ultraviolet divergences cancel in the difference of the gauge actions.  The
important observation is, that the necessary statistics to reach a certain level of
precision on $b_Q$'s does not depend on the temperature. 

To obtain the relative weights $Z_Q/Z_0$, we just have to integrate
Equation \eqref{eq:bQ} in the temperature.  For that we start from a temperature
$T_0$, where the standard approach (or the eigenvalue reweighting for the case of full QCD) is still feasible and determine the relative
weights of the sectors $Z_Q/Z_0$ from a direct measurement.  Then by measuring the
$b_Q$'s for higher temperatures, where the direct measurement would become
prohibitively expensive, we can use the following integral to obtain the
$Z_Q/Z_0$'s:
\begin{align}
    \label{eq:int}
    Z_Q/Z_0|_{T}=
    \exp \left( \int_{T_0}^T d\log T'\ b_{Q}(T') \right)
    Z_Q/Z_0|_{T_0}.
\end{align}

If the temperature is high enough,
the contribution of $Q>1$ sectors can be neglected and the
susceptibility is given by $\chi=2 Z_1/(Z_0N_s^3N_ta^4)$. Then the rate of change
of the susceptibility $b$ is given by:
\begin{align}
    \label{eq:bsusc}
    b\equiv \frac{d\chi}{d\log T}= b_1-4,
\end{align}
where the term $-4$ takes into account, that the physical volume also changes with the
temperature.  To derive the Stefan-Boltzmann limit of Equation \eqref{eq:bsusc},
we can use that for large temperatures $\beta= 33\log a/(4\pi^2)$. The gauge
action difference is given by the classical action of one instanton $\langle
S_g \rangle_{1-0}= 4\pi^2/3$. Up to lattice artefacts we get $b=7$ in the
Stefan-Boltzmann limit.

As we have already mentioned, the statistics can be kept constant with
increasing the temperature to reach the same level of precision on $b_1$.
However, with increasing the spatial size $N_s$, the statistics has to be
increased as $N_s^3$, the computer time as $N_s^6$. This can be understood as
follows: the gauge action difference between sectors $1$ and $0$ will be
approximately given by the action of one instanton, which remains constant with
increasing volume. The gauge action $S_g$ itself however increases with the
volume and the cancellation in Equation \eqref{eq:bQ} gets more severe. This
volume squared scaling problem can be mildened by putting more and more
topological charge into the box with increasing box size.  If the topological objects are
localized, then for large volumes the action difference between sectors $1$ and $0$ can be
obtained from the difference between sectors $Q$ and $0$:
\begin{align}
    \label{eq:eqspa}
    \langle S_g\rangle_{1-0}= \langle S_g\rangle_{Q-0}/|Q|.
\end{align}
This results in $Q$-fold increase in the signal-to-noise ratio, which
translates into a $Q^2$-fold decrease in the necessary computing time.  The
total gain in computer time over the ``brute force'' approach will be discussed
in the next subsection.

\subsection{Numerical tests}

We have carried out several numerical simulations to test the new approach. We
used the Wilson-plaquette action. For scale setting we took the $r_0$-scale
parameterized as in \cite{Durr:2006ky}. To convert the temperatures into units
of $T_c$ we took $r_0T_c=0.75$.  Configurations were generated by
overrelaxation/heatbath steps. To implement the fixed topology setup we added a
Metropolis step to the end of each update, that rejected configurations if the
topological charge escaped from a predefined region (see later).  Since the
Metropolis step is a global update, one has to make sure, that the
overrelaxation/heatbath steps separately satisfy detailed balance, see
\cite{Hasenbusch:1998yb}.

\begin{figure}[h]
    \centering
    \includegraphics*{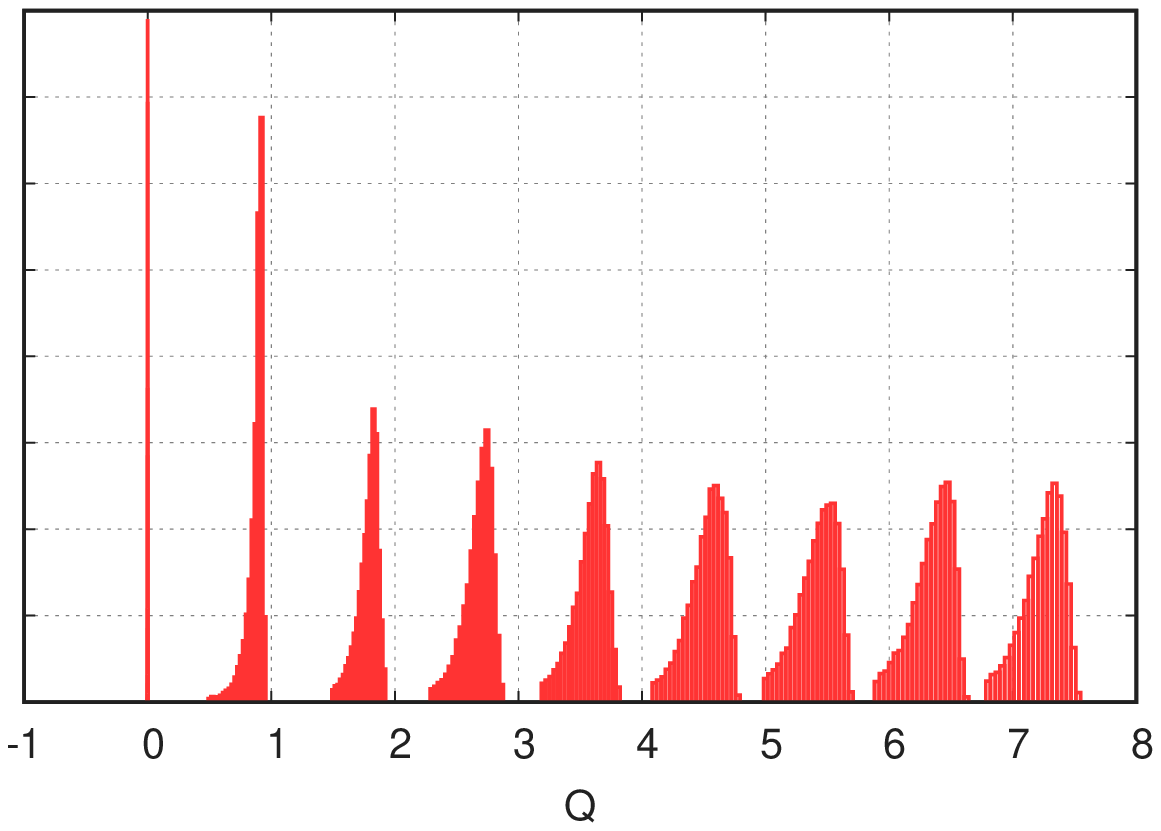}
    \caption
    {
	\label{fi:hist}
	Histograms of the topological charge from fixed sector simulations. The relative
	weight of the sectors is determined by measuring the gauge action difference, see text.
	The plot shows pure gauge theory simulations on $8\times 16^3$ lattices at $T=5T_c$ temperature.
    }
\end{figure}

The topological charge was defined using the standard clover definition after
applying a Wilson-flow for a flow time of $(8T^2)^{-1}$. There is a certain
degree of ambiguity in defining the topological sectors, this ambiguity
disappears in the continuum limit. For each $N_t$ we first explored the density
of states of $Q$, by running simulations, that were not allowed to go below some
fixed value of $Q$. For small $Q$ values we found sharp peaks, for an example
see Figure \ref{fi:hist}, which was made on $8\times 16^3$ lattices.  To define
the $Q=0,1,2\dots$ sectors, we constrain the $Q$ such, that the zeroth, first,
second, \dots peaks are in the middle of the allowed regions.  We performed fixed
topology simulations between these boundaries using the above Metropolis step.
Usually it was enough to fix only the boundary, which is closer to 0, since the
system did not attempt to cross the other boundary. For small $Q$'s we achieved
an acceptance ratio, which was around 70\% or better.

With increasing $Q$ the peaks get broader and for large $Q$'s the distributions
in the above defined sectors do not show a peak any more.  It can also happen,
that a simulation gets trapped on the sector boundary with a small acceptance
ratio. We discarded such $Q$'s and simulations in our analysis. With
approaching the continuum limit the peaks get sharper. Simulations that are
trapped on a sector boundary, no longer occur. This can also be achieved by
using gauge actions that suppress the topological tunneling, like the
tree-level Symanzik, Iwasaki or DBW2 actions \cite{DeGrand:2002vu}.

\begin{figure}[h]
    \centering
    \includegraphics*{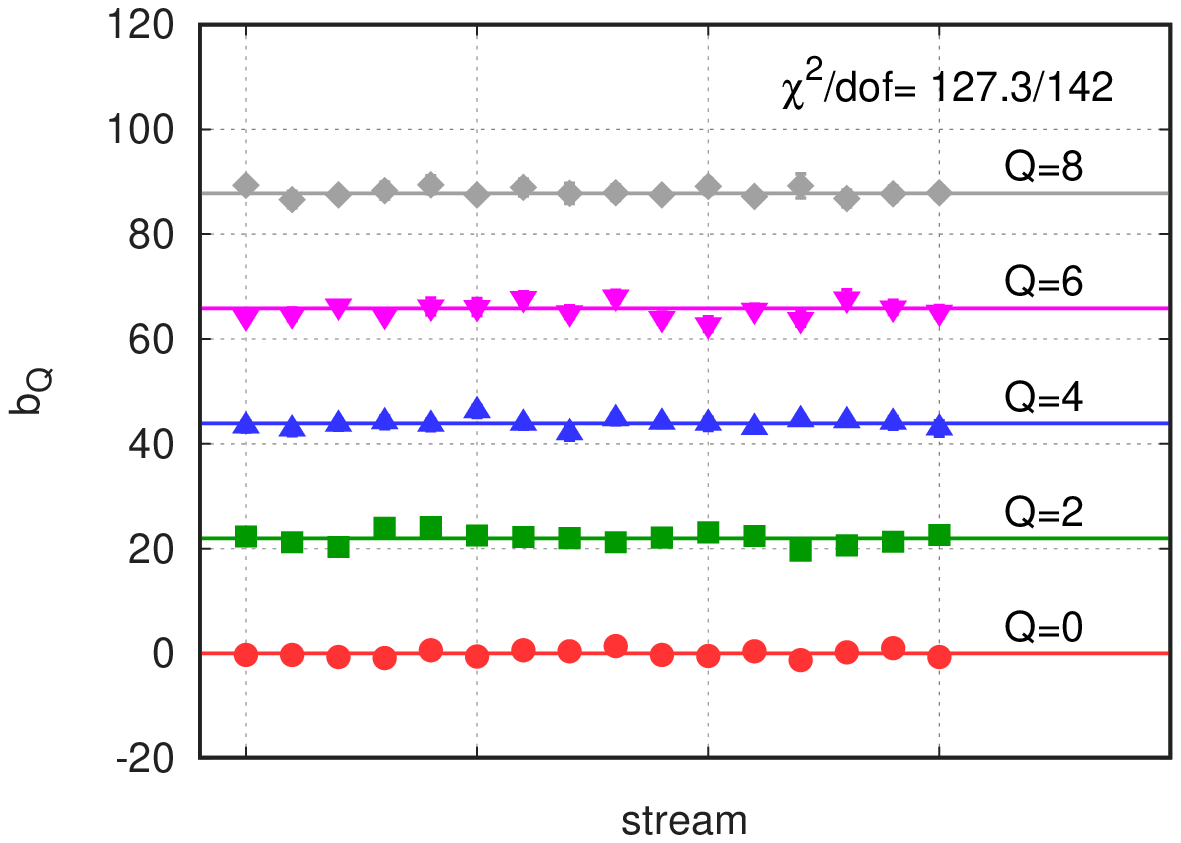}
    \caption
    {
	Gauge action difference as defined in Equation \eqref{eq:bQ}. The different points correspond to independent
	simulations and different topological sectors. A good fit can be obtained assuming ergodicity and
	that the action difference scales linearly with the topological charge, see Equation \eqref{eq:eqspa}.
	The plot shows pure gauge theory simulations on $8\times 16$ lattices at $T=5T_c$ temperature.
	\label{fi:bqs}
    }
\end{figure}

An important issue in fixed topology simulations is the ergodicity. Usually we
ran the simulations in 16 streams. The starting configurations were picked from a
simulation at a low temperature, where topology decorrelated on a timescale of
few updates. Therefore the streams can be regarded as independent. After
sufficiently many updates the gauge action was consistent between the different
streams. As an example, in Figure \ref{fi:bqs} we show the result of fixed topology runs on an
$8\times16^3$ lattice. Plotted is $b_Q$ from Equation
\eqref{eq:bQ}. The odd-$Q$ sectors are not shown. The result was obtained from
20k updates per stream. The results obtained from different streams are
all consistent with each other. The $Q$-dependence is consistent with a linear
increase of the gauge action difference with $Q$, see Equation \eqref{eq:eqspa}.
The lines represent the fit to all streams and charges assuming Equation
\eqref{eq:eqspa}.

\begin{figure}[h]
    \centering
    \includegraphics*{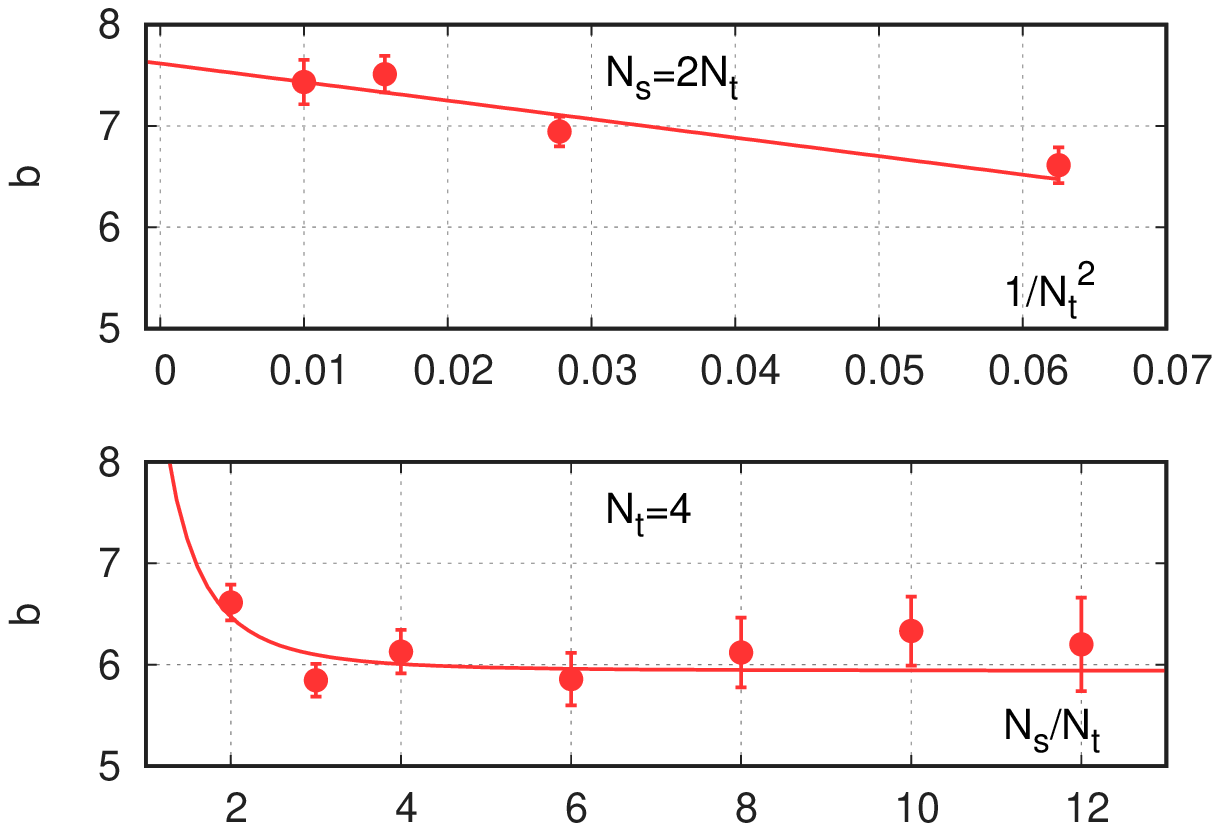}
    \caption
    {
	Lattice spacing (top) and finite volume (bottom) dependence of the
	decay exponent of the topological susceptibility $b$. The lines are
	fits taking into account leading order artefacts.  The plot shows pure
	gauge theory simulations at $T=6T_c$ temperature.  \label{fi:scan}
    }
\end{figure}

Let us now consider the finite spacing and size effects. The upper plot in
Figure \ref{fi:scan} shows $b_{1}-4$ as a function of the lattice spacing
squared in a fixed physical volume, whereas the middle panel shows it as a
function of the aspect ratio $N_s/N_t$.  Starting from aspect ratio $\approx
3$, we see no significant finite size effects.  Note, that starting from aspect
ratio $6$, the boxes are large enough to accommodate non-perturbative length
scales.  We see no difference between boxes with perturbative and
non-perturbative size.  The runs of the finite spacing and volume scans are
fitted jointly with a formula, that takes into account both effects linearly.
The fits are shown with solid lines on the plots. For the exponent we obtain
$b=7.1(3)$ in the continuum and infinite volume limit at $T=6T_c$. This is in
with our previous estimate from the direct method
\cite{Borsanyi:2015cka}.

\begin{figure}[h]
    \centering
    \includegraphics*{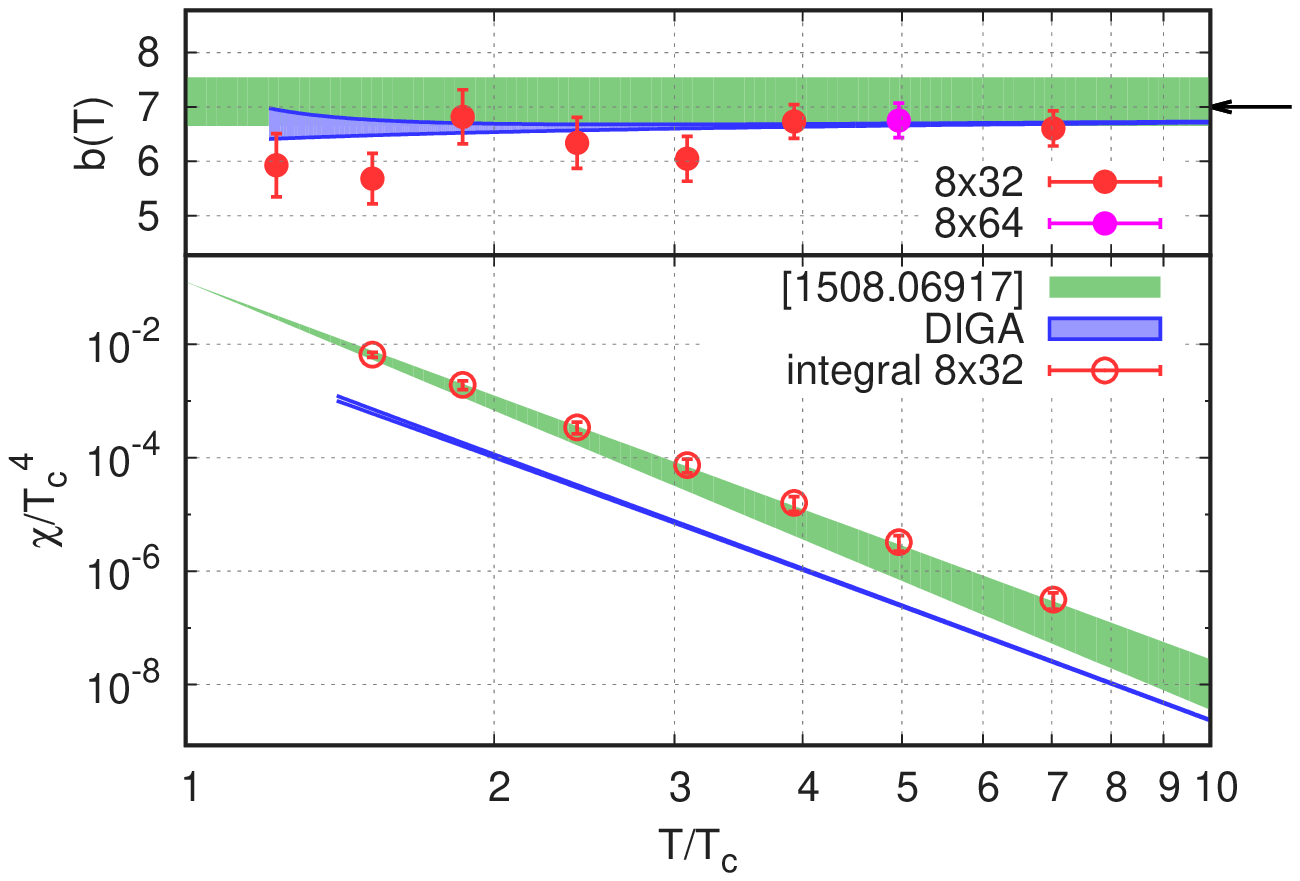}
    \caption
    {
	\label{fi:z1z0} Topological susceptibility in the pure gauge theory.
	Results shown from an earlier direct simulation \cite{Borsanyi:2015cka}, from the DIGA
	and from the
	novel fixed Q integral method.  Upper plot is the decay exponent $b$, the lower
	the susceptibility itself. The arrow indicates the Stefan-Boltzmann limit.
    }
\end{figure}

Figure \ref{fi:z1z0} shows the temperature dependence of the decay exponent
$b=b_1-4$ obtained from $8\times32^3$ simulations. Again we find agreement
between the new data and the direct approach. At one temperature we did a
simulation on an $8\times64^3$ lattice, where the exponent was obtained from
measuring the difference between the $Q=8$ and $0$ sectors, $b=b_8/8-4$. We see
no significant finite size effect. To get the $Z_1/Z_0$ ratio we performed a
direct simulation at a temperature of $T_0=1.2T_c$.  From this temperature we
integrated up the $b$ curve to obtain the $Z_1/Z_0$ as the function of
temperature, up to $7T_c$, see Figure \ref{fi:z1z0}. The result can be compared
to the lattice result obtained from the direct method \cite{Borsanyi:2015cka}
and we find a good agreement both for the exponent and the susceptibility itself.

We also calculated the prediction of the dilute instanton gas approximation
(DIGA). The necessary formulas can be found in eg. \cite{Ringwald:1999ze}. To
convert the result into units of $T_c$ we used
$T_c/\Lambda_{\overline{MS}}=1.26$ from \cite{Borsanyi:2012ve}. Three different
renormalization scales were used to test the scheme dependence: $1$, $2$ and
$1/2$ times $\pi T$. For the exponent $b$ we see a good agreement for
temperatures above $\sim 4T_c$, for smaller temperatures the lattice tends to
give smaller values than the DIGA. In case of the susceptibility the DIGA
underestimates the lattice result by about an order of magnitude, this was
already observed in \cite{Borsanyi:2015cka}. The ratio at $T=2.41T_c$ is
$K=11.1(2.6)$, where the error is dominated by the error of the lattice
calculation.

Figure \ref{fi:z1z0} was made using 30 million $8\times32^3$ and 1 million
$8\times64^3$ update sweeps. The cost of a simulation at $T=7T_c$ using the
standard method can be estimated from \cite{Borsanyi:2015cka}: it would require
about 250 million updates on $8\times64^3$ lattices or about 2 billion on
$8\times32^4$, two orders of magnitude more, than with the novel method.

\section{Fixed sector integral with staggered fermions}
\label{se:som_st}

The method presented in Section \ref{se:som_ym} can be trivially generalized in
the presence of fermions.  The definition of $b_{Q}$ is still given by Equation
\eqref{eq:bQ}. In the fixed-$N_t$ approach changing the temperature is achieved
via changing the lattice spacing, which requires a simultaneous change of
$\beta$ and the mass parameters $m_f$, to keep the system on the LCP, see
Equation \eqref{eq:oldlcp}.
Then for $b_Q$ we obtain:
\begin{align}
    \label{eq:bQst}
    b_{Q}\equiv\frac{d \log Z_Q/Z_0}{d\log T}=
	\frac{d\beta}{d\log a} \langle S_g \rangle_{Q-0}
	+ \sum_f \frac{d\log m_f}{d\log a} m_f \langle \overline{\psi}\psi_f \rangle_{Q-0}.
\end{align}
Besides the gauge action $S_g$, we also have to measure the chiral
condensate $\overline{\psi}\psi_f$ of each flavor. The full expression
is a renormalized quantity, and so is the chiral condensate difference multiplied
by the quark mass. To obtain the susceptibility we have to apply the same
integral as in the pure gauge case, see Equation \eqref{eq:int}.

Now let us look at the Stefan-Boltzmann limit of the decay exponent of the
susceptibility. We can neglect the contribution of the $Q>1$ sectors, so the
decay exponent is $b=b_1-4$. The gauge action difference is the same as in the
pure gauge case $4\pi^2/3$. The gauge parameter depends on the lattice spacing
as $\beta= (33-2n_f)\log a/(4\pi^2)$ and the mass parameter as $\log m_f= \log
a$ up to logarithmic corrections in $a$. The difference in the chiral condensate
between sectors $Q=1$ and $0$ comes entirely from the presence of the zero
mode, which gives a $\langle \overline{\psi}\psi_f \rangle_{1-0}=1/m_f$.
Altogether we have $b=(33-2n_f)/3-4+n_f$ in the high temperature limit. 

The statements of Section \ref{se:som_ym} about the computer time scaling with
the volume and the possibility of using $Q>1$ sectors also apply in the case of
dynamical fermions. We used $Q=0$ and $1$ in this work, this is sufficient,
since the topological susceptibility with dynamical fermions is tiny.

In numerical simulations the statistical noise on the gauge action difference
is much larger than on the chiral condensate difference.  This is very similar
to, what was already observed in the context of the equation of state
\cite{Borsanyi:2010cj}. This inspired us to use the following strategy:
evaluate the $b_Q$ and the susceptibility at a quark mass, where the simulation
is less expensive than at the physical point. We choose a point, the so-called
three-flavor symmetric point, where the two light-quark masses were set to the
physical strange mass: $m_{ud}\equiv m_{s,{\rm phys}}$.  At this point we
determined $\chi$ using the eigenvalue reweighting method (see later). Then we
carried out an integration in the light-quark mass from $m_{s,{\rm phys}}$ down
to the physical light-quark mass $m_{ud,{\rm phys}}=m_{s,{\rm phys}}/R$. In
this way we could avoid calculating the expensive gauge action difference at
the physical point.

\begin{figure}[p]
    \centering
    \includegraphics*[width=10.5cm]{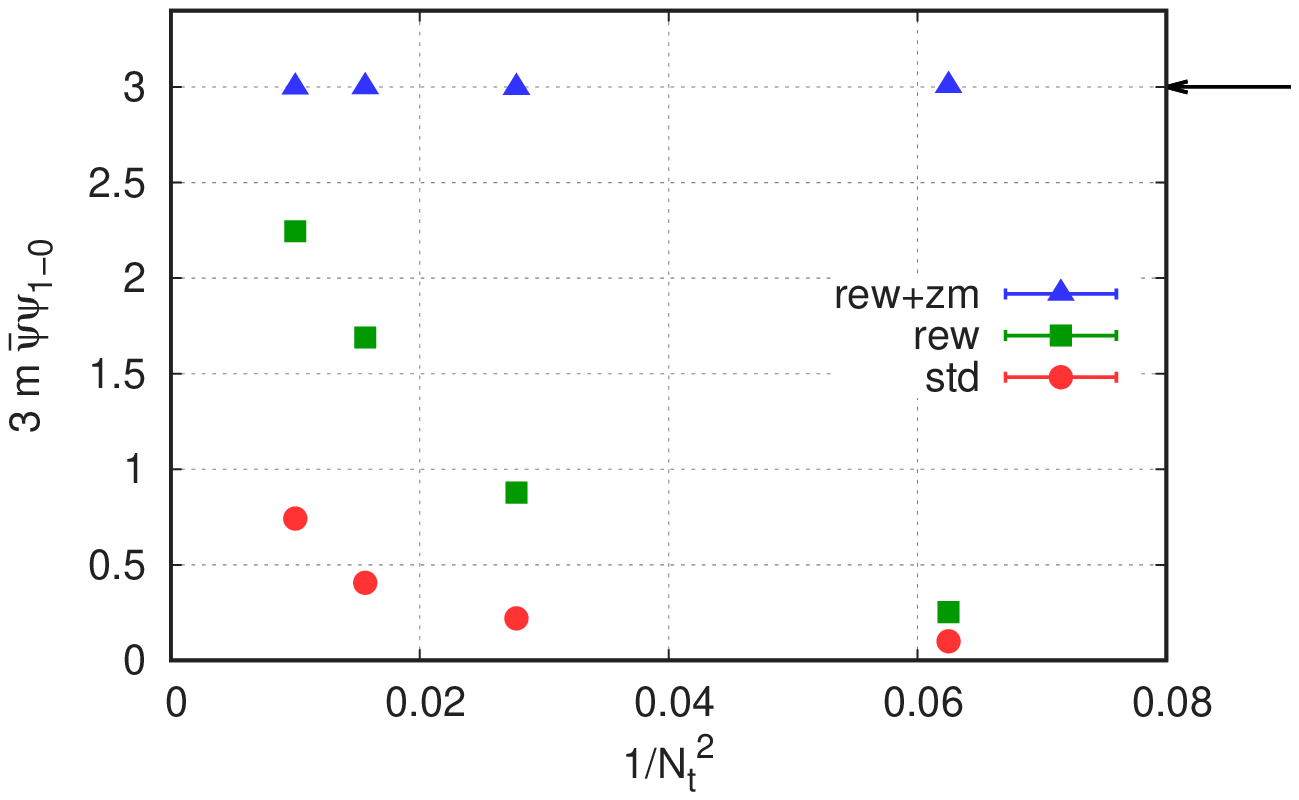}
    \includegraphics*[width=10.5cm]{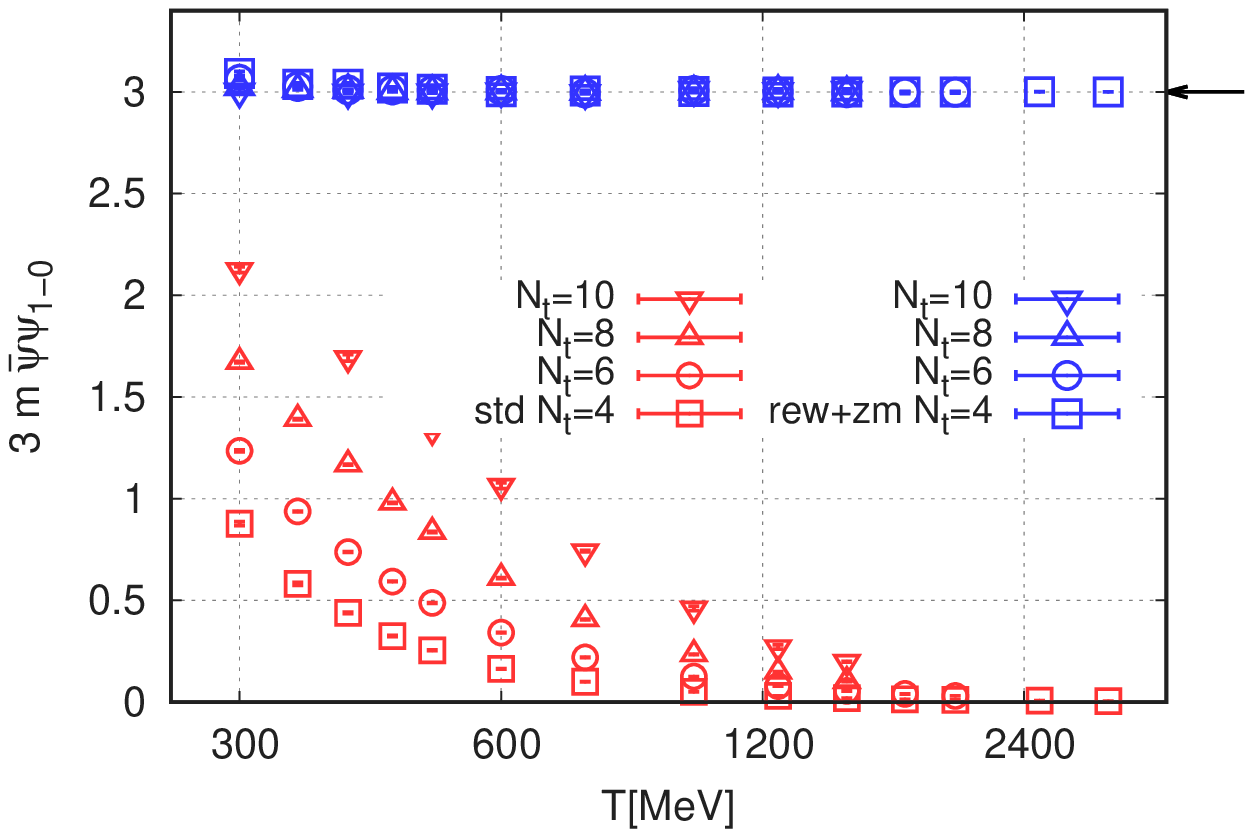}
    \includegraphics*[width=10.5cm]{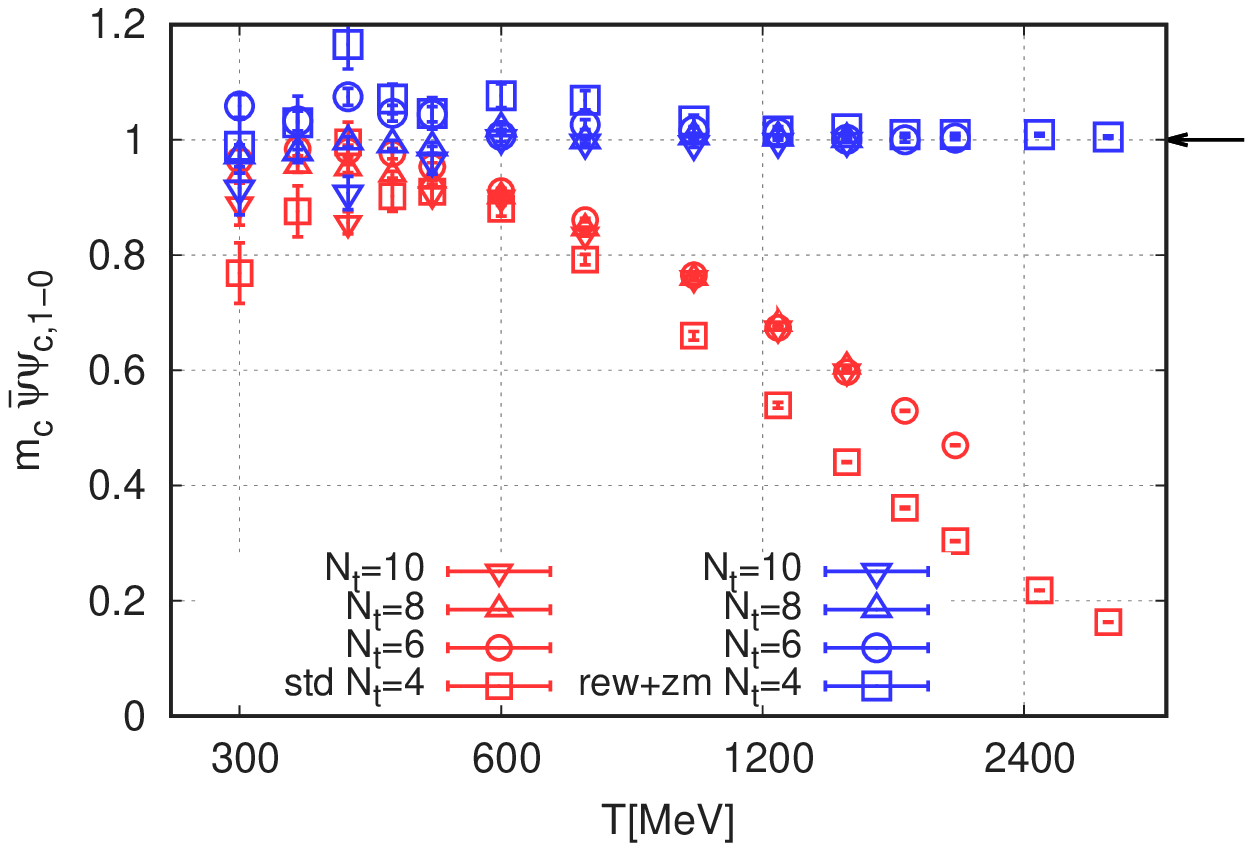}
    \caption
    {
	Chiral condensate difference between sectors $Q=1$ and $Q=0$ multiplied
	by the quark mass.  The data labeled by ``std'' denotes the value
	calculated from the standard chiral condensate. The ``rew'' data is
	obtained by reweighting with the weights in Equation \eqref{eq:reweight}.
	The ``rew+zm'' data includes the contribution of the zero modes, i.e.
	the mass dependence of the weight factors, see Equation
	\eqref{eq:bQstrw2}. The arrows indicate the Stefan-Boltzmann limit. The upper plot shows the difference as a function of
	the lattice spacing squared at $T=750$ MeV temperature. The middle
	plot shows the difference as a function of the temperature, whereas the lower plot
	is the same for the charm quark.
	The plots show $n_f=3+1$ flavor staggered
	simulations on $N_t=4,6,8$ and $10$ lattices.
	\label{fi:pbpa}
    }
\end{figure}

We observed that there are huge lattice artefacts on the chiral condensate
contribution, if a non-chiral fermion discretization is used. In the absence of
exact zero modes the chiral condensate difference needs very fine lattices to
reach the continuum limit. The lattice spacing dependence of the three flavor
chiral condensate difference is shown on the data labeled by ``std'' in the upper panel Figure \ref{fi:pbpa}.
We used $3+1$ flavor staggered quarks in the simulation at a temperature of
$T=750$~MeV.  There is an order of magnitude increase in the condensate by
going from the coarsest to the finest lattice spacing.  In the middle panel of
Figure \ref{fi:pbpa} the temperature dependence of the three flavor condensate
is shown for different lattice spacings.  As the temperature increases the
condensate, which is calculated in the standard way, approaches zero
contrary to the expectation in the high temperature limit. This also happens
for the charm quark condensate, although at a somewhat smaller pace, see lower
panel. The vanishing of the chiral condensate difference decreases the decay
exponent of the susceptibility by $n_f$ in the standard approach, which largely
explains the unexpectedly small exponent obtained in the recent lattice
calculation \cite{Bonati:2015vqz}.

We present two independent approaches to solve this problem.  One is to use a
chiral fermion discretization to evaluate the chiral condensate difference,
this is explained in a separate section, Section \ref{se:som_ov}.  The other is
the modification of the path integral by the reweighting technique, that we
already introduced in Section \ref{se:som_rw}. Let us present the details of
the reweighting here. The introduction of the reweighting factors $w[U]$ in
Equation \eqref{eq:reweight} means, that our simulation corresponds to a modified
partition function:
\begin{align}
    Z^{\mathrm{rw}}=\int [dU]\ \exp(-\beta S_g) \cdot \prod_f \det (D_{st}+2m_f)^{1/4} \cdot w[U].
\end{align}
Note, that reweighting affects only the sectors with
non-trivial topologies.
This results in a modification of the expression for $b_{Q}$ in Equation \eqref{eq:bQst} as:
\begin{align}
    \label{eq:bQstrw}
    b_{Q}^{\rm rw}=\frac{d\log Z_{Q}^{\rm rw}/Z_{0}^{\rm rw}}{d\log T}=
	\frac{d\beta}{d\log a} \langle S_g \rangle^{\rm rw}_{Q-0}
	+ \sum_f\frac{d\log m_f}{d\log a} m_f \langle \overline{\psi}\psi_f \rangle^{\rm rw+zm}_{Q-0},
\end{align}
where $\langle . \rangle^{\rm rw}_Q$ denotes the fixed $Q$ expectation value
including the weights $w[U]$.  The fermionic contribution has two parts, one
coming from the reweighted chiral condensate and another from the mass
dependence of the weight factors, which we call zero mode contribution:
\begin{align}
    \label{eq:bQstrw2}
    \langle\overline{\psi}\psi_f \rangle^{\rm rw+zm}_{Q-0}=
    \langle\overline{\psi}\psi_f \rangle^{\rm rw}_{Q-0} + \frac{|Q|}{m_f} - 
    \left\langle \frac{1}{2m_f}\sum_{n=1}^{2|Q|} \frac{4m_f^2}{\lambda^2_n[U] + 4m_f^2}\right\rangle^{\rm rw}_Q.
\end{align}
In the end we have to measure three observables, the gauge action, the chiral
condensate and the zero mode contribution, on the reweighted configurations.

The effects of reweighting and including zero mode contribution can also be seen in
the panels of Figure \ref{fi:pbpa}. In the upper panel we see, that reweighting
already improves the estimate of the fermionic contribution significantly and
including the additional zero mode contribution reduces the lattice artefacts
even further. In the middle panel of the figure we see that the reweighted
condensate together with the zero mode contribution approaches the
Stefan-Boltzmann limit for high temperatures, as expected.  Indeed for the
strange quark at our smallest temperature $T=300$ MeV
the Stefan-Boltzmann value is already reached, whereas the charm contribution
is about $20$\% lower. The temperature dependence of the latter is plotted
in the lower panel of Figure \ref{fi:pbpa}.

\begin{figure}[h]
    \centering
    \includegraphics*{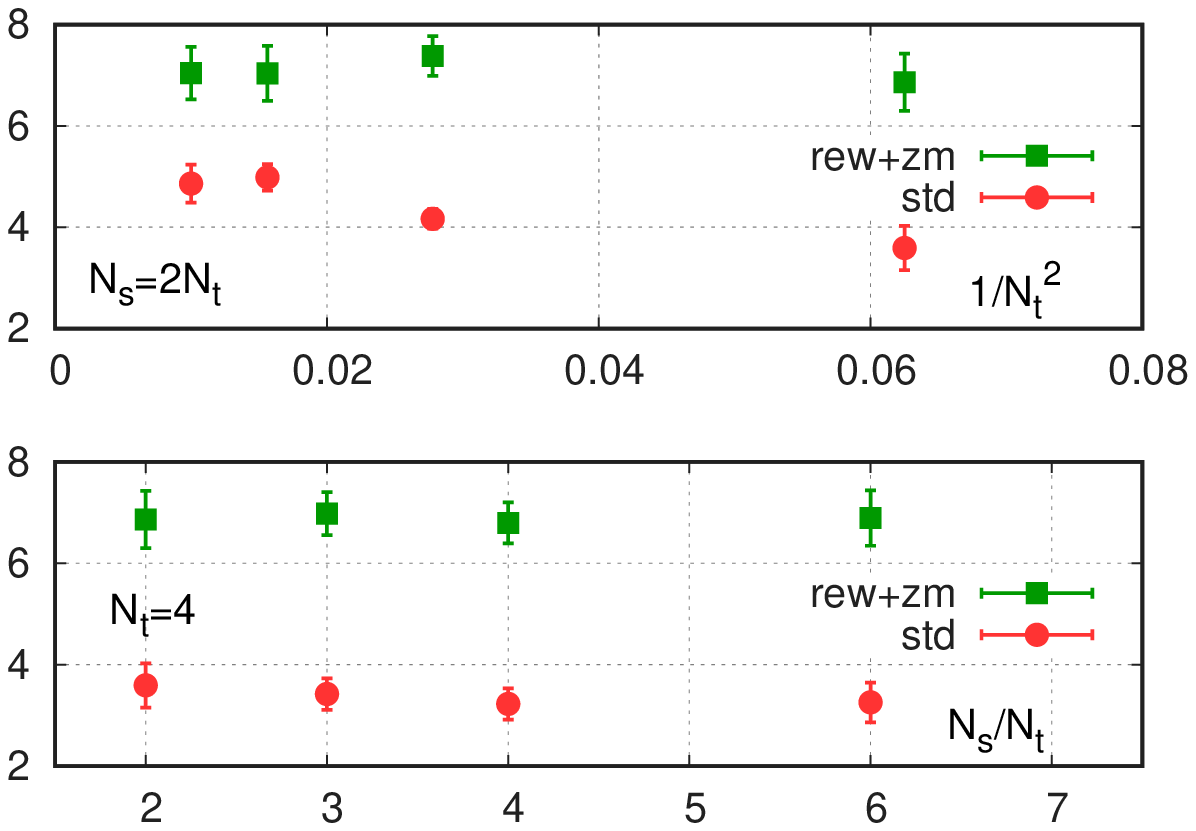}
    \caption
    {
	Lattice spacing (upper panel) and finite size (lower panel) dependence
	of the rate of change of the topological susceptibility $b$.  The
	lattice spacing dependence is shown for $N_s=2N_t$, while the finite
	volume dependence for $N_t=4$. The labels ``std'' and ``rew'' indicate,
	whether the data set was obtained with the standard or with the
	reweighting method including the zero mode contribution. The plot shows
	$n_f=3+1$ flavor staggered simulations at $T=750$ MeV temperature.
	\label{fi:baV}
    }
\end{figure}

Now let's come to the lattice artefacts and finite size effects on the total
decay exponent $b=b_1-4$, including not only the fermionic but also the gauge
contribution, see Equation \eqref{eq:bQstrw}. We did dedicated simulations at
$T=750$ MeV temperature: first we varied the lattice spacing in a fixed volume,
secondly we changed the volume at a fixed lattice spacing. The upper panel of
Figure \ref{fi:baV} shows the lattice spacing dependence of $b$ with and
without reweighting. The lattice artefacts are larger without reweighting. The
continuum extrapolations of the two data sets differ significantly.  This
difference is due to the problematic behaviour of the fermionic contribution
without reweighting, as we explained before in detail.  The lower panel of
Figure \ref{fi:baV} shows the finite size dependence of $b$. We did
simulations with an aspect ratio up to and including $6$.  The box size
corresponding to the largest aspect ratio is $1.6$~fm, which is large enough to
accommodate all non-perturbative length scales. Note, that the pion mass is
$m_\pi^{(3)}\approx 710$ MeV in these $n_f=3+1$ flavor simulations.  We see no
significant finite size effects even for an aspect ratio as small as $2$.
Although finite size and finite lattice spacing seem to be not significant at
a first glance, these effects will be properly taken into account in our final
analysis, where we perform a global fit over our data set with different
$N_t$'s and aspect ratios, see Section \ref{se:som_ana}.

\section{Fixed sector integral with overlap fermions}
\label{se:som_ov}

As demonstrated in Sections \ref{se:som_rw} and \ref{se:som_st} staggered
fermions produced huge lattice artefacts in the topological susceptibility and
in the high temperature chiral condensate.  An obvious remedy is to carry out
simulations in a chirally symmetric discretization.

Determining the topological susceptibility with overlap fermions has a long
history. The direct measurement is numerically difficult, since one has to deal
with the non-analyticity of the overlap operator on the topological sector
boundary. Though solutions exist \cite{Fodor:2003bh,Cundy:2005pi,Egri:2005cx},
they are somewhat cumbersome. Alternatively one can perform simulations in
fixed topology \cite{Fukaya:2006vs} and determine the topological
susceptibility from the long distance behaviour of the topological charge
correlator \cite{Aoki:2007ka}. This is a viable approach at zero temperature,
but for high temperatures, where the susceptibility is small, one needs to
measure the correlator with a very high precision. 

Our new approach, presented in Section \ref{se:som_st}, also requires simulations
with fixed topology. However we need to determine only the chiral condensate
difference, the rest can be taken from direct simulations at parameters, where
the direct approach is feasible.  As we have seen the use of staggered fermions
is complicated and difficult for this purpose. The difference, as we will show in this
section, can be nicely measured in the overlap formulation.  For algorithmic
and other technical details we refer the reader to Section \ref{se:som_ovalgo}. Before
showing results for the chiral condensate, we start with a previously unknown
subtlety in fixed topology simulations with overlap fermions, which is related
to configurations with a pair of an instanton and an anti-instanton.

\subsection{Instanton--anti-instanton (IA) configurations}

The overlap topological charge is defined as the difference of the number of left and
right handed zero modes of the overlap Dirac operator. A smooth
instanton/anti-instanton produces a left/right handed zero mode in the
spectrum. In practical simulations, one never encounters a configuration, where
simultaneously left and right handed zero modes are present. This of course
does not mean, that configurations with an IA pair are not allowed.  If we
look at smooth configurations, which contain a well-separated instanton and an
anti-instanton, the overlap operator has a complex conjugate pair of overlap
modes with very small but non-zero eigenvalues\footnote
{
    One can prepare an artificial configuration with simultaneous left and
    right handed overlap zero modes, where an instanton is placed in the first half
    of the volume and the second half is obtained by $CP$ transforming the
    first. Such configurations are expected to be a zero measure subset
    in the configuration space.
}.

\begin{figure}[h]
    \centering
    \includegraphics*{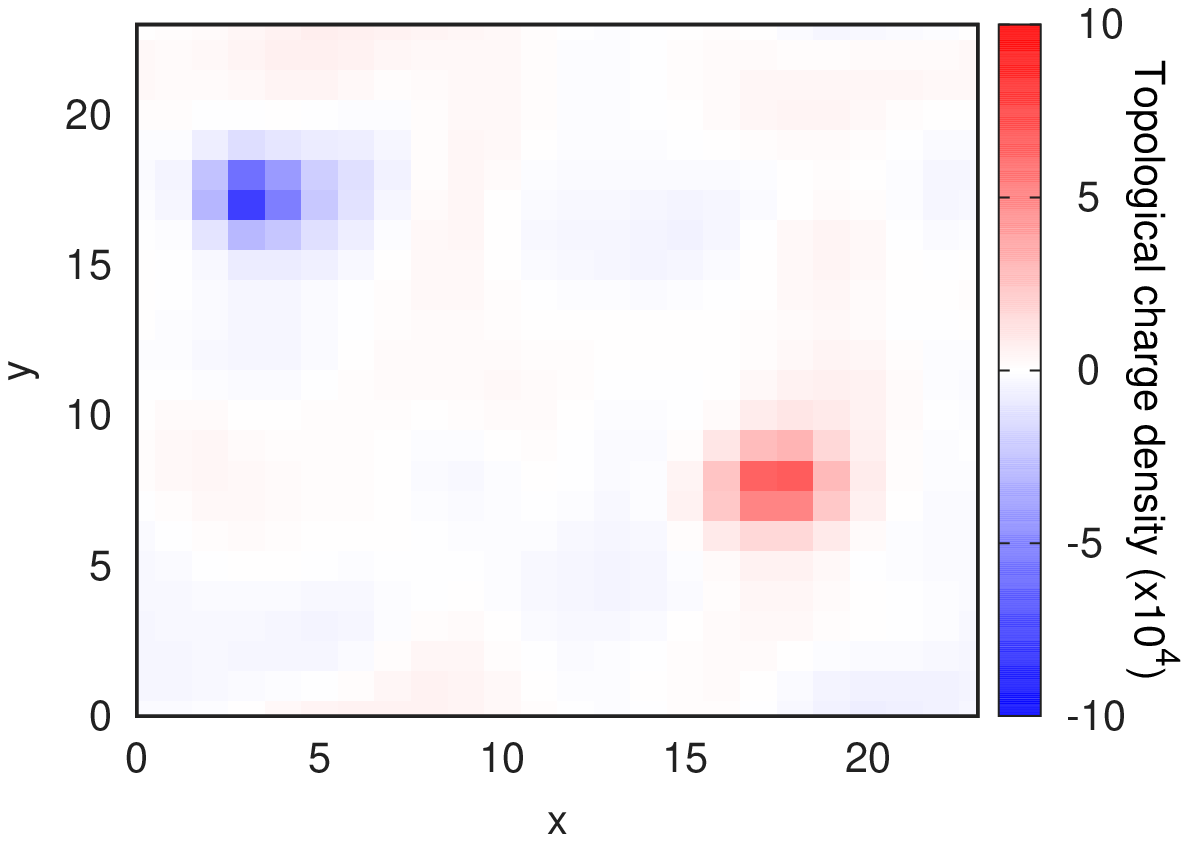}
    \caption
    {
	\label{fi:ia} Topological charge density $q(x)$ distribution on an IA
	configuration. The $q(x)$ is averaged over the $z$ and $t$ coordinates
	and scaled by $10^4$. The plot shows a configuration from an $n_f=3$ flavor
	overlap simulation on a $6\times 24^3$ lattice at $T=300$ MeV
	temperature.
    }
\end{figure}
For general configurations the definition of an IA pair or the number of IA
pairs is of course not unambiguous. However for sufficiently high temperatures
we observe very small modes, that are well separated from the rest of the
non-zero modes, the latter being on the scale of the temperature. We looked at
the topological charge distribution of such configurations and indeed observed
the concentration of the charge into a positive and a negative lump, see Figure \ref{fi:ia}. Since
such objects produce small complex conjugate pairs in the overlap operator
spectrum, the value of the chiral condensate depends strongly on the presence
of IA pairs. Therefore it is important to know, what is the fraction of
configurations with IA pairs.

\begin{figure}[h]
    \centering
    \includegraphics*{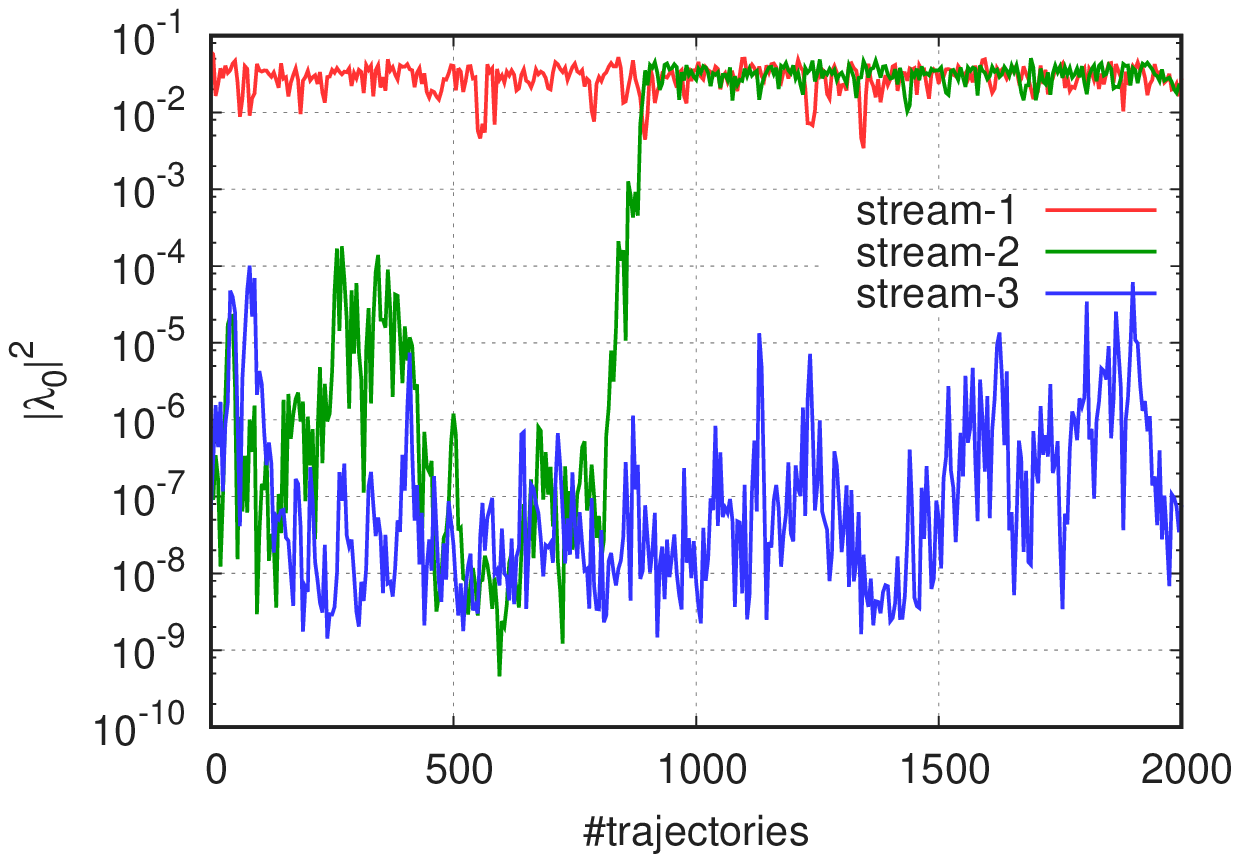}
    \caption
    {
	\label{fi:ovhist} Lowest eigenmode squared $|\lambda|^2$ of the overlap
	Dirac operator in three different Monte-Carlo streams in the trivial
	topological sector. The plot shows $n_f=3$ flavor overlap simulations
	on a $6\times24^3$ lattice at $T=300$ MeV temperature.
    }
\end{figure}
For an overlap fermion with a topology fixing term, configurations with a
well-separated IA pair pose the following problem. Annihilating such a pair
cannot proceed by simply removing the instanton and the anti-instanton
one-by-one, since this would change the topological sector. Either they have to
be removed simultaneously or they have to be brought to the same position, where
they can annihilate. If
the volume is large this latter can be difficult to achieve. In unfortunate cases
we are stuck with an IA pair, and do not sample the probability distribution
correctly.  In our concrete numerical simulations we encountered this problem
at only one parameter set: at the mass of the strange quark, $T=300$ MeV and an
aspect ratio of $N_s/N_t=4$. For the history of the lowest eigenvalue in three
different Monte-Carlo streams see Figure \ref{fi:ovhist}. ``stream-1'' contains
no, ``stream-3'' one IA pair, in ``stream-2'' there was an IA annihilation
after about 900 trajectories.  For smaller masses and larger temperatures the
runs always ended up without having IA pairs after a short thermalization time.

\begin{figure}[h]
    \centering
    \includegraphics*{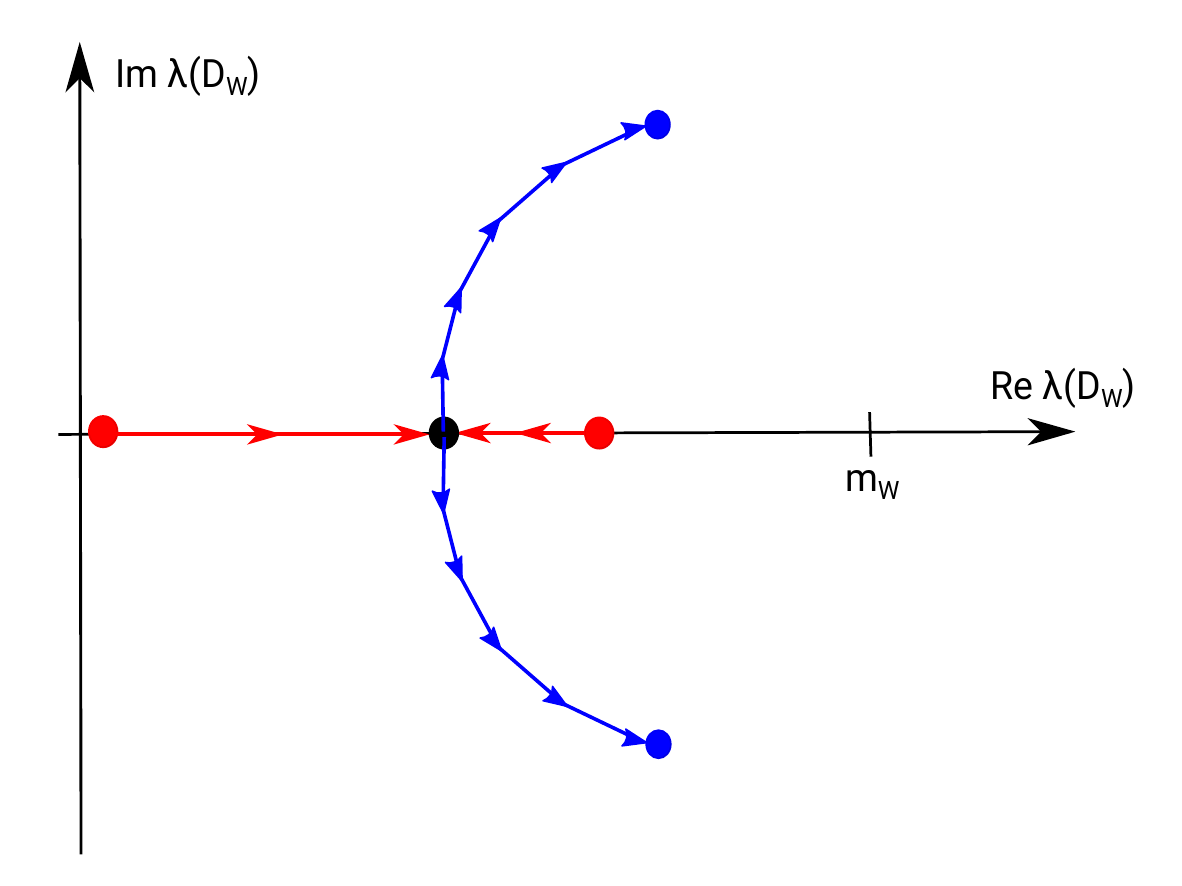}
    \caption
    {
	\label{fi:Dw} Illustration of an IA annihilation in the spectrum of the
	Wilson-Dirac operator, which is used in the kernel of the overlap Dirac
	operator.
    }
\end{figure}

It is interesting to look at the eigenvalues of the overlap kernel operator on
IA configurations.  In our case the kernel is a Wilson-Dirac operator with a
negative mass: $D_W-m_W$.  In the presence of an IA pair $D_W$ will have two
real modes lying between $0$\dots$m_W$ and with opposite chiralities.  To
annihilate the IA pair, the two real modes have to be placed to a different
region in the spectrum. For an illustration see the plot in Figure \ref{fi:Dw}. 
The simplest way, i.e. to move them towards larger real values, is not possible.
This is because, the effect of the topology fixing term is to forbid real
modes to go through the point $m_W$.  So the only way they can disappear is, to
move into the complex plane. However due to the $\gamma_5$-hermiticity of the
$D_W$ operator, complex eigenvalues have to come in complex conjugate pairs.
So in order to leave the real axis, the two real eigenvalues first have to
become degenerate, it is only then possible for them to go into the complex plane.

\begin{figure}[h]
    \centering
    \includegraphics*{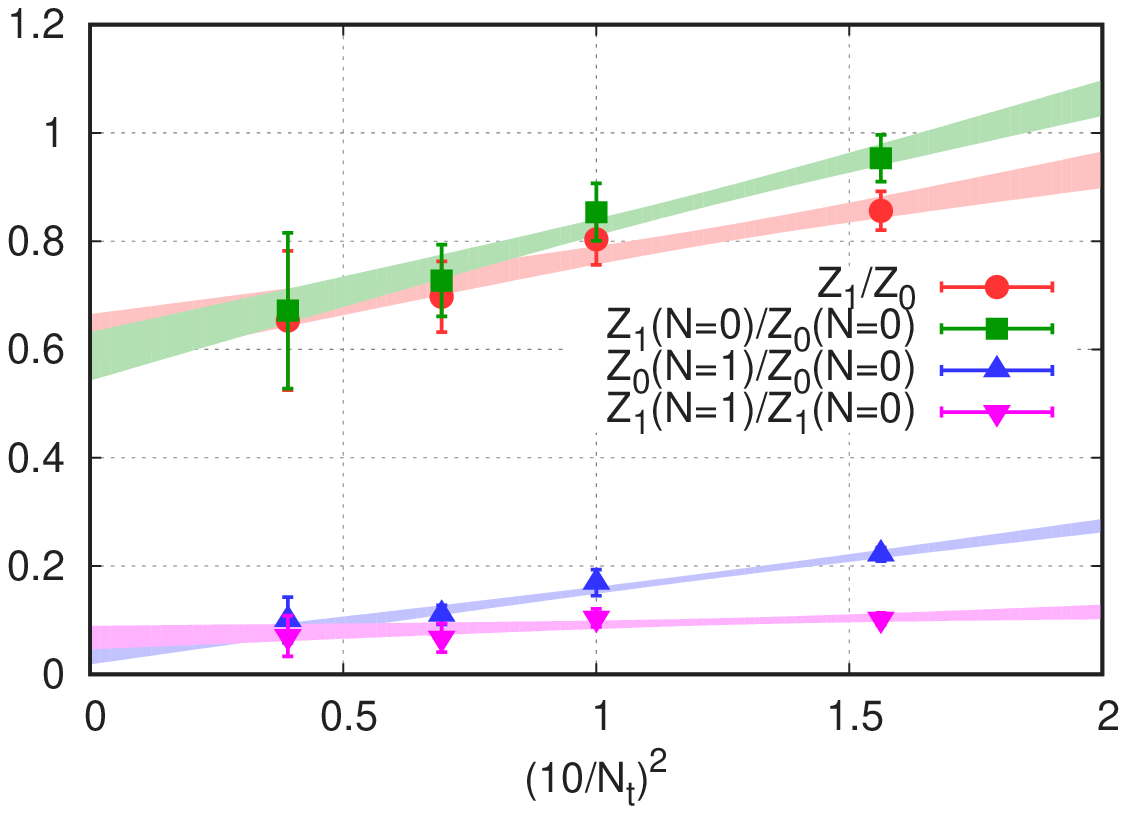}
    \caption
    {
	\label{fi:zstagg}Continuum extrapolation of partition function ratios.
	The lower index in $Z_Q$ stands for the topological charge, $N$ for the
	number of IA pairs. The plot shows $n_f=3+1$ flavor staggered simulations
	on lattices with $N_s/N_t=4$ at $T=300$ MeV temperature.
    }
\end{figure}
For the single problematic overlap run, mentioned above, we calculate the
contribution of the IA configurations as follows. We measure the weight of
such configurations with staggered fermions in the continuum limit and with the
same physical parameters ($m=m_s$, $T=300$ MeV and $LT=4$). We use the direct
approach without fixing the topology to generate configurations, which is still
efficient at these parameters. To measure the number of topological objects we
use a smeared overlap operator with kernel mass parameter $m_W=-1.3$. To define
the number of IA pairs, $N$, we counted the number of complex conjugate pairs,
for which the eigenvalue satisfied $|\lambda|^2<10^{-4}$. We checked that the
results are not sensitive to small variations of the upper bound exponent. The
results as a function of the lattice spacing can be seen in Figure
\ref{fi:zstagg}. Although the probability of $N>0$ configurations is
non-negligible, somewhat less than 10\% at this particular parameter
set, their contribution drops out of the ratio $Z_1/Z_0$ and
\begin{align}
    \label{eq:z10pz00}
    \frac{Z_1}{Z_0}=\left.\frac{Z_{1}}{Z_{0}}\right|_{N=0}
\end{align}
holds to a very good accuracy in the continuum limit.  We also find, that the
weight of configurations depends only on the total number of topological
objects, i.e. only on $Q+2N$.  Based on this, as we increase the temperature or
decrease the mass, we expect that in a given topological sector the IA pairs
will have a decreasing contribution, like the contribution of
non-trivial topological sectors decreases compared to the trivial sector. Since
all our overlap runs use either a larger temperature or a smaller mass, than
this particular simulation, we can safely neglect the contribution of IA
configurations in all of our overlap simulations and use Equation
\eqref{eq:z10pz00} to calculate sector weights.

\subsection{The chiral condensate difference}

The chiral condensate can be decomposed into the eigenmodes of the Dirac-operator.
The contribution of each topological mode is $1/m_f$. In the infinite temperature
limit the rest of the spectrum does not contribute and therefore
\begin{align}
    \label{eq:pbp1}
    m_f \langle \overline{\psi}\psi_f\rangle_{1-0} = 1.
\end{align}
For finite temperatures we expect corrections to this result.  We also expect,
that with decreasing quark mass the topological contribution will dominate and
the corrections to Equation \eqref{eq:pbp1} will get smaller.  To investigate the
size of these corrections we have carried out overlap simulations for a wide
range of parameters. These are given in Table \ref{tab:ov}. 

\begin{table}[p]
    \centering
    \begin{tabular}{|c|c|c|c|c||c|}
	$\beta$ & $N_{s}\times N_{t}$ & $m_{ud}$ & $m_{s}$ & \# ktraj & $\frac{1}{2}m_{ud} \langle \overline{\psi}\psi_{ud}\rangle_{1-0}$\\
	\hline
	\multicolumn{6}{|c|}{$m_{ud}$-scan at $T=300$ MeV}\\
	\hline
	3.99 & $12\times6$ & 0.0690 & 0.0690 & 10 & 1.00(1)\\
	3.99 & $12\times6$ & 0.0460 & 0.0690 &  5 & 0.99(1)\\
	3.99 & $12\times6$ & 0.0172 & 0.0690 &  8 & 1.00(1)\\
	3.99 & $12\times6$ & 0.0069 & 0.0690 & 10 & 1.00(1)\\
	\hline
	\multicolumn{6}{|c|}{$m_{ud}$-scan at $T=450$ MeV}\\
	\hline
	4.19 & $12\times6$ & 0.0389 & 0.0389 & 10 & 1.00(1)\\
	4.19 & $12\times6$ & 0.0291 & 0.0389  & 6 & 1.00(1)\\
	4.19 & $12\times6$ & 0.0259 & 0.0389  & 3 & 1.00(1)\\
	4.19 & $12\times6$ & 0.0195 & 0.0389  & 3 & 1.00(1)\\
	4.19 & $12\times6$ & 0.0097 & 0.0389  & 3 & 1.00(1)\\
	4.19 & $12\times6$ & 0.0049 & 0.0389  & 3 & 1.00(1)\\
	\hline
	\multicolumn{6}{|c|}{$m_{ud}$-scan at $T=650$ MeV}\\
	\hline
	4.38 & $12\times6$ & 0.0242 & 0.0242 & 5 & 1.00(1)\\
	4.38 & $12\times6$ & 0.0181 & 0.0242 & 5 & 1.00(1)\\
	4.38 & $12\times6$ & 0.0161 & 0.0242 & 3 & 1.00(1)\\
	4.38 & $12\times6$ & 0.0121 & 0.0242 & 2 & 1.00(1)\\
	4.38 & $12\times6$ & 0.0060 & 0.0242 & 2 & 1.00(1)\\
	\hline
	\multicolumn{6}{|c|}{$N_t$-scan}\\
	\hline
	3.99 & $12\times6$   & 0.0690 & 0.0690 & 12 & 1.00(1) \\
	4.13 & $16\times8$   & 0.0458 & 0.0458 & 29 & 1.02(2) \\
	4.24 & $20\times10$  & 0.0342 & 0.0342 & 80 & 1.00(1) \\
	\hline
	\multicolumn{6}{|c|}{$N_s$-scan}\\
	\hline
	3.99 & $12\times6$  & 0.0690 & 0.0690 & 12 & 1.00(1) \\
	3.99 & $16\times6$  & 0.0690 & 0.0690 & 20 & 1.00(1) \\
	3.99 & $20\times6$  & 0.0690 & 0.0690 & 32 & 1.02(1) \\
	3.99 & $24\times6$  & 0.0690 & 0.0690 & 48 & 1.00(1) \\
    \end{tabular}
    \caption
    {
	\label{tab:ov}
	Gauge coupling parameter, lattice size, quark masses and number of thousand
	trajectories for the 2+1 flavor overlap simulations at finite temperature.
	Last column contains the chiral condensate difference.
    }
\end{table}

First we calculated the corrections to Equation \eqref{eq:pbp1} for light-quark
masses in the range $m_{ud}/m_s=0.1$\dots$1$, we refer to it as
``$m_{ud}$-scan'' in the table.  We fixed the strange mass and used lattices of
fixed size $6\times12$ and used three different temperatures. Then we looked at
the lattice spacing dependence of the results, at $T=300$ MeV at the three
flavor point, we call it ``$N_t$-scan''. Finally at the same temperature and
quark mass we investigated the finite size effects, these runs are called
``$N_s$-scan''. In all cases we found, that Equation \eqref{eq:pbp1} holds with
an accuracy of about one percent.  The results are given in the last column of
Table \ref{tab:ov}.

\section{Analysis for the topological susceptibility}
\label{se:som_ana}

We combine all approaches developed in the other sections to obtain our final
result for the continuum extrapolated topological susceptibility at the
physical point. Figure \ref{fi:chi} in the main text shows this result
including statistical and systematic error estimates.

\begin{figure}[h]
    \centering
    \includegraphics*{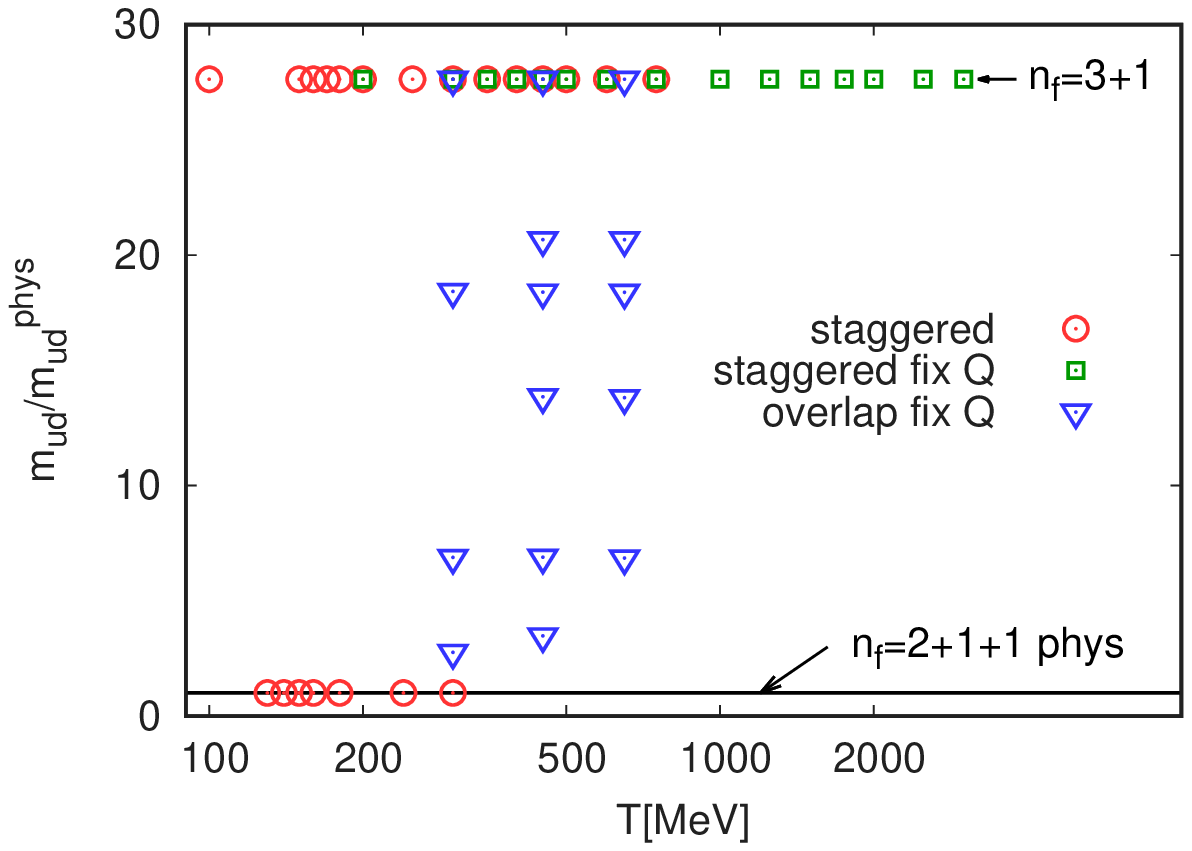}
    \caption
    {
	\label{fi:sims}Summary plot of simulation points to determine $\chi$.
	There are staggered simulations in the $n_f=3+1$ and $n_f=2+1+1$ flavor
	theories, which are then connected by overlap simulations.
    }
\end{figure}

In Figure \ref{fi:sims} we show the simulation points that were used in the
analysis. The plot shows the temperature -- light-quark mass ($m_{ud}$) plane. Four
simulation sets can be distinguished:
\begin{enumerate}
    
    \item $n_f=3+1$ flavor staggered simulations in the region $T=100\dots750$ MeV

    \item $n_f=3+1$ flavor staggered simulations at fixed topology in the
	region $T=200\dots3000$ MeV

    \item $n_f=2+1$ flavor overlap simulations at fixed topology for
	temperatures $T=300,450$ and $650$ MeV building a bridge between the
	three flavor and the physical theories

    \item $n_f=2+1+1$ flavor staggered simulations at the physical point for
	temperatures $T=130\dots300$ MeV
 
\end{enumerate}
The main feature of our strategy is that the majority of the simulations are
done in the three flavor symmetric theory (1. and 2.) instead of gathering
statistics at the physical point. Starting from a temperature of $T\sim300$ MeV
the difference between the two can be taken into account by rescaling the
topological sector weights, which is justified by the results of the overlap
simulations (3.) connecting the two theories. 
The main observation is, as expected, that for large temperatures $\chi$ scales with the mass. Since
it works on the one percent level already at $300$~MeV, there is no need to
go beyond $650$~MeV for these bridging simulations.
In the transition region the
scaling behavior of the susceptibility with the quark mass is expected to change, so for
temperatures $T\lesssim300$ MeV we still resort to direct simulations at the
physical point (4.).

\subsection{Topological susceptibility for $n_f=3+1$ flavors}

In the region between $T=100$ and $750$ MeV we have performed new simulations at six
different lattice spacings, $N_t=6,8,10,12,16$ and $20$ for direct measurements
of $\chi$.  The aspect ratio was set to $N_s/N_t=4$.  The simulation points
together with the statistics are given in Table \ref{ta:nf3}. On these
configurations we calculated the low-lying eigenvalues of the Dirac operator.

\begin{table}[h]
    \centering
    \begin{tabular}{|c|c|c|}
	\hline
	$N_t\times N_s$ & T[MeV] & ktraj \\
	\hline
	$6\times24$ & 200 & 170\\
	            & 250 & 260\\
	            & 300 & 360\\
	            & 350 & 380\\
	            & 400 & 480\\
	            & 450 & 490\\
	            & 500 & 490\\
	            & 600 & 490\\
	            & 750 & 490\\
	\hline
	$8\times32$ & 200 & 100\\
	            & 300 & 100\\
	            & 350 & 70\\
	            & 400 & 70\\
	            & 450 & 140\\
	            & 500 & 100\\
	            & 600 & 480\\
	            & 750 & 460\\
	\hline
    \end{tabular}
    \hspace*{3cm}
    \begin{tabular}{|c|c|c|}
	\hline
	$N_s\times N_t$ & T[MeV] & ktraj \\
	\hline
	$10\times40$ & 200 & 140\\
	             & 250 & 80\\
	             & 300 & 150\\
	             & 350 & 400\\
	             & 400 & 400\\
	             & 450 & 860\\
	             & 500 & 1300\\
	\hline
	$12\times48$ & 200 & 110\\
	             & 250 & 200\\
	             & 300 & 270\\
	             & 350 & 410\\
	             & 400 & 470\\
	             & 450 & 620\\
	             & 500 & 340\\
	\hline
	$16\times64$ & 200 & 10\\
	             & 250 & 30\\
	             & 300 & 180\\
	\hline
	$20\times80$ & 200 & 10\\
	\hline
    \end{tabular}
    \caption
    {
	\label{ta:nf3}$n_f=3+1$ flavor staggered simulations points for direct
	measurements of $\chi$. Lattice geometry, temperature and number of
	trajectories are given (in thousands).
    }
\end{table}

\begin{figure}[h]
    \centering
    \includegraphics{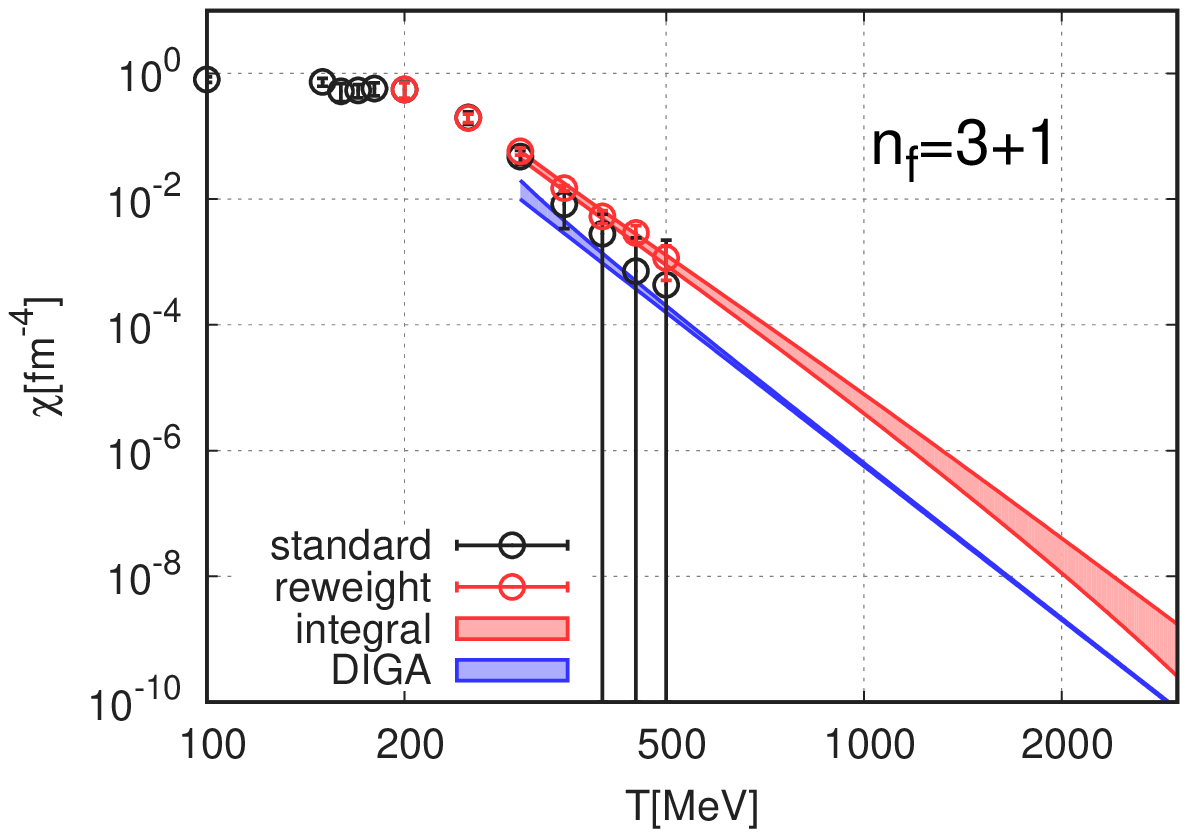}
    \caption
    {
	\label{fi:chi3}
	Continuum extrapolated topological susceptibility as a function of the temperature in the three flavor
	symmetric theory. The result was obtained from $n_f=3+1$ flavor staggered
	simulations after performing the continuum limit. Lattice results of two direct
	methods (standard and reweighted) and of the fixed sector integral are shown. Also
	shown is the prediction of the DIGA.
    }
\end{figure}

We employ the reweighting method, as described in Section \ref{se:som_rw}, to
decrease the lattice artefacts.  The topological charge was measured using
Wilson-flow.  The systematic errors were determined from varying the definition
of the charge (plain or rounded to the nearest integer), from the choice of
points considered in the continuum extrapolation (including a change of the fit
range), and from the parametrization of the lattice artefacts.  This gave us in
total 20-64 fits per temperature. We used them in our histogram method to
determine the systematic uncertainties~\cite{Durr:2008zz,Borsanyi:2014jba}. In
Figure \ref{fi:chi3} we show the continuum extrapolated results both with and
without reweighting. The large errors without reweighting are coming from the
continuum extrapolation.

To reach the temperature region that is needed for axion phenomenology we
employ the fixed sector integral method developed in Sections \ref{se:som_ym}
and \ref{se:som_st}. In this way no extrapolation in the temperature is needed.
For this we generated configurations in topological sectors $Q=0$ and $1$, at
four different lattice spacings $N_t=4,6,8$ and $10$ and aspect ratios
$N_s/N_t=2\dots6$. The simulation points are given in Table \ref{ta:nf3int}.
The number of trajectories are given for all temperatures together.

\begin{table}[h]
    \centering
    \begin{tabular}{|c|l|c|}
	\hline
	$N_t\times N_s$ & T[MeV] & Mtraj \\
	\hline
	$ 4\times 8$ & 300, 350, 400, 450, 500, 600, 750, 1000, 1250, 1500, 1750, 2000, 2500, 3000 & 1\\
	$ 4\times12$ & 750, 1500 & 2\\
	$ 4\times16$ & 750, 1500 & 3\\
	$ 4\times24$ & 750, 1500 & 9\\
	\hline
	$ 6\times12$ & 300, 350, 400, 450, 500, 600, 750, 1000, 1250, 1500, 1750, 2000 & 30\\
	$ 8\times16$ & 300, 350, 400, 450, 500, 600, 750, 1000, 1250, 1500 & 56\\
	$10\times20$ & 300, 400, 500, 600, 750, 1000, 1250, 1500 & 20\\
	\hline
    \end{tabular}
    \caption
    {
	\label{ta:nf3int}$n_f=3+1$ flavor staggered simulation points with
	fixed topology, $Q=0,1$ for measuring $\chi$
	with the fixed sector integral method. Lattice geometry, temperature values and
	total number of trajectories are given (in millions).
    }
\end{table}

\begin{figure}[h]
    \centering
    \includegraphics{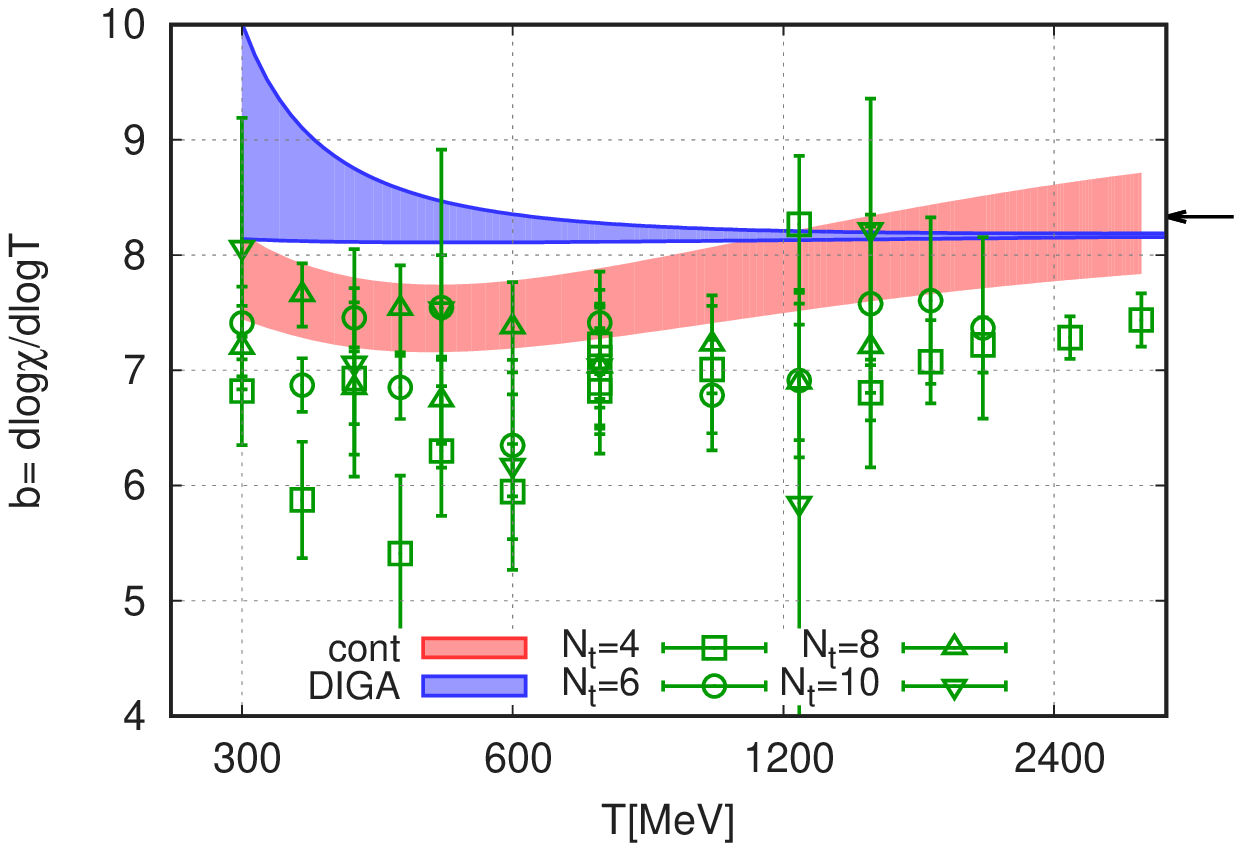}
    \caption
    {
	\label{fi:baT}
	Temperature dependence of the exponent $b=d\log\chi/d\log T$. The plot
	shows $n_f=3+1$ flavor staggered simulations on $N_t=4,6,8$ and $10$
	lattices. The red band is the continuum extrapolation. 
	The prediction of the $n_f=3+1$ flavor DIGA
	is given by the blue band.
	The arrow shows
	the Stefan-Boltzmann limit. 
    }
\end{figure}

We first determined the continuum extrapolation of the exponent $b=
d\log\chi/d\log T$. For this we use a combined fit to all of the data which
includes also infinite volume extrapolation.  We use 16 ans\"atze which are
combinations of four choices of the temperature dependence (polynomials in
$1/T$ up to third order), two choices for the parametrization of the lattice
artefacts (none or linear in $1/N_t^2$) and two for the parametrization of
finite volume effects (none or linear in $(N_t/N_s)^3$). The different fits are
combined using the Akaike Information Criterion (AIC) and also yield a
systematic error estimate. For a more detailed exposition of this procedure,
see \cite{Borsanyi:2014jba}.  The result is shown in Figure \ref{fi:baT}. We
also calculated the prediction of the DIGA for $n_f=3+1$ flavors. For this we
took the strange mass from \cite{Durr:2010vn} and $m_c/m_s=C$ from
Equation \eqref{eq:oldlcp}, since this ratio was used in the simulations. We
took $\Lambda^{(4)}_{\overline{\mathrm{MS}}}$ from \cite{Agashe:2014kda} to
convert the perturbative results to physical units.  The renormalization scale
dependence was estimated by using three different scales: $1$, $1/\sqrt{2}$ and
$\sqrt{2}$ times $\pi T$. The continuum extrapolated
lattice results agrees with the DIGA for temperatures above $T\sim 1000$ MeV,
whereas for smaller $T's$ the lattice gives somewhat smaller exponents. This behaviour is qualitatively similar to the quenched case.

To obtain $\chi$ an integration in the temperature has to be performed, see
Equation \eqref{eq:int}. We start with a continuum extrapolated value from the
low temperature region and then use the continuum extrapolation of the data for
the exponent $b$ to perform the integration. The systematic error is derived
from the systematic error at the starting point of the integration and from
three choices of the starting point ($300$, $350$ and $400$ MeV). The result is
then plotted together with the direct simulations at lower temperature in
Figure \ref{fi:chi3} and also with the prediction of the DIGA. Similarly
to the quenched case, Figure \ref{fi:z1z0}, the lattice result
is considerably larger than the DIGA predicition.
At $T=500$ MeV the ratio
of the two is $K=6.1(1.1)(0.7)$, where the first error comes from the lattice,
the second from the scheme dependence of the DIGA.

\subsection{Topological susceptibility for $n_f=2+1+1$ flavors}

For the zero temperature susceptibility we applied a method based on leading
order chiral perturbation theory to remove the lattice artefacts. This is
described in Section \ref{se:som_zero}.

At finite temperature we can start from the topological sector weights in the $n_f=3+1$ theory.
The result for $n_f=2+1+1$ flavors is obtained by performing an integration in the
light-quark mass. For example, the relative weight of sectors $Q=0$ and $1$ can be calculated as:
\begin{align}
    \label{eq:mint}
    \left.\frac{Z_1}{Z_0}\right|_{2+1+1}=
    \exp\left( \int_{m_{ud}^{phys}}^{m_s^{phys}} d\log m_{ud}\ m_{ud}\langle \overline{\psi}\psi_{ud}\rangle\right)\cdot
    \left.\frac{Z_1}{Z_0}\right|_{3+1}
\end{align}
The overlap simulations in Section \ref{se:som_ov} provided ample
evidence, that above $T=300$ MeV, to a very good precisions the integrand
is given by the number of light flavors $m_{ud}\langle
\overline{\psi}\psi_{ud}\rangle=2$. Thus the sector
weights at the physical point are given by the following scaling:
\begin{align}
    \label{eq:rescale}
    \left.\frac{Z_1}{Z_0}\right|_{2+1+1}= R^2\cdot\left.\frac{Z_1}{Z_0}\right|_{3+1},
\end{align}
with $R$ given in Equation \eqref{eq:oldlcp}.

Equation \eqref{eq:mint} can of course be used at any temperature, but the simple
scaling with $R^2$ will not work e.g. at low temperatures or in the transition
region. To be on the safe side for temperatures below $T=300$ MeV we decided to
fall back on the direct measurement at the $n_f=2+1+1$ physical point, which
turned out to be feasible even on the already existing statistics from
\cite{Bellwied:2015lba}. As was described in Section \ref{se:som_rw}
reliable continuum extrapolations could only be performed after applying the
reweighting procedure.

\subsection{Topological susceptibility - full result}

There are two effects, that are missing in the $n_f=2+1+1$ flavor theory, and
have to be taken into account to obtain the full result for $\chi$: the presence
of the bottom quark and the mass difference between the up and the down quarks.

As we have seen in Section \ref{se:som_st}, the charm contribution to the decay
exponent $b$ has almost reached the high temperature limit at $T=300$ MeV.  We
also found that the charm starts to contribute to the equation of state at
$T\sim 250$ MeV (see Section \ref{sec:eosresult}).  We therefore expect that the
bottom contribution starts to be appreciable at temperatures above
$\sim m_b/m_c\times 250$ MeV.  To take into account the bottom contribution, we
added $1/3$ to the $n_f=2+1+1$ flavor exponent for temperatures higher than
some threshold temperature.  The value $1/3$ is the contribution of an extra
flavor to the high temperature limit. Then we integrated the so obtained $b$
and finally rescaled the results according to Equation \eqref{eq:rescale}. We
have chosen three different threshold temperatures: $T=1.0, 1.5$ and $2.0$ GeV.
The resulting variation in $\chi$ was added to the systematic error.

Although isospin violating effects are typically on the level of 1\%, the
topological susceptibility is a notable exception. This is because, the
susceptibility is proportional to the product of the quark masses. Therefore the topological
susceptibility is a factor of
\begin{align}
    \frac{4m_um_d}{(m_u+m_d)^2} \approx 0.88
\end{align}
smaller than in the isospin symmetric, $n_f=2+1+1$ flavor case. The quark mass
values were taken from \cite{Fodor:2016bgu}. To take isospin violation into
account we scaled the isospin symmetric results by this factor for all
temperatures.

Our final result for $\chi(T)$ is shown in Figure \ref{fi:chi} of the main
text. We also tabulate the base-10 logarithm $-\log_{10}\chi(T)$ for a couple
of temperature values in Table \ref{ta:chi}.

\begin{table}
    \centering
    \begin{tabular}{|c|c|}
	\hline
	$T$[MeV] & -$\log_{10}(\chi$[fm$^{-4}$]) \\
	\hline
	100 & -1.66(5)\\
	120 & -1.65(7)\\
	140 & -1.75(9)\\
	170 & -2.18(5)\\
	200 & -2.72(6)\\
	240 & -3.39(6)\\
	290 & -4.11(6)\\
	350 & -4.74(6)\\
	420 & -5.34(7)\\
	500 & -5.90(8)\\
	600 & -6.49(9)\\
	720 & -7.08(11)\\
	860 & -7.67(13)\\
	1000 & -8.17(15)\\
	1200 & -8.79(17)\\
	1500 & -9.56(20)\\
	1800 & -10.20(23)\\
	2100 & -10.75(26)\\
	2500 & -11.38(28)\\
	3000 & -12.05(33)\\
	\hline
    \end{tabular}
    \caption
    {
	\label{ta:chi} Topological susceptibility of QCD taking into account the effect
	of the up, down, strange, charm and bottom quarks.
    }
\end{table}

\section{
Axion dark matter from misalignment}
\label{se:som_axion}

The details of axion production via the misalignment mechanism are
well described in the literature (see e.g. \cite{Wantz:2009it}) but for
completeness we briefly discuss our calculations.

In order to calculate the amount of axions produced we have to solve
the equation of motion for the $A(x)$ axion field or equivalently
for the $\theta(x)=A(x)/f_A$ axionic angle in an expanding universe:
\begin{equation}\label{eq:axion}
\frac{d^2\theta}{dt^2}+3H(T)\theta\frac{d\theta}{dt}+\frac{dV(\theta)}{d\theta}=0,
\end{equation}
where $V(\theta)=m_A^2(T)(1-\cos\theta)=\chi(T)/f_A^2(1-\cos\theta)$ 
is the temperature dependent axion potential. Since we focus on the misalignment
mechanism, we assume that $\theta$ changes slowly in space on the relevant
scales. Spatial fluctuations and defects lead to a string contribution which
we do not discuss here.

The expansion is governed by the Friedmann equations:
\begin{eqnarray}
H^2&=&\frac{8\pi}{3M_{Pl}^2}\rho \\
\frac{d\rho}{dt}&=&-3H(\rho+p)=-3HsT
\end{eqnarray}
where $\rho,p$ and $s$ are the energy density, pressure and entropy
density of the early universe and $M_{Pl}$ is the Planck mass. 
At the temperatures where axion production happens, the contribution of axions to these densities
can be neglected. $\rho$
and $s$ can be expressed as:
\begin{equation}
\rho=\frac{\pi^2}{30}g_\rho T^4 \quad\quad s=\frac{2\pi^2}{45}g_s T^3
\end{equation}
using the effective number of degrees of freedom of Figure \ref{fi:eos}.
Since we determined the energy density and entropy density for a wide
temperature range, the solution of these equations yields the following
relation between the age of the universe ($t$) and its temperature:
\begin{equation}
\frac{dt}{dT}=-M_{Pl}\sqrt{\frac{45}{64\pi^3}}\frac{1}{T^3g_s(T)\sqrt{g_\rho(T)}}
\left(T\frac{dg_\rho(T)}{dT} +4g_\rho(T)\right)
\end{equation}
With the help of this expression, Equation \eqref{eq:axion} can be rewritten in terms of
temperature derivatives:
\begin{equation}\label{eq:axionT}
\frac{d^2\theta}{dT^2}+\left[3H(T)\frac{dt}{dT}-\frac{d^2t}{dT^2}/\frac{dt}{dT}
\right]\frac{d\theta}{dT}+\frac{\chi(T)}{f_A^2}\left(\frac{dt}{dT}\right)^2 \sin\theta=0
\end{equation}

\begin{figure}[h]
    \centering
    \includegraphics*{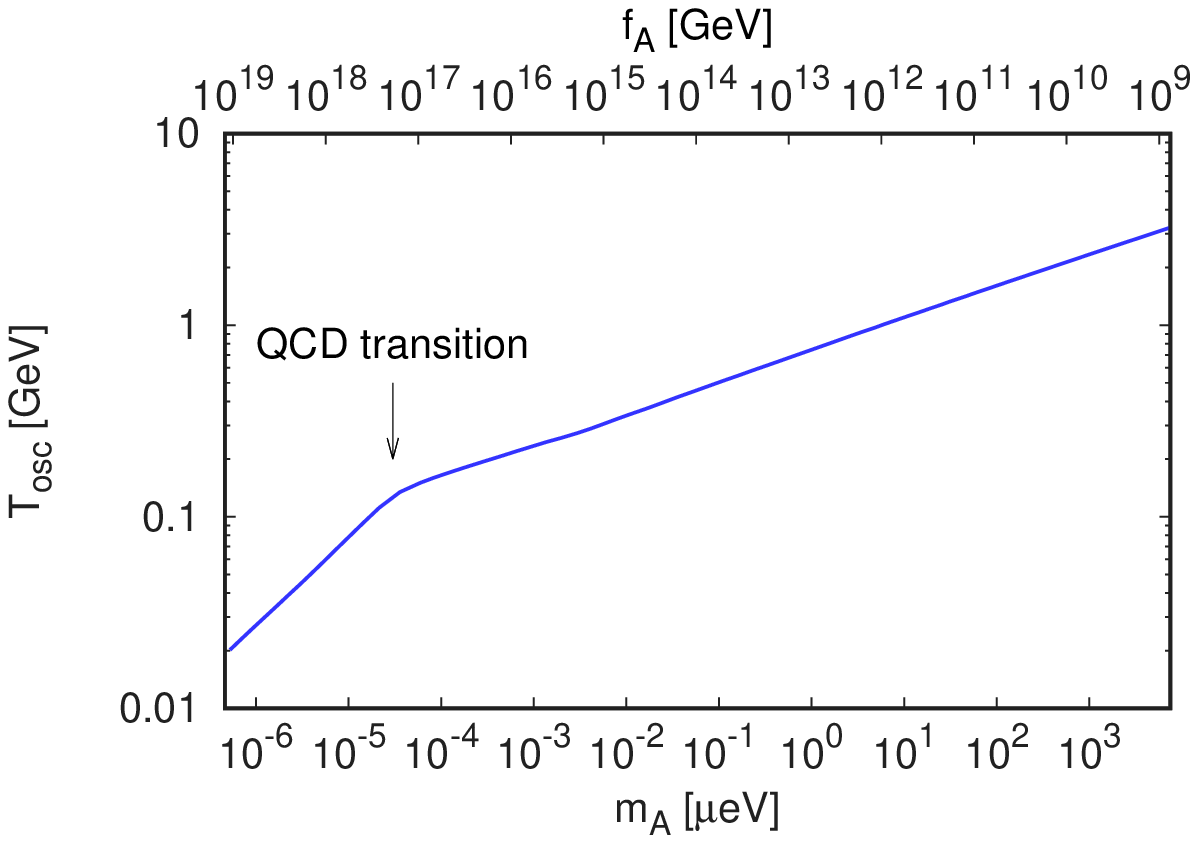}
    \caption
    {
	\label{fi:T_f}
	The oscillation temperature as a function of the axion mass. For this figure we assume that all the observed dark matter comes from the axions.
	Within this pre-inflation scenario the roll-down comes from a single $\theta_0$ angle. The bend on the figure represents the QCD transition
	temperature. It reflects the very different behaviour of $\chi(T)$.
	Above the QCD transition $\chi$ rapidly drops, whereas below the QCD transition it has a much milder --almost constant-- behaviour.
    }
\end{figure}

We solve this equation by numerical integration with some initial angle $\theta_0$
and vanishing first derivative. When the temperature is large, the $\theta$
angle is frozen to its initial value. It starts to roll down the potential
around $T_{\rm osc}$ which is defined as $3H(T_{\rm osc})=m_A(T_{\rm osc})$. 
At the same time the axion number density ($n_A$) starts to increase. After
a few oscillations its ratio to the entropy density ($n_A/s$) converges
to a finite value which is then conserved for the rest of the evolution.
Figure \ref{fi:T_f} shows $T_{\rm osc}$ for a large range of axion 
masses/couplings. Note that a coupling close to the Planck scale results
in a $T_{\rm osc}$ below the QCD phase transition which emphasizes the 
need for the equation of state and $\chi(T)$ even for 
these low temperatures. Below the transition $T_{\rm osc}\propto m_A^{0.47}$
while above the transition $T_{\rm osc}\propto m_A^{0.17}$.

We start the numerical solution at $T=5T_{\rm osc}$. The 
oscillation starts around $T_{\rm osc}$. We detect this by 
looking for the first sign change of $\theta$ which happens at $T_s$. We then 
extract $n_A/s$ by averaging 
\begin{equation}
\frac{n_A}{s}(T)=\frac{45}{2\pi^2}\frac{f_A^2}{m_Ag_sT^3}\left[\frac{1}{2}\left(\frac{d\theta}{dT}/\frac{dt}{dT}\right)^2+
\frac{\chi(T)}{f_A^2}(1-\cos\theta)\right]
\end{equation}
for the temperature range $0.8$ -- $0.2T_s$. Throughout the solution of Equation
\eqref{eq:axionT} we fix $f_A$ (or equivalently $m_A$) and
use our results for $\chi(T)$ and $\rho(T)$. The present axion energy density is obtained
by using the conservation of $n_A/s$:
\begin{equation}
    n_{A;{\rm today}}=\frac{n_A(T)}{s(T)}s_{\rm today}\quad\quad
    \rho_{A;{\rm today}}=m_A n_{A;{\rm today}}
\end{equation}
The current entropy of the universe is dominated by photons and neutrinos:
\begin{equation}
s_{\rm today}=\frac{2\pi^2}{45}(2T_\gamma^3+6\frac{7}{8}T_\nu^3)=
\frac{2\pi^2}{45}\frac{43}{11}T_\gamma^3
\end{equation}
where $T_\gamma=2.725K$ is the cosmic microwave background temperature.
This axion energy density has to be compared to the critical density or its 
dark matter component:
\begin{equation}
\Omega_{\rm axion}=\frac{\rho_{a;{\rm today}}}{\rho_{\rm crit}}\quad\quad R_A=\frac{\Omega_{\rm axion}}{\Omega_{DM}}
\end{equation}

In the pre-inflation scenario a single $\theta_0$ and $m_A$ (or $f_A$)
determines $R_A$ uniquely. Assuming $R_A=1$ results in the curve in
Figure \ref{fi:axion} of the main text. In the post-inflation scenario all $\theta_0$ angles
are present with equal probabilities after the Peccei-Quinn transition and
we have to average over them:
\begin{equation}
\overline{R_A}(m_A)=\frac{1}{2\pi}\int_{-\pi}^\pi R_A(\theta_0,m_A)d\theta_0
\end{equation}
For the whole $m_A$ range which is relevant for the post-inflation
scenario, we found that up to a few per mil the average angle of $\theta_0=2.155$ can be used:
\begin{equation}
\overline{R_A}(m_A)=R_A(\theta_0=2.155,m_A)
\end{equation}

\renewcommand\({\left(}
\renewcommand\){\right)}
\renewcommand\[{\left[}
\renewcommand\]{\right]}

\section{Experimental searches for the axion in the predicted mass region}
\label{se:som_exp}

Using lattice QCD to determine the cosmological equation of
state (EoS) and the temperature dependence of the topological
susceptibility $\chi(T)$ the present paper showed that post-inflation
dark matter axions have a mass between 50 and 1500~$\mu$eV (see
Figure~\ref{fi:axion}). In this section we show what should be done
experimentally to approach and explore this mass region and detect dark
matter axions. We conclude that, though presently operating and planned
next generation experiments are not able to cover the predicted mass
region, it is possible to design experiments, which offer discovery
potential already in the near future.

Axions in the predicted 50-1500 $\mu$eV mass range are extremely challenging to detect.  
Recently, their theoretical appeal has been increasingly recognised and new techniques and experiments have been proposed. These include axion dark matter searches exploiting the excitation of atomic transitions in tuneable Rydberg atoms~\cite{Sikivie:2014lha} and electron spin precession~\cite{Barbieri:2016vwg}, but also purely laboratory searches for virtual axion long-range mediated forces \cite{Arvanitaki:2014dfa,Crescini:2016lwj}. Unfortunately, these are model dependent. The success of the former depends on a large axion-electron coupling and the latter upon the existence of new sources of CP violation beyond the Standard Model. 

The most promising venue is to exploit the axion coupling to 
photons, ${\cal L}_{a\gamma} = - \theta {\bf E}\cdot {\bf B} \alpha  C_{A\gamma} /(2\pi)$, with $C_{A\gamma}$ being an ${\cal O}(1)$ model-dependent constant. 
The simplest model compatible with the post-inflation scenario is the
Kim-Shifman-Vainshtein-Zakharov (KSVZ) axion~\cite{Kim:1979if,Shifman:1979if},
which has $C_{A\gamma}=-1.92$  and we take it as our benchmark. 
The local axion dark matter field oscillations $\theta(t)\sim \theta_0 \cos(m_a t)$ in a homogeneous magnetic field ${\bf B}_e$ generate an electric field ${\bf E}_\theta = \theta(t) {\bf B}_e C_{A\gamma}\alpha/(2\pi)$. 

The haloscope experiment of Sikivie~\cite{Sikivie:1983ip} uses this field to drive the resonant mode of a microwave cavity when the oscillation frequency ($\nu_A=m_A/2\pi$) coincides with its resonant frequency. 
Several collaborations have already employed this technique.  
The local dark matter density $\rho_{\rm dm}\simeq 0.3$\,GeV/cm$^3$ fixes the amplitude of the oscillations $\theta_0=\sqrt{2\rho_{\rm dm}/\chi(0)}\simeq 3.6 \times 10^{-19}$ and the electric field $|{\bf E}_\theta|=1.2\times 10^{-12} (|{\bf B}_e|/10{\rm \,T})$ V/m. Since the precise axion mass still remains unknown, the cavity has to be tuned to scan over the desired mass range. The bandwidth of the signal follows from the velocity dispersion of dark matter particles in the galactic halo $\Delta\nu\sim \nu_A/Q_A$ with $Q^{-1}_A=\langle (v/c)^2\rangle/2\sim 10^{-6}$. In the mass range of interest, there are still $10^6\log(1500/50)\sim 3.4 \times 10^6$ channels to be explored in the frequency range $\nu_a=$12-363 GHz. The power extracted from the cavity on resonance is given by 
\be
P_A =  \kappa V m_A   {\rm max}\{Q,Q_A\} {\cal G} |{\bf E}_\theta|^2/2, 
\ee
where ${\cal G}=\(\int dV {\bf E}_m\cdot {\bf B}_e\)^2/(|{\bf B}_e|^2V\int dV|{\bf E}_m|^2)$ is the geometric overlap between the electric field of the cavity mode ${\bf E}_m$ with the background B-field, $Q$ the quality factor, $V$ the volume of the cavity, and $\kappa$ the coupling coefficient (ratio of the power extracted to the full cavity losses), optimally set to $\sim 0.5$. The integration time required to find this signal with a given signal-to-noise-ratio $S/N$ within the thermal and amplifier noise fluctuations is given by Dicke's radiometer equation
\be
\Delta t = \Delta \nu \(\frac{T_{\rm sys}}{P_A}\frac{S}{N}\)^2
\ee 
where $T_{\rm sys}$ is the system noise. 

\begin{figure}[htbp]
\begin{center}
\includegraphics[width=10cm]{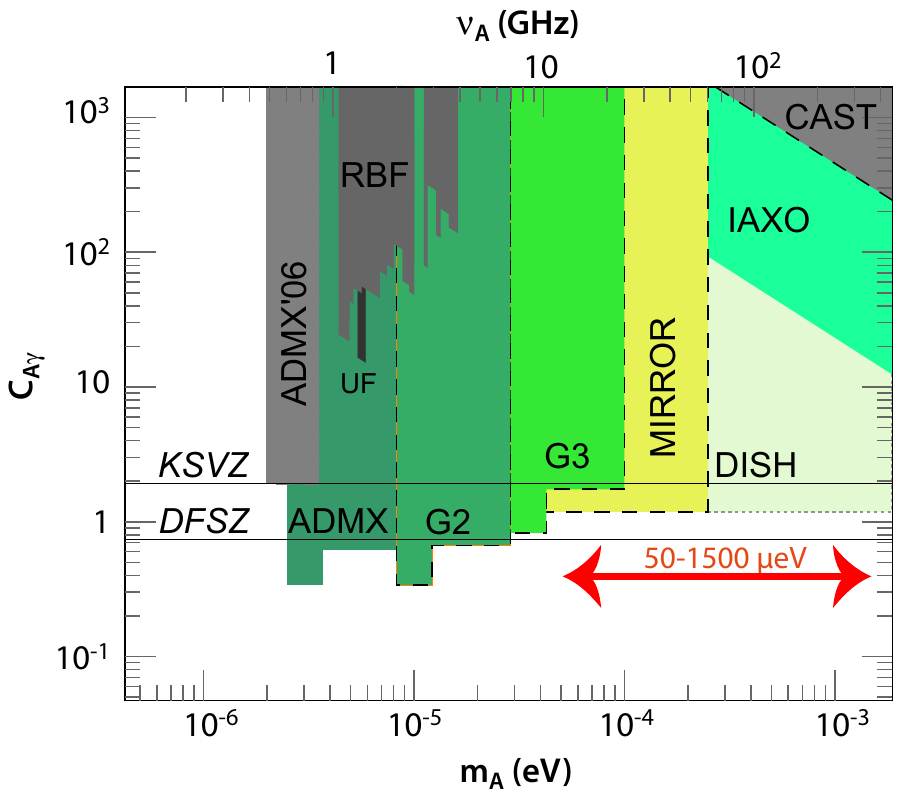}
\caption{\label{fi:sensi}
Sensitivity reach of the experiments discussed in the text as well as
the International Axion Observatory (IAXO)~\cite{Armengaud:2014gea} 
(colored regions)
with the current exclusion limits from previous cavity experiments: ADMX, 
RBF, UF and CAST (grey regions). We also show $C_{A\gamma}$ for the KSVZ 
and Dine-Fischler-Srednicki-Zhitnitsky 
(DSFZ)~\cite{Dine:1981rt,Zhitnitsky:1980tq} models most interesting in the post-inflation and pre-inflation 
scenarios, respectively, and the range of $m_A$ that could fit the dark 
matter abundance in the post-inflation scenario as follows from this work.}
\end{center}
\end{figure}

As an example, ADMX  is a state of the art and only fully commissioned experiment \cite{Asztalos:2009yp}. It utilises a cylindrical cavity (1\,m long, 0.5\,m diameter, $Q\sim 10^5$) in an 8\,T magnetic solenoid in a dilution refrigerator reaching 100\,mK and SQUID amplifiers with noise close to the quantum limit. A measurement campaign of three years is being started and has the sensitivity to find dark matter axions in the pre-inflation scenario in the region labelled ADMX in Fig.\,\ref{fi:sensi}.
Generation 2 (G2) experiments to reach higher frequencies are currently under preparation by the ADMX HF-group and the Center of Axion and Precision Physics (CAPP) in South Korea. Our estimated sensitivities with  $Q\sim10^6$ and with cavities operated in fields of up to 20\,T may discover axions in the pre-inflation scenario up to $m_A=30\,\mu$eV, see G2 region in Fig.\,\ref{fi:sensi}. 
The post-inflation scenario predicted in this paper may only be partially explored by presently envisaged Generation 3 (G3) experiments. Still, this would require magnetic fields as strong as 40\,T and combining signals of several tuneable cavities. At this moment it is not clear at all if the required technologies will ever be available for such a search. Nevertheless we include this G3 region in Fig.\,\ref{fi:sensi}. 

An alternative method was proposed in a recent paper~\cite{Horns:2012jf}. A spherical mirror in a strong magnetic field was shown to emit electromagnetic waves of frequency $\nu_A$ that focus at the center of curvature in response to the oscillating axion dark matter field. 
The power per unit dish area is
\be
\frac{P_A}{A} = \frac{|{\bf E}_\theta|^2}{2}=2.2 \times 10^{-27}\frac{\rm W}{\rm m^2}\(\frac{C_{A\gamma} |{\bf B}_e|}{10\, \rm T}\)^2,  
\ee 
too small for a wide-band search. 
However, it has been pointed out that the power can be enhanced by exploiting a dielectric planar mirror made of a sequence of $N$ dielectrics~\cite{Jaeckel:2013eha}. An equivalent power per area is emitted by each dielectric interface and can be added up coherently. 
This increases the power by a factor $4N^2$ (using a mirror at one end) which can be focused into a microwave receiver by a parabolic dish like the one used in~\cite{Suzuki:2015vka}. Detuning the dielectric thickness from $\lambda/2$, the dielectrics become partially reflecting, the power stored builds up like in a resonant cavity and the boost factor can be increased significantly so that a realistic axion search becomes feasible \cite{MADMAX}.  

Here, we envision a variable set of 20-40 sapphire dielectric slabs of $\sim 1$\,mm thickness and 1\,m$^2$ transverse area placed in a 10\,T  magnetic field with a planar mirror at one side.  The distance between the dielectrics can be adjusted to have boost factors of order $\beta\sim 10^5$ in a relatively broad band ${\cal O}(50)$ MHz~\cite{MADMAX}. Typically, one needs $d\sim \lambda/2$, which ranges between $3.1$ cm and $2.4$ mm in the axion mass range $40-250 \, \mu$eV. 
For $m_A=250\, \mu$eV, the coherence length of the axion field reaches  $(m_A v)^{-1}\sim 1$\,m. Thus, coherent detection with such large axion-photon transducers can be severely hampered for larger masses. 
A 3 year measurement campaign with such an apparatus may scan the 50-100 $\mu$eV range with sensitivity to KSVZ axions with commercial 
low-noise high-electron-mobility transistor (HEMT) amplifier technology and up to 250 $\mu$eV with quantum limited detection. The reach of such a tunable dielectric mirror is shown in Fig.~\ref{fi:sensi} as a green-yellow band. The feasibility of an 
experiment of this type is currently being assessed at the Max-Planck-Institute for Physics in Munich.

This problem of coherence suggests us to reconsider the spherical dish antenna idea. Reaching sensitivity to KSVZ axions at 1\,meV with a plain dish in a magnetic field of 10\,T requires a sensitivity of 2.2 photons/(m$^2$ day). Again, the yield can be increased by a factor $4N^2$ when a few dielectric slabs are mounted (at adjustable relative distances) in front of each (planar) mirror element. With a boost of only ${\cal O}(100)$ and a total mirror area of 5\,m$^2$ the photon rate would increase to $10^{-2}$\,Hz with appears technically feasible for devices cooled down to temperatures of 10\,mK and operating near the quantum limit \cite{Barbieri:2016vwg}. Still, it will be very challenging to shield the entire setup sufficiently against thermal noise.

In summary, we have shown that the region of high axion masses predicted by the QCD lattice calculations in this paper remains largely unexplored by presently operating and planned next generation experiments. This is mostly because of technical and practical limitations, particularly when searches over large ranges of axion masses are attempted. However, new experimental directions, some of which have been discussed in this paper and still being very challenging, may offer discovery potential already in the near future.

\printbibliography[prefixnumbers=S]

\end{refsection}

\end{document}